\documentclass[]{aa}  
\usepackage{graphicx}
\usepackage{caption}
\usepackage{subcaption}
\usepackage{txfonts}
\usepackage{diagbox}
\usepackage[colorlinks=true, citecolor=blue]{hyperref}

\usepackage{physics}

\begin{document} 

\title{Morphology and dynamical stability of self-gravitating vortices:}
\subtitle{Numerical simulations}

\author{S. Rendon Restrepo \inst{1,2}
\thanks{\email{steven.rendon-restrepo@lam.fr}}
\and
P. Barge \inst{1}
\thanks{\email{pierre.barge@lam.fr}}
}

\institute{Aix Marseille Univ, CNRS, CNES, LAM, Marseille, France
\and 
Aix Marseille Univ, CNRS, Centrale Marseille, IRPHE, Marseille, France}

   \date{Received March 10, 2022; accepted June 20, 2022}

 
\abstract
{Theoretical and numerical studies have shown that large-scale vortices in protoplanetary discs can result from various hydrodynamical instabilities. Once produced, such vortices can survive nearly unchanged over a large number of rotation periods, slowly migrating towards the star.  Lopsided asymmetries recently observed  at sub-millimetre and millimetre wavelengths in a number of transition discs could be explained by the emission of the solid particles trapped by vortices in the outer disc. However, at such a distance from the star, disc SG may affect the vortex evolution and must be included in models.}
{Our first goal is to identify how vortex morphology is affected by its own gravity. Next, we look for conditions that a self-gravitating disc must satisfy in order to permit vortex survival at long timescales. Finally, we characterise as well as possible the persistent self-gravitating vortices we have found in isothermal and non-isothermal discs.}
{We performed 2D hydrodynamic simulations using the \textit{RoSSBi 3.0} code. The outline of our computations was limited to Euler's equations assuming a non-homentropic and non-adiabatic flow for an ideal gas. A series of 45 runs were carried out starting from a Gaussian vortex-model; the evolution of vortices was followed during 300 orbits for various values of the vortex parameters and the Toomre parameter. Two simulations, with the highest resolution (HR) thus far for studies of vortices, were also run to better characterise the internal structure of the vortices and for the purpose of comparison with an isothermal case.}
{
We find that SG tends to destabilise the injected vortices, but compact small-scale vortices seem to be more robust than large-scale oblong vortices.
Vortex survival critically depends on the value of the disc's Toomre parameter, but may also depend on the disc temperature at equilibrium. 
disc SG must be small enough to avoid destruction in successive splitting and an approximate  `stability' criterion is deduced for vortices.
The self-gravitating vortices that we found persist during hundreds of rotation periods and look like the quasi-steady vortices obtained in the non-self-gravitating case.
A number of these self-gravitating vortices are eventually accompanied by a secondary vortex with a horseshoe motion.
These vortices reach a new rotational equilibrium in their core, tend to contract in the radial direction, and spin faster.
}
{
We propose an approximate `robustness criterion', which states that, for a given morphology, a vortex appears stable provided that the disc's Toomre parameter overcomes a fixed threshold.
Global simulations with a high enough numerical resolution are required to avoid inappropriate decay and to follow the evolution of self-gravitating vortices in protoplanetary discs. 
Vortices reach a nearly steady-state more easily in isothermal-discs than in non-isothermal discs.
}


\maketitle
%

\section{Introduction}

The formation of planetesimals from the dust-grains embedded in the gas of protoplanetary discs (hereafter PPDs) remains one of the key problems of planet formation. Standard scenarios start from the sub-layer that the dust forms in the mid-plane of the disc after settling under the vertical component of the star's gravity.
They are based on the collisional growth of the solid particles through coagulation and/or sticking mechanisms, generally followed by the formation of self-gravitating dust-clumps that are thought to compact into kilometre-sized planetesimals \citep{Safronov1972, Goldreich1973}. 
However, these models are known to stumble over a number of `barriers' (metre-size, fragmentation, bouncing) \citep{1972fpp..conf..211W,1977MNRAS.180...57W,2005A&A...434..971D, 2008ARA&A..46...21B,2010A&A...513A..57Z} that prevent them from giving a complete picture of the problem. 
More recent models are based on resonant drag instabilities such as the streaming instability \citep[]{2005ApJ...620..459Y,Jacquet2011}, in which pressure bumps can marry with dust-clumps to deliver self-gravitating dust-clouds that are dense enough to form big planetesimals (up to Ceres-sized planetesimals)
\citep{Johansen2007}.
Once planet cores have grown due to the accumulation of planetesimals or the sweeping of residual pebbles \citep{Lambrechts2012}, another important requirement for models is to reach the critical mass for run-away gas accretion before the solid material is accreted by the star and the gas is dissipated by winds and photoevaporation.

Another scenario proposes that anti-cyclonic vortices, formed during the early history of PPDs, could play an important role in the disc evolution, with a redistribution of angular momentum and a clumping of the solid material that (possibly) results in the formation of planetesimals or planetary cores \citep{1995A&A...295L...1B}.
Indeed, such vortices are known to very efficiently trap the dust particles, rapidly increasing the dust-to-gas ratio in the core of the vortices (this mass-ratio $\sim 1$ in a few orbital periods for an optimal Stokes parameter, e.g. \citep{2015ApJ...804...35R}).

Theoretical works and numerical simulations have shown that gaseous vortices in PPDs can be the outcomes of hydrodynamical instabilities, such as the Rossby wave instability \citep[hereafter RWI;][]{1999ApJ...513..805L, 2000ApJ...533.1023L}.
This instability can grow when the generalised potential vortensity distribution reaches a local extrema\footnote{This quantity is $\mathcal{L}(r)=( \sigma \Omega / \kappa^2 ) \left( \frac{P}{\sigma^\gamma}\right)^{2/\gamma}$ and the inverse of the vortensity, $\sigma / (\vec{\nabla} \times \vec{v}) \cdot e_z$, for non-homentropic and for homeontropic 2D discs, respectively.}, which  has also been studied in 3D \citep{Meheut2010, Lin-MK2012, Richard2013}.
Such a situation is favoured at the boundaries of the magnetically inactive region \citep{2006A&A...446L..13V,2008ApJ...689..532T,2012MNRAS.419.1701R} or at the edges of the gap carved by a giant planet 
\citep{2006MNRAS.370..529D,2007A&A...471.1043D,2009A&A...493.1125L,Lin2011,2019MNRAS.486..304B}.
Besides RWI, many other instabilities were found to generate vortices, such as the baroclinic instability {\citep{2003ApJ...582..869K,2004ApJ...606.1070K, Petersen2007a, Petersen2007b, Lesur2010, Barge2016}}, the convective overstability 
\citep{Teed2021}, the vertical shear instability \citep{Richard2016}, or the zombie vortex instability \citep{2015ApJ...808...87M}. 
\citep{2018haex.bookE.138K} give a detailed review of instabilities occurring on PPDs, and further details about the origin and evolution of vortices can be found in \citep{2012A&A...545A.134M,2012MNRAS.419.1701R,Meheut2013,2013MNRAS.433.2626R,Fu_2014a,2017MNRAS.466.3533H,2019MNRAS.486..304B,2020A&A...641A.128R,2021A&A...656A.130R}.

On the other hand, observations with high-resolution instruments, such as the Atacama Large Millimeter/submillimeter Array (ALMA) \citep{2018ApJ...869L..41A}, have revealed strong azimuthal asymmetries in the dust thermal emission of transition discs around young stars such as Oph IRS 48 \citep{2013Sci...340.1199V},
MWC 758 \citep{2018ApJ...860..124D}, HD 135344B \citep{2018A&A...619A.161C}, HD 143006 \citep{2018ApJ...869L..50P}, 
TW Hya \citep{2019ApJ...878L...8T}, and HD 142527 \citep{2019PASJ...71..124S}. 
It is striking that such asymmetries can be interpreted as the thermal emission of dust confined in an undetected gaseous vortex. 
The possibility that such asymmetries are actually tracks of vortices in PPDs has been explored by various authors, but more solid arguments will have to wait for observations from instruments like the James Webb space telescope (JWST) (with the Mid-InfraRed Instrument (MIRI) instrument in the thermal mid-IR region) or the future Extremely Large Telescope (ELT) (using HIgh-REsolution Spectrograph (HIRES) in synergy with ALMA).

From a theoretical point of view, strong dust concentrations in PPDs are known to be the location of instabilities due to the dust feedback. 
This generally happens after the formation of the dust layer with two different occurrences: (i) a Kelvin-Helmoltz instability at the interface between the gas and the dust-rotating layers \citep{Weiden1993,Johansen2006}, (ii) a streaming instability in the flow of radially drifting solid particles \citet{2005ApJ...620..459Y, 2007ApJ...662..627J}.
Similar situations also occur in the strong dust concentrations produced by a vortex, as observed in numerical 2D simulations \citep{Fu_2014, 2016ApJ...831...82S}.
In this case various dust and gas instabilities are at work, such as the Kelvin-Helmoltz instability or specific resonant-drag instabilities \citep{Squire2018}.
This evolution is complex and seems to end in a dissipation of the vortex. 
Of course, a more realistic description of the problem should include the vortex generation mechanism;
for example, \citet{2017ApJ...835..118M} claim that dust-laden vortices could last for thousands of orbits, which would be sufficient for dust growth or gravitational collapse.
In any case, in 2D models, the dissipation timescales are difficult to estimate since the back-reaction of the dust particle becomes problematic at a high dust-to-gas ratio.
3D simulations are scarce and seem to show that vortices remain stable and behave as Taylor columns, even if the mid-plane region is perturbed by a thin dust-layer \citep{2021ApJ...913...92R}.
These authors also find that dust densities in these vortices can be higher than Roche's density, possibly leading to a gravitational collapse.
Another issue of the vortex scenario is the possible growth of the elliptical instability, which is known to destroy 3D-vortices \citep{Lesur2009}.
However, vortices with a large enough aspect ratio could survive due to a weaker growth rate of this instability \citep{Richard2013}, and furthermore, the exact role of compressibility on the growth rate also appears unclear.

The impact of SG (hereafter SG) on the formation and evolution of giant vortices in PPDs is key in the vortex scenario, particularly in order to explain the observed disc asymmetries.
This was first addressed in a number of two-dimensional studies: \citet{2013MNRAS.429..529L} have shown that SG can affect RWI only if the Toomre parameter is less than the inverse of the non-dimensional geometric aspect ratio $h$;
other authors have claimed that 2D vortices are either stretched by gravitational torques \citep{2017MNRAS.471.2204R}, strengthened by the indirect potential induced by the vortex itself \citep{2015ApJ...798L..25M,2016MNRAS.458.3918Z}, or weakened by thermal diffusion
\citep{2020MNRAS.493.3014T}.
On the other hand, three-dimensional studies \citep{2018MNRAS.478..575L} seem to show that 3D vortices with a turbulent core could survive the growth of the elliptic instability.
Thus, the exact role of SG in the vortex evolution is still an open question.

In this paper we revisit the problem using the same strategy as \citet{Surville2015}, looking for approximate numerical vortex solutions of the compressible Euler equation when it is coupled to the Poisson equation.
Thus, our simulations consisted of following up a Gaussian vortex initially superimposed on a disc at equilibrium.
The vortex persistence was checked all along the evolution, particularly against splitting into secondary vortices.
Following this procedure, a series of simulations were carried out and our first task was to organise the results to build up a `stability'
map of the vortices, which we used to identify the most suitable conditions for vortex survival.
Finally, high-resolution (HR) simulations were carried out to characterise as well as possible the internal structure of a self-gravitating vortex hosted in an isothermal and a non-isothermal disc.

This work is organised as follows.
Section \ref{sec:Theory} is devoted to setting up the disc model and presenting the methodology used in the paper.
Then, in Section \ref{sec:evolution}, we compare the evolution of a fiducial quasi-stationary vortex for different values of the Toomre parameters, which permits us to distinguish between three different cases according to the importance of SG.
In Section \ref{sec:vortex stability}, a large parametric study is performed in order to evaluate the stability of vortices with respect to the axisymmetric disc's Toomre parameter and the vortex internal structure. 
This enabled us to choose, in Section \ref{sec:high resolution simulations}, the most promising structures and compare the HR simulations  for vortices 
evolving in isothermal and non-isothermal discs.
Finally, in Section \ref{sec:discussion} we suggest a discussion on observational constraints, numerical resolution, and spiral waves, followed by a conclusion.
  
\section{Theoretical context}\label{sec:Theory}

\begin{table}
\caption{SG parameter and surface density at $r_0=$50 AU} 
\label{tab:Q0 and sigma0}                      
\centering                                     
\begin{tabular}{c c c c c c}                   
\hline                                         
$q_0$ & 0.25 & 0.5 & 1 & 2 & 4 \\ 
\hline                                         
$\sigma_0 $ ( g.cm$^{-2}$ ) & 9.38 & 4.69 & 2.35 & 1.17 & 0.59 \\  
\hline
\end{tabular}
\end{table}

\begin{table}
\caption{Vortex parameters at $t=50\,t_0$ with respect to the initial Gaussian model parameters. \\
\emph{Column and row header:} Gaussian model parameters at $t=0$.
}   
\label{tab:vortex features}  
\centering                                     
\begin{tabular}{@{}c| c c c}                   
\hline                                         

\diagbox[height=2.em,width=3.em]{$\chi$}{$\delta$} &       1.0           &        1.5          &          2.5         \\ 
\hline                                         
5  &  ( 7.50, 0.79)  &  (6.21, 1.43) &  ( 5.65, 2.78)  \\  
8  &  ( 8.78, 1.02)  &  (9.07, 1.68) &  ( 9.00, 3.06)  \\  
14 &  (10.73, 1.12)  &  (13.6, 1.65) &  (12.62, 2.69)  \\  
\hline
\end{tabular}
\end{table}

This section is devoted to introducing the main theoretical background of this paper.
We begin with the description of the equations governing the disc evolution.

\subsection{Disc model and evolution equations}

We investigate the evolution of an ideal gas (molecular dihydrogen, $\mu\sim2.34$) in a thin (2D),
non-homentropic disc orbiting a solar-type star. 
Compressibility must be taken into account since the vortex size often exceeds the scale height of the disc.
On the other hand, viscosity is neglected since we are working in the magnetically dead zone of the protoplanetary discs.
The equations governing gas dynamics in a 2D protoplanetary disc are thus the continuity equation, Euler's equation, and the conservation of energy equation, which are: 
\begin{eqnarray}
\partial_t \sigma 
          + \vec{\nabla} \cdot ( \sigma \vec{v} ) 
      & = & 0 ,\label{Eq:continuity equation}\\
\partial_t \vec{v} + (\vec{v} \cdot \vec{\nabla}) \vec{v}
      & = & 
          - \frac{\vec{\nabla} P}{\sigma} 
          - \vec{\nabla} \left(\Phi_\odot + \Phi_{\mathrm{SG}} + \Phi_{\mathrm{ind}} \right), \label{Eq:Euler equation}\\
\sigma \left[\partial_t e + \left( \vec{v} \cdot \vec{\nabla} \right) e \right] & = & - P \, \vec{\nabla} \cdot \vec{v} \label{Eq:conservation energie},
\end{eqnarray}
where $P$, $\sigma$, and $\vec{v}$ are the vertically integrated pressure, density, and the gas velocity, respectively,
$e=\frac{1}{\gamma-1} \frac{P}{\sigma}$ is the internal energy of an ideal gas, and $\gamma=1.4$ is the adiabatic index for a diatomic gas.
The gravitational potential has three different contributions: 
$\Phi_{\odot}$, $\Phi_{SG}$, and $\Phi_{ind}$ due to the central object, the disc SG and the vortex itself (indirect term), respectively.
The last term is a virtual potential that accounts for the offset between the centre of the frame of reference, located at the star's position, and the $\{$star+vortex$\}$ centre of mass.
No viscosity is expected in the non-active region of the PPDs, although \citet{2014A&A...572A..77S} noted that, even in this region, vertical-shear instability can generate turbulence associated with an effective viscosity.
According to \citet{2020MNRAS.499.1841M} and \citet[Formula (14)]{2021MNRAS.508.5402M}, our simulations correspond to an infinite cooling time and to $\alpha/(10^{-4})<<1$, so that the viscosity is very small and compatible with our inviscid disc assumption.
In the case of a finite cooling time, we should account for a non-negligible viscosity in the computations. \\

At equilibrium the background flow is assumed to be axisymmetric with radial profiles for the temperature $T_0(r)$, the surface density $\sigma_0(r),$ and the pressure $P_0(r)$ chosen in the form of simple power laws. 
The model only differ from the minimum mass solar nebulae (MMSN) hypothesis \citep{1981PThPS..70...35H} by the choice of the local values of the surface density index $\beta_\sigma$ and the temperature index $\beta_T$ to account for the observational constraints.
With these assumptions, the unperturbed axisymmetric disc's Toomre parameter, $Q_0(r)={\Omega_K c_{s,0}}/{(\pi G \sigma_0)}$, also obeys power laws that depend on the surface density in the centre of the simulation box.
Thus, the main variables defining the stationary, axisymmetric flow are:
\begin{equation}
\begin{array}{lccl}
\sigma_0(r)        & = & \sigma_0 \, \left(r/r_0\right)^{\, \beta_\sigma} 
                   & \mathrm{with} \ \  \beta_\sigma=-1.1,  \\ [2pt]
T_0(r)             & = & T_0 \, \left(r/r_0\right)^{\,  \beta_T}, 
                   &                                       \\ [2pt]
H/r                & = & 0.05 \,     \left(r/r_0\right)^{\, \beta_H-1},        
                   & \mathrm{with} \ \   \beta_H=\frac{\beta_T+3}{2},   \\ [2pt]
\mathrm{Q}_0(r)    & = & 0.0279      \left( \frac{1700 \mbox{ g.cm}^{-2} }{\sigma_0} \right) \, \left(r/r_0\right)^{\, \beta_Q} ,
                   & \mathrm{with} \ \   \\
                   &  &
                   &  \beta_Q=\beta_H-3-\beta_\sigma,   
\end{array}
\end{equation}
where $r$ is the normalised radius at 1 AU, $r_0=50$, $\sigma_0(r)$ and $\sigma_0$ are the surface density (in g.cm$^{-2}$) at radial positions $r$ and $r_0$, respectively, and $H$ is the pressure scale height.
$T_0(r)$ and $T_0=6.2$ K refer to the temperature at radial positions $r$ and $r_0$, respectively.
In particular, the temperature was chosen such that the disc flaring, $h$, in the centre of the simulation box is equal to 0.05. For an ideal gas, the axisymmetric pressure also obeys a power law with index $\beta_p=\beta_T+\beta_\sigma$.
In Sections \ref{sec:evolution} and \ref{sec:vortex stability}, and for the non-isothermal disc studied in Section \ref{sec:high resolution simulations}, we have set $\beta_T=-0.5$.
On the other hand, we have chosen $\beta_T=0$ for the isothermal disc considered in Section \ref{sec:high resolution simulations}.

\subsection{Gravity terms}
   
In polar coordinates the SG potential reads :  
\begin{equation}\label{Eq:self gravity potential}
   \Phi_{\mathrm{SG}}(\vec{r}) = - \, G \int\limits_{\mathrm{disc}} 
                                 \frac{\sigma_0(r')}{|\vec{r}-\vec{r}'|}
                                 d^2\vec{r}' 
                                 - \, G \int\limits_{\mathrm{disc}}
                                 \frac{\sigma(r',\theta')-\sigma_0(r')}{|\vec{r}-\vec{r}'|} 
                                 d^2\vec{r}',  \\
\end{equation}
where the first term is the axisymmetric contribution of the disc, while the second is the contribution of all other symmetry deviations. 
For example, the last term can correspond to the presence of an anti-cyclonic vortex or to a density wave.
In order to account for the disc thickness in SG computations, we considered a softening length equal to $0.6\,H$ \citep{2012A&A...541A.123M}.
The unperturbed disc potential can be expressed analytically as an infinite sequence \citep{2008A&A...490..477H}, which, truncated at second order, allows the axisymmetric disc potential to be estimated at an accuracy better than 99\% \citep[in French]{survi2013}.
The presence of a massive vortex in the outer disc can lead to an offset
of the mass barycentre, which may strengthen the vortex itself \citep{2017A&A...601A..24R,2016MNRAS.458.3918Z}. 
To account for this possible effect, the gradient of the indirect potential has been taken into account; this additional term can be written as:
\begin{eqnarray}
    \Vec{\nabla} \Phi_{ind} & = & 
    \left(
    \begin{matrix} 
       \ A \cos{\theta}  + B \sin{\theta}  \\ 
       - A \sin{\theta}  + B \cos{\theta} 
    \end{matrix}\right),
\end{eqnarray}
where $A=\int\limits_{disc}\frac{ G \ \cos{\theta'} dm'}{r'^2}$ and $B= \int\limits_{disc}\frac{ G \ \sin{\theta'} dm'}{r'^2}$.

For convenience, all of our simulations are referenced and labelled with the Toomre parameter of the unperturbed disc. 
Following \citet{2013MNRAS.429..529L}, it is also useful to introduce in the discussion the critical parameter $Q_c=1/h$.
Indeed, RWI is significantly changed by SG when $q_0 = Q_0/Q_c \leq 1$.
The normalised value of the Toomre parameter, $q_0$, will be used through the rest of the paper and will be called the SG parameter.
Our investigation was restricted to $q_0 =[0.25, 0.5, 1, 2, 4]$ because below the lower limit, any vortex is unstable, whereas beyond $q_0=4$ (the upper limit), SG has no observable effect on the vortex.
The relationship between $q_0$ and $\sigma_0$, the surface density at 50 AU, is presented in Table \ref{tab:Q0 and sigma0}.

\subsection{Numerical method and physical parameters}\label{subsec:Methodology and numerical setup}
   
The system of equations \eqref{Eq:continuity equation}-\eqref{Eq:Euler equation} was solved numerically thanks to the latest version of the RoSSBi (Rotating Systems Simulation for Bi-fluids) code, which was specifically developed to study the evolution of  PPDs \citep{Inaba2005,Inaba2006, survi2013}.
This new version has been extended to 3D and structured for high-performance parallelism;
it is presented in detail elsewhere \citep{unpublishedrendon} with many tests and validations.
The code includes SG, which is computed using the Fastest Fourier Transform in the West 3 library (FFTW3) \citep{FFTW05} with a logarithmic and a linear grid in the radial and azimuthal direction, respectively.\\

In the absence of both SG and viscosity,  a number of authors \citep{Godon1999,Barranco2005} started numerical simulations from crude vortex models.
Later, other studies started from approximate vortex models of the steady state equations \citep{Lesur2009, Surville2015}. 
A Gaussian vortex model is particularly suitable in the case of compressible simulations; the injected vortices were found to rapidly relax to long-lived vortices, which can then be studied in detail \citep{Surville2015}.
Here, the same strategy is used to address the problem of vortices in a disc that has SG.
Thus, instead of generating vortices through RWI or baroclinic instability, we preferred to inject a Gaussian solution as modelled in \citet{Surville2015} since this allows a large variety of vortices to be explored by tuning their control parameters (radial width, Rossby number, and aspect ratio).
This method allows the most appropriate vortex parameters to be selected, but also to save computational time since the timescale necessary to generate a steady vortex from a hydrodynamic instability generally exceeds hundreds of orbits.
Since the geometrical parameters of the vortex can change during the relaxation phase, we gathered in Table \ref{tab:vortex features} all of the parameters used to initialise our numerical simulations, as well as their value at $t=50\,t_0$, the time at which SG is plugged in.
It is interesting to note that in presence of viscosity, $\alpha=10^{-4}$, and simulations with or without SG have shown that vortices decay at a rate proportional to the disc mass \citep{2017MNRAS.471.2204R}.

\subsubsection{A smooth activation of SG} 

In the presence of SG, numerical tests (at low-resolution, hereafter LR) have shown that persistent vortices can also survive, but after a transition period that is longer and more complex than in the non-self-gravitating case.
This transient evolution is due to the difference between the injected and the exact vortex-solutions; of course it reduces to the relaxation already discussed in the previous section for vanishing SG.
For these reasons, and to keep the same strategy as previously used, it was decided to gradually introduce SG during the computations, in order to avoid spurious vortex bursts, which occur  when SG is introduced with a Heaviside time step function.
To this end we have implemented in the code the time function:
\begin{equation}\label{Eq:smooth function gravity activation}
g(t)= \left\{
\begin{matrix}
0                                                   &\mbox{ if } & t \leq 60 \, t_0 \\    
1-\exp(-\left(\frac{t-60 \, t_0}{T_{act}}\right)^2) &\mbox{ otherwise} &
\end{matrix} \right..
\end{equation}
\noindent
The time constant was chosen to be equal to the vortex relaxation time in the non-SG case; that is $T_{act}\simeq 15 \, t_0$, where $t_0=353$ years is the orbital period at $r_0=50$ AU.
With this procedure, SG reaches 98\% of its final value in $\simeq 30$ orbital periods.

\subsubsection{Choice of computational box}

Besides the smooth activation of SG, another requirement is to correctly account for the vortex migration.
In a self-gravitating disc, this migration not only results from the asymmetry of the vortex wake between the inner and outer regions \citep[e.g.][]{Paardekooper2013} but also from all the Lindblad torques produced by the density waves that are excited by the vortex.

Since in this work we are mainly interested by variations in vortices' structure and not in their type I migration, the initial vortex should keep its radial position during the whole run.
This is possible provided that the computational box includes the same number of inner and outer Lindblad resonances.
Therefore, prior to the main study, for each Gaussian vortex structure, we ran a preliminary study at LR, ($N_r, N_\theta$)=(360, 720), during 150 orbits and for a large computational window (r $\in[15, 100]$ AU) in order to catch all Lindblad resonances and forecast their location for the main study.
This allowed us to choose the suited simulation window, r $\in[25,72]$ AU, such that there was the same number of inner and outer Lindblad resonances during the whole simulation, to keep the vortex in the centre of the simulation box and avoid spurious vortex migration. 
Figure \ref{fig:extent box} illustrates the extent of the full computational box and shows an example of compressibility waves emitted by a quasi-steady vortex. 


%
\begin{table}
\caption{Parameter definitions}             
\label{table:parameters definition}      
\centering          
\begin{tabular}{c c l}     
\hline\hline       
Symbol & Definition/Value & Description                                        \\ 
\hline                    
$r$                   &   & Normalised radius at 1 AU                          \\
$\sigma$              &   & Gas surface-density                                \\  
$P$                   &   & Vertically integrated pressure                     \\
$T$                   & $\displaystyle\frac{\mu m_u}{k_B} \frac{P}{\sigma}$ & Gas temperature  \\
$\vec{v}$             &   & Gas velocity-field                                 \\
$\vec{v}'$            &  $\vec{v}-\Omega_\mathrm{K} r \vec{e}_\theta$ 
                          & Relative gas-velocity (with respect                \\ 
                      &   & to the angular gas velocity)                       \\
$\Phi_\odot$          &   & Central object  potential                          \\
$\Phi_{\mathrm{SG}}$  &   & Gas SG potential                         \\
$\Phi_{\mathrm{ind}}$ &   & Indirect potential (offset of $\Phi_\odot$)        \\
$\Omega_\mathrm{K}$   & $\displaystyle\sqrt{\frac{GM_\odot}{r} + r \frac{\partial_r P_0}{\sigma_0}} $
                      & Pressure-supported rotation-rate                       \\
$\gamma$              & 1.4 & Adiabatic index                                  \\
$c_s$                 & $\sqrt{\gamma \frac{P}{\sigma}}$ & Gas sound-speed     \\
$H$                   & $\sqrt{\frac{2}{\gamma}}\frac{c_{s}}{\Omega_K}$  & Pressure scale-height \\
\rule{0pt}{2.5ex}$h(r)$                & $\displaystyle\frac{H}{r}$ = $0.05\left(r/r_0\right)^{\beta_H-1}$ & Disc aspect ratio \\
\rule{0pt}{2.5ex}$Q$                   &  ${\Omega_K c_s}/{(\pi G \sigma)}$  & Local Toomre's parameter    \\
\rule{0pt}{2.5ex}$Q_0$                 &  ${\Omega_K c_{s,0}}/{(\pi G \sigma_0)}$ &   Toomre's parameter at 50 AU \\
                                       &           & for the Axis-symmetric disc                                  \\
\rule{0pt}{2.5ex}$Q_c$& $1/h = 20$     & Critical Toomre parameter             \\
\rule{0pt}{2.5ex}$q_0$& $Q_0/Q_c$      & SG parameter                \\
\rule{0pt}{2.5ex}$Ro$ & ${\nabla \times \vec{v}'} / {2 \Omega_K }$ & Rossby number \\
$\overline{Ro}$       & $<Ro>_{core} $ & Vortex spin                           \\
$\chi$                &   & Vortex aspect-ratio                                \\
$\Delta \mathrm{r}$   &   & Vortex radial-width                                \\
$\delta$              & $\Delta r/H$  & Vortex radial-extent                   \\
$t_0$                 & 353 years  & Orbital period at 50 AU                   \\
$C_\sigma$            & $\sigma/\sigma_0-1$               & Density contrast   \\
$C_P$                 & $P/P_0-1$                         & Pressure contrast  \\
\hline                  
\end{tabular}
\tablefoot{Quantities defined for the unperturbed axis-symmetric gas disc are denoted with a 0 subscript. The values provided for $t_0$, h, and $\mathrm{Q_c}$ correspond to the values at 50 AU.}
\end{table}

\subsubsection{Suitable physical parameters}\label{subsec:suitable parameters}

A number of physical parameters have been chosen to characterise as well as possible the coherent gaseous vortices and to make easier the discussion.
It is, for example, convenient to distinguish between morphological and global parameters such as the radial position, the bulk-mass, and the spin.
Since elliptical streamline models \citep[e.g.][]{1981Kida,1987MNRAS.225..695G, Chavanis2000} approximately describe the central region of 2D vortices, the most natural morphological parameters are the ellipse parameters such as the vortex aspect ratio $\chi$ and its radial extent (scaled to the pressure scale height), $\delta=\Delta r/H$.
It is useful to recall that when the radial extent lies below unity, the flow is subsonic (incompressible), while when it lies above unity, the flow is supersonic (compressible). 
These structure parameters are determined by fitting ellipses on the iso-contours of the Rossby number at the $Ro = -0.08$ level, where Ro $={\nabla \times \vec{v}'} / {2 \Omega_K }$ and $\vec{v}'$ is the relative gas velocity with respect to the local pressure supported rotation rate (see Table \ref{table:parameters definition} and Figure \ref{fig: definiton aspect ratio and radial extent}).
Appendix \ref{app:computation aspect ratios} gives the details of the procedure and explains why the Rossby number is chosen instead of density or pressure.

Global parameters are well suited to describe the bulk-vortex evolution once a nearly steady-state is reached, such as a long-term dynamical evolution.
In particular, we numerically computed the distance from the vortex to the star, tracking the pressure maximum as a function of time.
A debated question is whether vortices can be long-lived and robust against instabilities, and for this reason we computed the mean Rossby number in the vortex core or spin, $\overline{\mathrm{Ro}}$, namely the vortex strength. 
According to Kida \citep{1981JFM...112..397K} and GNG \citep{1987MNRAS.225..695G} models, the vortex strength is inversely proportional to the aspect ratio. Numerical simulations by \citet{Surville2015} have also shown that, even if the dependence of the Rossby number as a function of the vortex aspect ratio differs from the standard GNG relation ($1/\chi$), the general trend is the same with a saturation for elongated vortices: $\lim\limits_{\chi \to \infty} Ro \simeq -0.1$.
This general trend motivated some authors to use the inverse of the aspect ratio as a measure of vortex strength. 
Yet, comforted by \citet{Surville2015} and by results shown in Sect. \ref{subsubsc:vortex shaped by SG}, we decided to use Ro as the vortex strength through the rest of the paper.
Finally, since we are mainly interested in SG effects, it is important to evaluate the mass that can be attributed to the vortex. 
This requires some caution since a vortex is a highly distributed object in contrast to planets, which are usually described as material points.
Furthermore, a vortex is also accompanied by spiral waves, an annular over-density (see Section \ref{sec:high resolution simulations}), and, eventually, some secondary eddies (see Sections \ref{subsec: strong SG} and \ref{sec:high resolution simulations}), which leads to a prudent calculation. 
We encourage the interested reader to refer to Appendix \ref{app: section - computed quantities}, where we explain in detail the way we define the mass of a gaseous vortex and also its mean Rossby number.
In Appendix \ref{app:section uncertainties} we estimated uncertainties.
The definition of all the parameters and key quantities used in this paper are gathered in Table \ref{table:parameters definition}.

\section{Vortex evolution }\label{sec:evolution}

%
\begin{figure}
\centering
\includegraphics[trim = 0.8cm 0cm 1.7cm 0.75cm, clip, width=0.98\hsize]{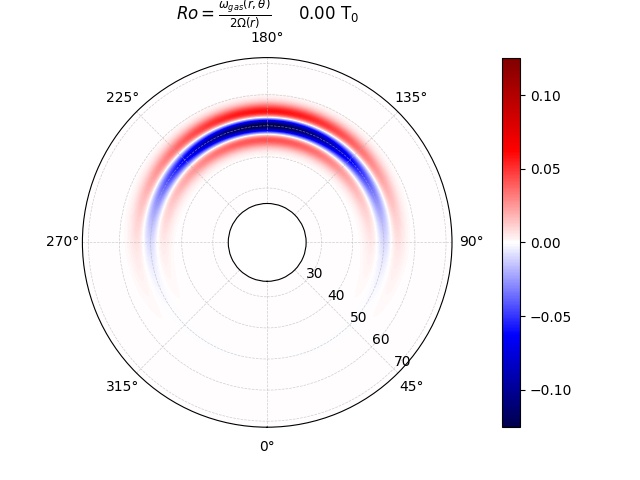}
\includegraphics[trim = 0.8cm 0cm 1.7cm 0.75cm, clip, width=0.98\hsize]{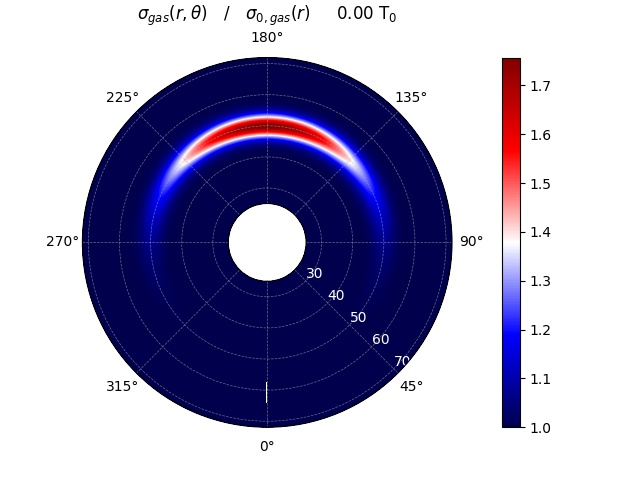}
\caption{\textbf{Gaussian vortex model injected at $t=0$.}\\
\emph{Top:} Rossby number \\ \emph{Bottom:} Normalised density ($\sigma(r,\theta)/\sigma_0(r)$)} 
\label{fig:gaussian model initial}
\end{figure}
%

\begin{figure*}
\centering
\begin{minipage}{0.86\hsize}
\resizebox{\hsize}{!}
          {\includegraphics[trim = 0.8cm 0cm 3.8cm 0.7cm, clip]{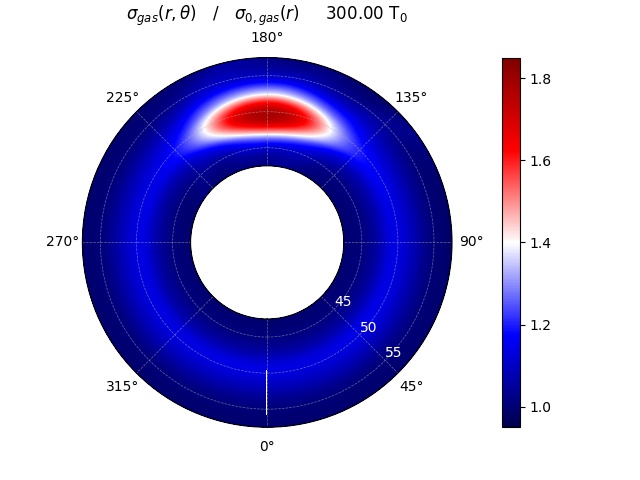}
           \includegraphics[trim = 0.8cm 0cm 3.8cm 0.7cm, clip]{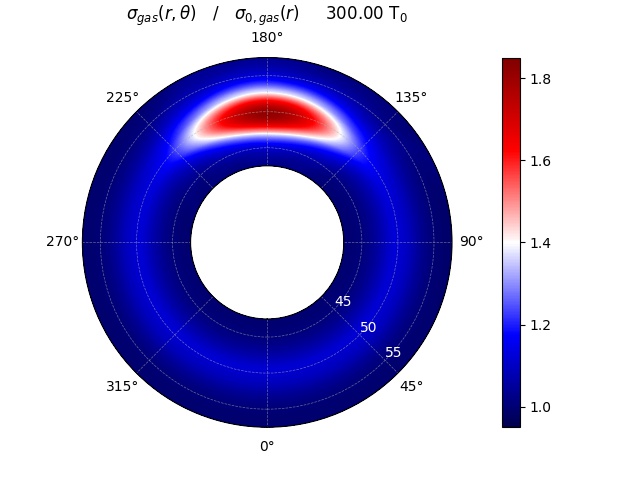}
           \includegraphics[trim = 0.8cm 0cm 3.8cm 0.7cm, clip]{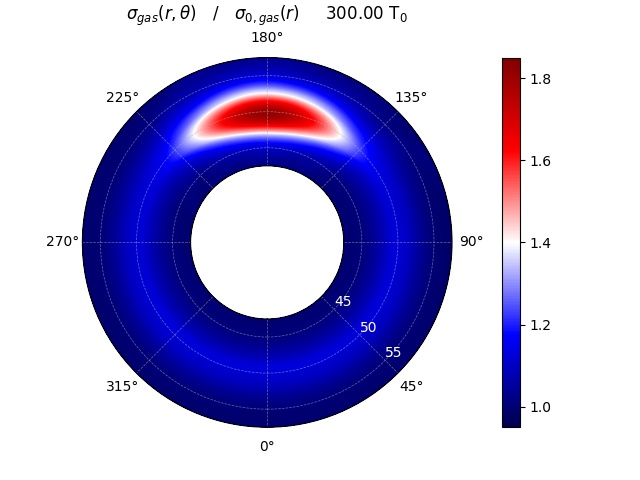}
            }
\resizebox{\hsize}{!}
          {\includegraphics[trim = 0.8cm 0cm 3.8cm 0.7cm, clip]{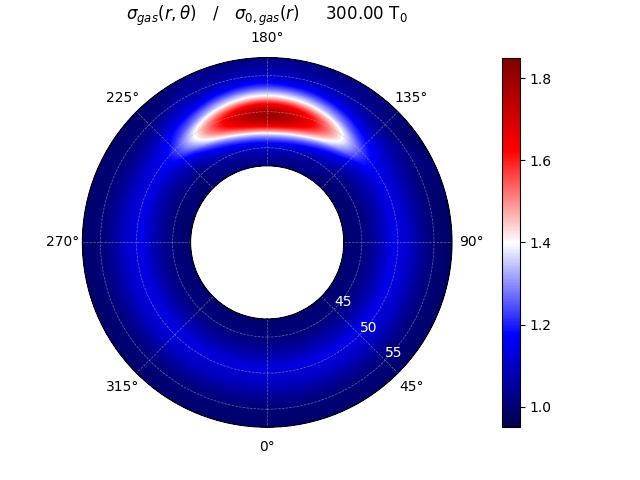}
           \includegraphics[trim = 0.8cm 0cm 3.8cm 0.7cm, clip]{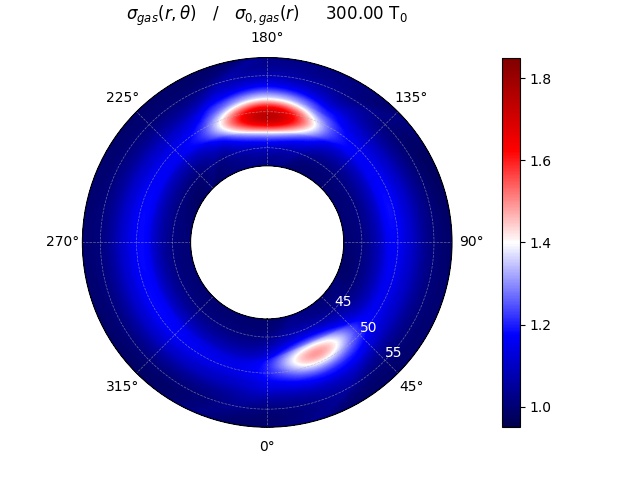}
           \includegraphics[trim = 0.8cm 0cm 3.55cm 0.7cm, clip]{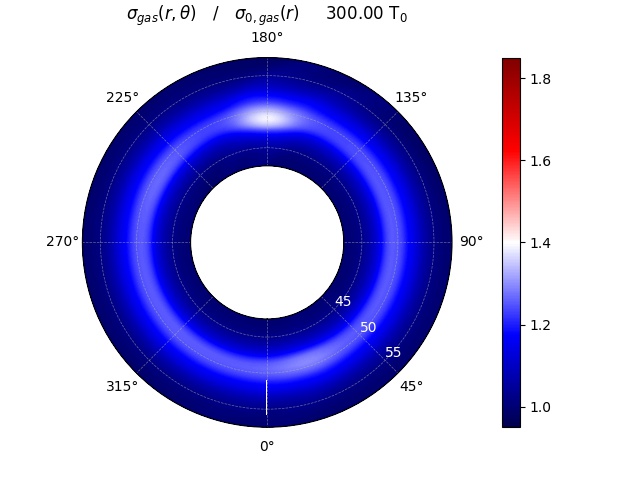}
            }
\end{minipage}\hfill
\begin{minipage}{0.14\hsize}
\includegraphics[trim = 0cm 0cm 0cm 0cm, clip, width=0.60\hsize]{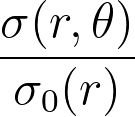}
\includegraphics[trim = 12.4cm 0cm 1.9cm 0.7cm, clip, width=0.60\hsize]{rho_T300_chi14_Qifty.jpg}
\end{minipage}
\caption{\textbf{Comparison of the end-of-run surface density for increasing SG ($t=300\,t_0$).} Top row and from left to right: $q_0=\infty$, 4, 2 ; bottom row and from left to right $q_0=$1, 0.5 , 0.25.
The initial vortex parameters are : $\delta = 1.5$ and $\chi=14$. The simulation window was centred in r $\in [42.5, 57.5]$ AU in order to better appreciate the inner vortex structure.
}
\label{fig: rho at t=300}
\end{figure*}
%

%
\begin{figure*}
\centering
\begin{minipage}{0.86\hsize}
\resizebox{\hsize}{!}
          {\includegraphics[trim = 0.8cm 0cm 3.8cm 0.73cm, clip]{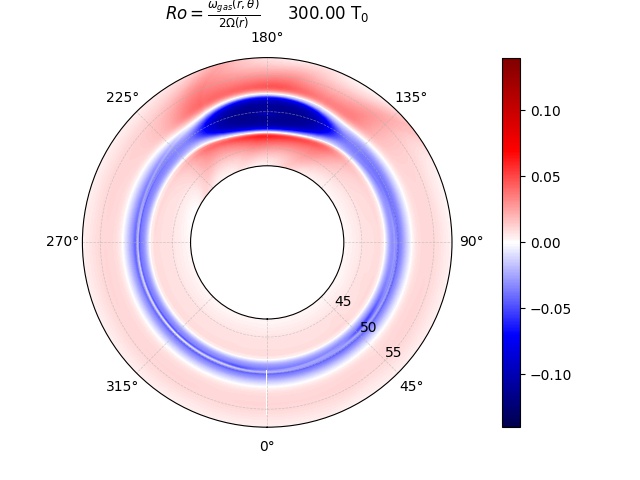}
           \includegraphics[trim = 0.8cm 0cm 3.8cm 0.73cm, clip]{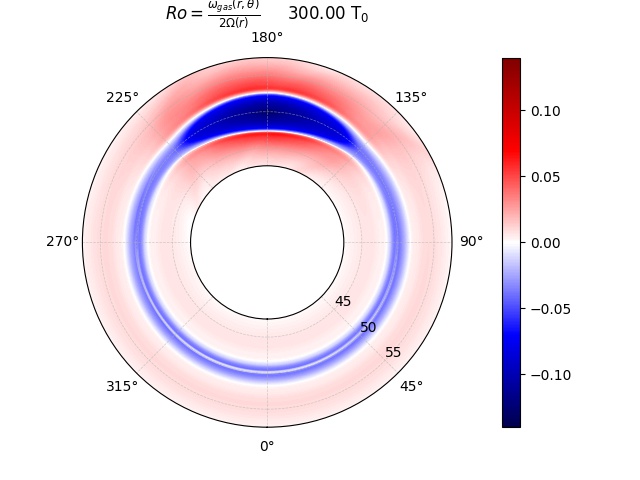}
           \includegraphics[trim = 0.8cm 0cm 3.8cm 0.73cm, clip]{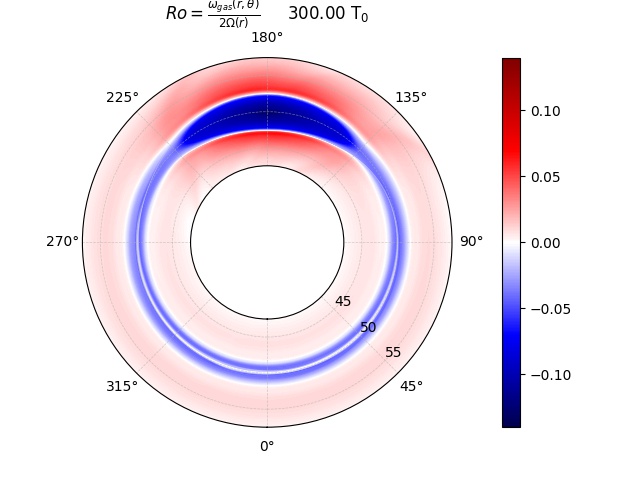}
            }
\resizebox{\hsize}{!}
          {\includegraphics[trim = 0.8cm 0cm 3.8cm 0.73cm, clip]{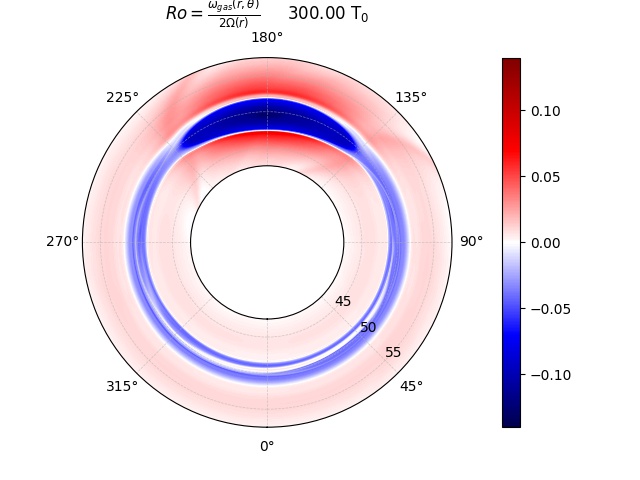}
           \includegraphics[trim = 0.8cm 0cm 3.8cm 0.73cm, clip]{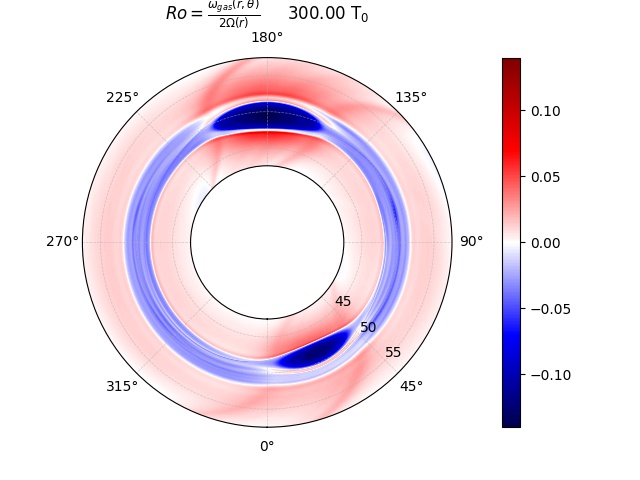}
           \includegraphics[trim = 0.8cm 0cm 3.8cm 0.73cm, clip]{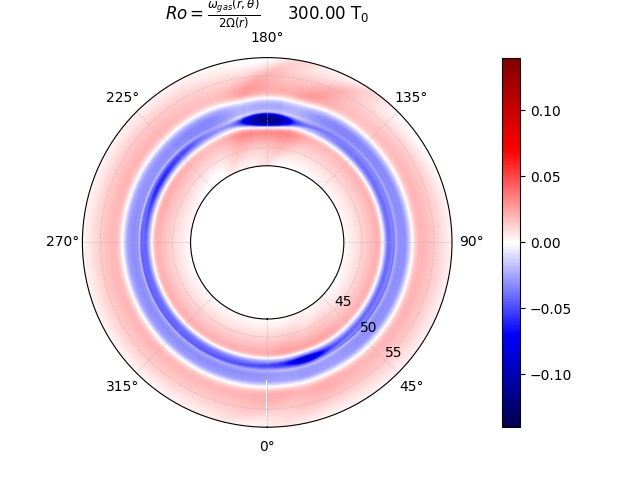}}
\end{minipage}\hfill
\begin{minipage}{0.14\hsize}
\includegraphics[trim = 0cm 0cm 0cm 0cm, clip, width=0.40\hsize]{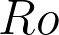}
\includegraphics[trim = 12.4cm 0cm 1.77cm 0.7cm, clip, width=0.60\hsize]{rossby_T300_chi14_Qifty.jpg}
\end{minipage}
\caption{\textbf{Comparison of the end-of-run Rossby number for increasing SG ($t=300\,t_0$).} Top row and from left to right: $q_0=\infty$, 4, 2 ; bottom row and from left to right $q_0=$1, 0.5 , 0.25.
The initial vortex parameters are : $\delta = 1.5$ and $\chi=14$.
The simulation window was centred in r $\in [42.5, 57.5]$ AU in order to better appreciate the inner vortex structure.
}
\label{fig: rossby at t=300}
\end{figure*}
%
\begin{figure}
\centering
\includegraphics[width=\hsize]{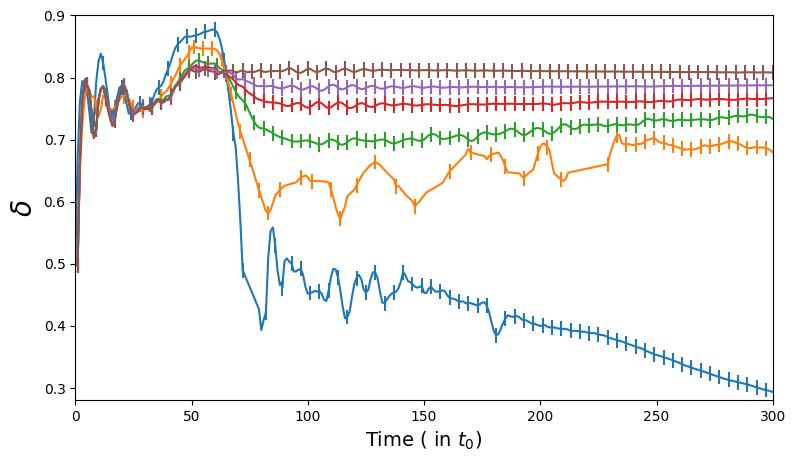}
\includegraphics[width=\hsize]{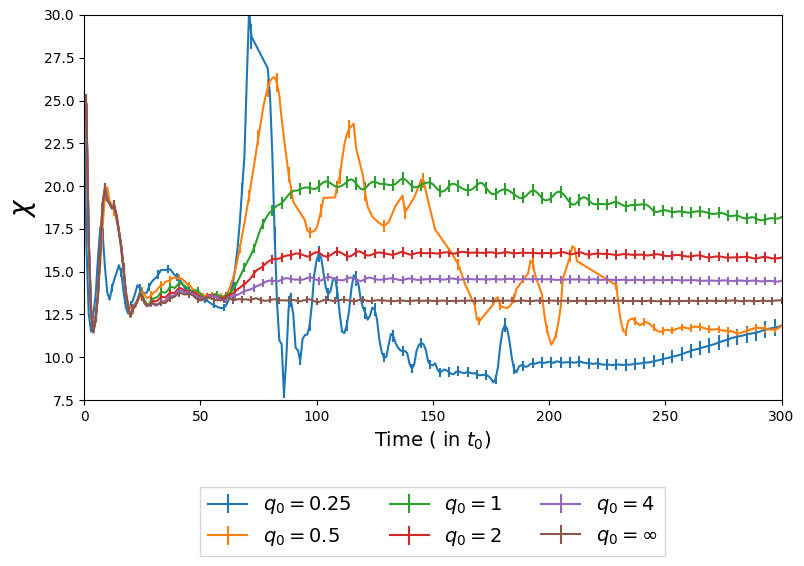}
\caption{\textbf{Morphological evolution for different values of the SG parameter.}
\emph{Top:} Semi-minor axis with respect to the pressure scale height at the vortex core for $q_0=$ 0.25, 0.5, 1, 2, 4, $\infty$.
\emph{Bottom:} Aspect ratio at the $Ro$=-0.08 level.  
Initial vortex parameters are : $\delta = 1.5$ and $\chi$=14.
For readability, uncertainties are plotted at every fourth time steps (see Appendix \ref{app:section uncertainties}). 
}
\label{fig: aspect ratios with respect to time}
\end{figure}
   
\begin{figure}
\centering
\includegraphics[width=\hsize]{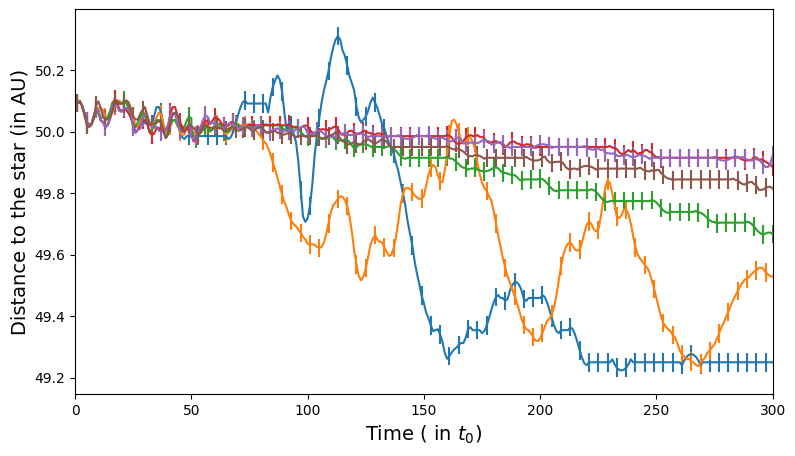}
\includegraphics[width=\hsize]{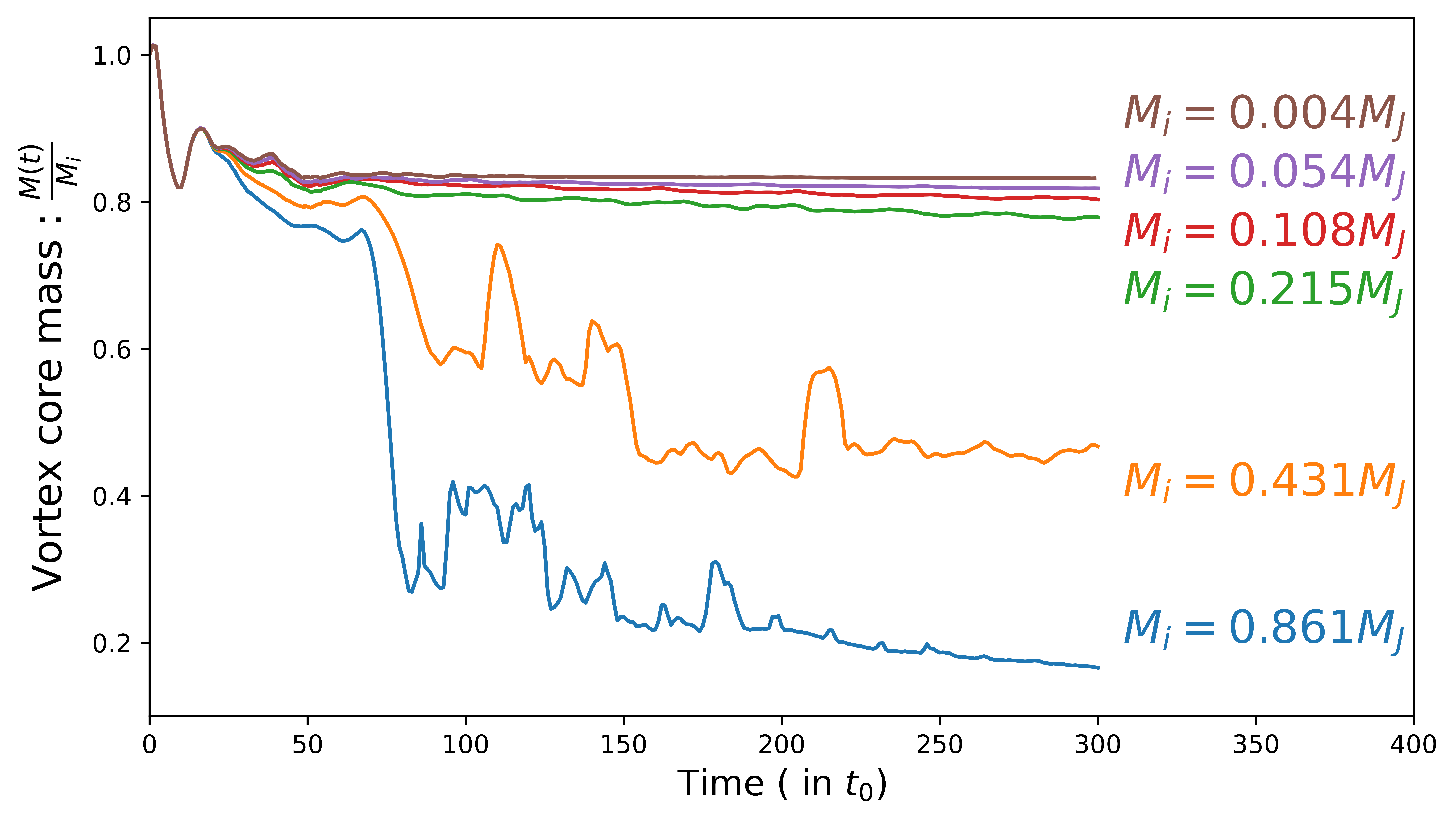}
\includegraphics[width=\hsize]{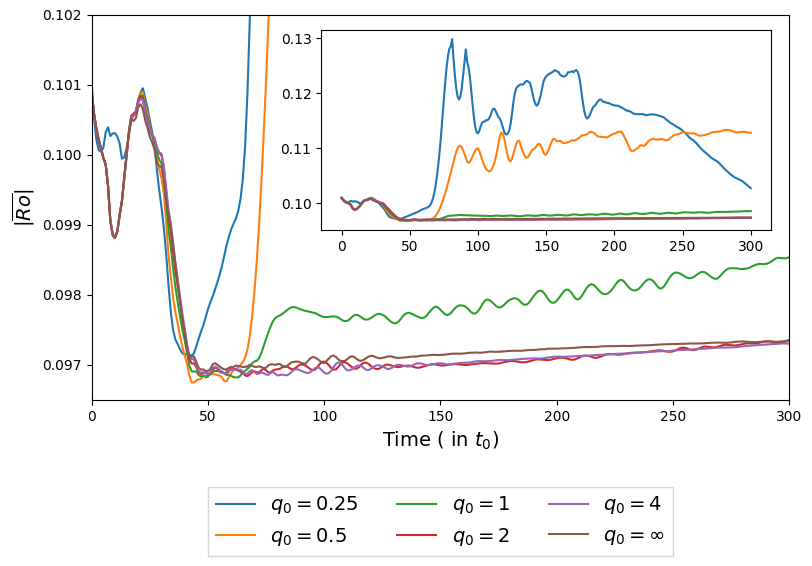}
\caption{\textbf{Bulk-vortex evolution for different values of the SG-parameter.}
\emph{Top:} Radial distance to the star.
\emph{Middle:} Bulk of the vortex mass; $M_i$ is the vortex mass at t=0.
\emph{Bottom:} Absolute value of the mean Rossby number $\overline{Ro}$.
The initial vortex parameters are: $\delta = 1.5$ and $\chi$=14. For readability, uncertainties are plotted at every fourth time steps (see appendix \ref{app:section uncertainties})}.
\label{fig: migration, mass, hill and rossby}
\end{figure}

Six numerical simulations were run to address the problem, starting from a single reference vortex that was followed over 300 orbital periods.
All runs performed in this section started from a Gaussian vortex with aspect ratio $\chi=14$ and radial extent $\delta=1.5$ which was injected at $r_0=50$ AU and $\theta_0=\pi$.
We exhibit in Figure \ref{fig:gaussian model initial} the initial Gaussian vortex.
Disc SG was quantified thanks to the SG parameter, $q_0$ , whose values were selected among six different values: $ 0.25, 0.5 , 1 , 2 , 4$  and $\infty$ (non-SG case).
The results of the simulations are presented in Figures \ref{fig: rho at t=300}, \ref{fig: rossby at t=300}, \ref{fig: aspect ratios with respect to time}, and \ref{fig: migration, mass, hill and rossby}, and the temporal evolution can be found in \textbf{online movies 1 and 2}.
The end-of-run snapshots of the surface density and the Rossby number, obtained for the six values of $q_0$, are presented in Figures \ref{fig: rho at t=300} and \ref{fig: rossby at t=300}.
Figures \ref{fig: aspect ratios with respect to time} and \ref{fig: migration, mass, hill and rossby} focus on the evolution of physical quantities such as migration, mass, and spin. 
In the whole section, the case ${q_0}= \infty$ and values at $t=50\,t_0$ are considered as the references on which comparisons are based. 
It is interesting to note that the slow increase in the vortex spin over the whole run is likely an artefact related to the numerical resolution as testified in Figure \ref{fig:numerical convergency 1}.

\subsection{Numerical procedure for the run series}\label{subsec:Numerical proc}

Computations started at $t=0$ with a LR defined by (N$_r$, N$_\theta$)=(500, 720).
During the first 50 orbits, the vortex morphological and global parameters did not stay constant but  slowly changed as a function of time.
Indeed, the injected vortex was not an exact solution of the fluid equations but less than 50 orbits were sufficient for the injected solution to relax and reach a quasi-stationary regime (see Table \ref{tab:vortex features}).
Then, at t=50 $t_0$, the resolution of the simulations was increased using a 2D cubic interpolation \footnote{With this interpolation, residuals are highly damped in less than two orbits.} into an intermediate grid resolution of (N$_r$, N$_\theta$)=(1500, 3000), which is equivalent to $\sim$ 71 cells/H and $\sim$ 24 cells/H in the radial\footnote{For a logarithmic mesh, the resolution at $r_0$ is $\delta_*=r_0 \left( \alpha^{1/2}-\alpha^{-1/2} \right)$ where $\alpha=\left( r_{out}/r_{in} \right)^{1/N_r}$.} and azimuth directions, respectively.
At $t=60\,t_0$, SG was activated gradually thanks to the function described in Equation \ref{Eq:smooth function gravity activation}.
Finally, all the simulations were stopped at the 300$^{th}$ orbit.
The successive steps of our procedure are recapped in a few lines:
\begin{equation*}
\begin{array}{lcl}
t=0         & : & \mbox{Gaussian vortex at LR,}             \\
            &   & \mbox{with : } (N_r, N_\theta)=(500, 720),            \\
0<t<50\,t_0 & : & \mbox{Vortex relaxes to a quasi-steady state,}        \\
t=50\,t_0   & : & \mbox{Resolution is increased to (1500, 3000),}       \\
t=60\,t_0   & : & \mbox{Plug-in of SG,}                       \\
60\,t_0<t<300\,t_0 & : & \mbox{Simulation at intermediate resolution.}
\end{array}
\end{equation*}
The rapid relaxation of the Gaussian solution was eased by the LR step, which introduced a significant numerical viscosity.
On the other side, during the last step, the choice of a higher resolution allowed a numerical convergence to be reached
Appendix \ref{subsec:convergency1}, stresses that LR can significantly speed up the decay of vortices.
More quantitatively, during a complete simulation with $t\in[0,300\,t_0]$, we found that the vortex mass-loss is $\sim 15\%$ for every 250 orbits, if the resolution is $(N_r, N_\theta)=(500, 720)$, while mass-loss is negligible if resolution is increased to $(N_r, N_\theta)=(1500, 3000)$.

\subsection{Weak SG : \texorpdfstring{\(q_0 =\)}{I} 2 , 4 }

There is no significant difference compared to the non-SG case where the vortex is slightly stretched by the shear.
Only a marginal contraction and flattening of the vortex is observed; 
its radial extent and aspect ratio change by less than 15\% and 5\%, respectively.
On the other hand, spin is not unchanged and vortex  mass is weakly affected (less than 5\% with respect to initial mass).
Finally, as desired, migration is very slow, only 0.1 AU in 300 orbits, which is nearly negligible with regards to the estimated numerical precision.

\subsection{Intermediate SG}

This case corresponds to the transition between stretching and splitting under gravitational forces.
In the first occurrence, the initial vortex is deformed but keeps a coherent structure, whereas in the second one, coherence is lost and the vortex tends to split into independent vortices.

\subsubsection{Vortex shaped by the SG : \texorpdfstring{\(q_0 =\)}{I} 1 }\label{subsubsc:vortex shaped by SG}

If $q_0 =1, $ the vortex is deeply modified by the SG.
It is stronger than the initial vortex with a 1\% increase in the mean Rossby number, but it is also: (i) stretched in the azimuthal direction with a 50\% increase in aspect ratio; and (ii) contracted in the radial direction with a 15\% decrease in $\delta$. 
In particular, the aspect ratio rapidly reaches  a maximum at t$\sim$100 $t_0$, followed by a slow decrease until the end of the simulation. 
It is important to stress that this increase in vorticity contrasts with the evolution of a vortex only stretched by the shear.
Migration is slightly accelerated to 0.15 AU in 300 orbits but, similarly to previous cases, it remains very weak.
Vortex mass declines by 5\%.
It is important to mention that, in this case, both vortex strength and aspect ratio increase simultaneously.
This observation, in disagreement with the statement that vortex strength is inversely proportional to $\chi$, confirms the choice we made in Sect. \ref{subsec:suitable parameters}.
Finally, the migration of the vortex is 50\% faster than in the previous case (which remains negligible over 300 orbits) and its mass declines by $\sim 5\%$.

\subsubsection{Vortex splitting: \texorpdfstring{\(q_0= \)}{I} 0.5 }\label{sec:vortex splitting}

At $t=60$ $t_0$, a sudden increase of 92\% occurs in the aspect ratio, and a sudden decrease of 30\%  occurs in the radial extent. 
These sharp variations are followed, at the 90$^{th}$ orbit, by a splitting.
Just before the splitting, $|\overline{Ro}|$ sharply varies as a function of time.
This likely indicates that the release of a secondary vortex results from the changes introduced by SG in the fluid-force equilibrium (pressure gradient, star attraction, and fictive forces); 
in the present case, ejection of secondaries could result from the predominance of the centrifugal forces.
In particular, this secondary vortex picks up mass to the primary which, at $t=90$ $t_0$, can lose more than 25\% of its initial mass.\\

After the first release, the secondary vortex remains in the co-orbital region and crosses the main vortex at  $\sim 117\, t_0$ and $\sim 145\, t_0$.
During each encounter, the main vortex can pick up part of the secondary mass and, at the end of the run, reaches 60\% of its initial value.
After the 220$^{th}$ orbit the flow stabilises in a no-encounter state.
The secondary, in gravitational interaction with the main vortex, has horseshoe oscillations of period $T_{hs}\sim 65 t_0$.
The exchange of angular momentum between the two vortices can easily explain the oscillatory behaviour observed in the migration rate with the same period (see the top panel of Figure \ref{fig: migration, mass, hill and rossby}). 
   
\subsection{Strong SG : \texorpdfstring{\(q_0= \)}{I} 0.25}\label{subsec: strong SG}

In this case the injected vortex is stiffly modified with a drastic increase in its spin and a strong radial contraction; it produces two secondary vortices, which tend to cascade into smaller vortices or transient eddies.

Similarly to the previous case, after t=75 $t_0$, the vortex is strongly deformed and loses more than 60\% of its mass during the ejection of two secondaries.
The evolution of the secondaries is also quite complex;
it is governed by three-body interactions with the primary-vortex, but also by successions of splitting, merging, or dissipation processes.
After the 190$^{th}$ orbit, the vortex strongly decays and finally disintegrates at the 300$^{\mathrm{th}}$ orbit in an annular overdensity and two residual vortices that are located at $[20^{\circ}, 180^{\circ}]$.
At this stage the density maximum in the annulus reaches half the peak density of the initial vortex.
At the end of the run, the main-vortex has lost 75\% of its mass.

\subsection{Summary of vortex evolution}

Three different regimes have been identified in our simulations.
The first evolution is dominated by \textbf{(A)} shear ($1\lesssim{q_0}$): there is no significant departure from the non-SG case. The outer regions of the vortex are slightly stretched by shear and SG.
The vortex reaches a quasi-steady regime with nearly constant mass and migration.
The second regime is the domain of \textbf{(B)} self-gravitating vortices ($0.5\lesssim {q_0}\lesssim 1$): SG is able to change the morphology and spin of the vortex: vortices spin slightly faster, contract radially, and elongate.
This is a new regime that vortices can reach for intermediate values of $q_0$.
Our simulations show that such vortices have a stable and long-term evolution, similar to the non-SG case.
This suggests the existence of quasi-steady vortex solutions of the Euler and Poisson equations, but further investigations are needed for firm conclusions on this point. 
We must stress that regime \textbf{(B)} includes the case of vortices releasing a single secondary vortex, such as  observed when $q_0=0.5$;
it corresponds, in fact, to a marginal case.
Indeed, no splitting of the main vortex is observed when the same run (at $q_0=0.5$) is performed using the methodology presented in Section \ref{sec:high resolution simulations}.
The last occurrence refers to \textbf{(C)} splitting by SG (${q_0} < 0.5$): SG strongly affects the vortex shape, up to destruction, but also the global flow.
The initial vortex releases more than one secondary vortex, which remain trapped in the co-orbital region of the primary.
In this regime the global parameters of the primary significantly change as a function of time.
For smaller values of $q_0,$ either the vortex directly releases multiple secondaries, or the process repeats on the primary with successive releases of weaker secondaries and/or eddies.
The result is a progressive fading of the main vortex that correlates with an increasing number of eddies and the growth of an annular overdensity.

In this last case $Q_0=5$ whereas, in the vortex core, the mean value of Toomre parameter is $Q\sim10$.
These values of $Q_0$ and $Q$ indicate that vortex splitting, although due to SG, cannot be the result of a standard gravitational instability with a criterion similar to that derived in the case of rotating discs.
In the following we suggest a possible way to extend this standard criterion to address vortex stability.

\section{Vortex stability}
\label{sec:vortex stability}
\begin{figure*}
\centering
\resizebox{\hsize }{!}
  {\includegraphics[trim = 3.0cm 1.0cm 2.5cm 2.5cm, clip, width=8cm]{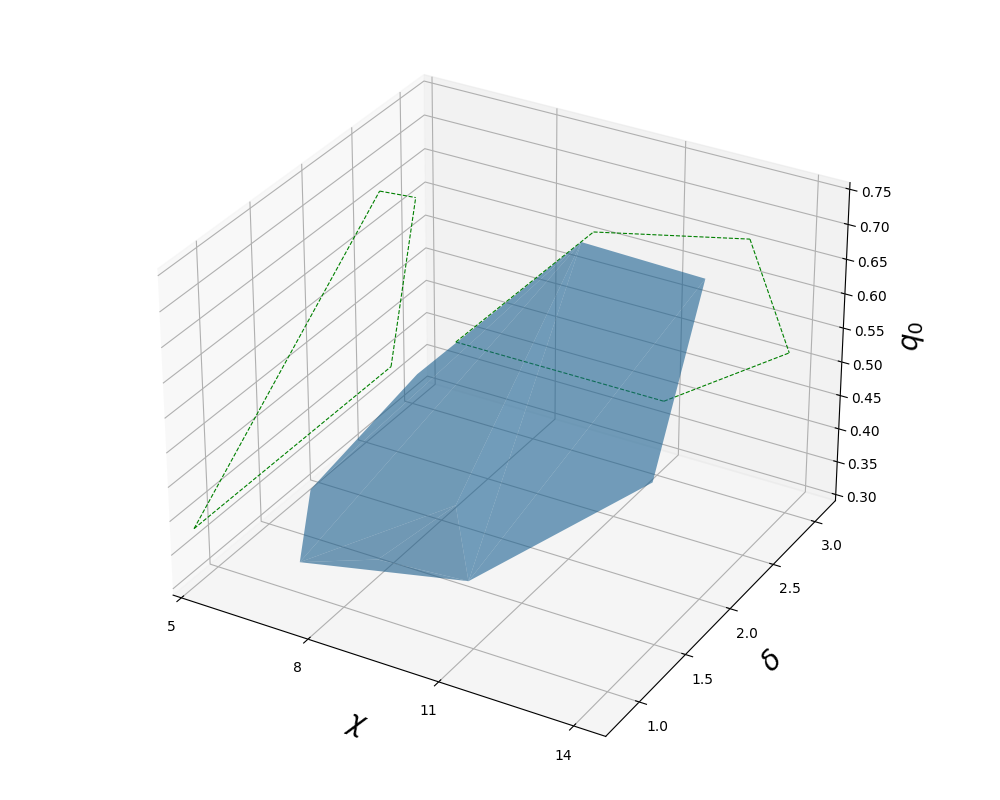}
  \includegraphics[trim = 3.0cm 1.0cm 2.5cm 2.5cm, clip, width=8cm]{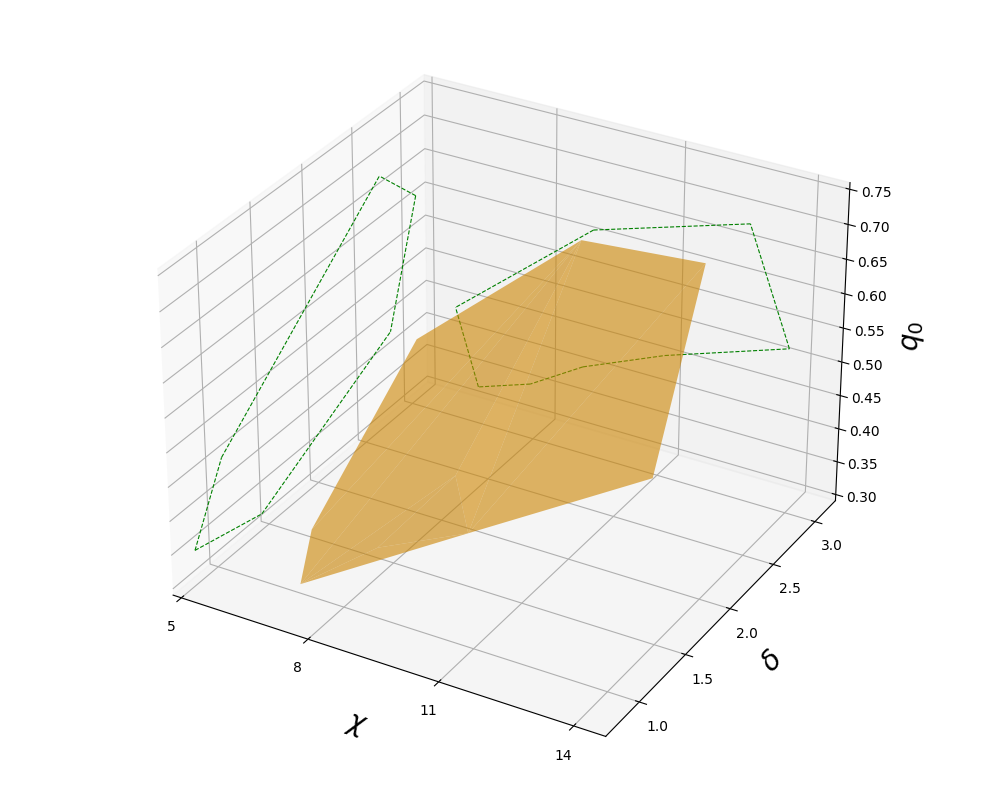}  }

\caption{\textbf{Stability domain of the self-gravitating vortices as a function of $q_0$.} \emph{Left,} when estimated from numerical simulations and \emph{right,} when deduced from Equation \ref{Eq:vortex stability final1}. Vortices are stable above the surface and unstable below (blue and orange, respectively). 
Both surfaces were projected on the left-hand side and the right-hand side of the box.
$\chi$ and $\delta$ are, respectively, the vortex aspect ratio and the radial extent at $t=50 \, t_0$.}
\label{fig: vortices stability surface}
\end{figure*}

This section is devoted to better characterising the stability of vortices submitted to their own gravity.
This was achieved thanks to two different methods: a study based on the compilation of an extended series of numerical simulations, and a theoretical estimate of the vortex parameters when constrained by Toomre's criterion analogue.

\subsection{Numerical study}\label{subsec:numerical study}

Here we used the same method as in Section \ref{sec:evolution},
with the same numerical resolution but with initial parameters selected among the range of values gathered in Table \ref{tab:vortex features}.
Our definition of vortex stability is a simple but empirical one: vortices are said to be `stable' if \textbf{(1)} they survive until the end of the run (300 $t_0$), \textbf{(2)} there is a possible ejection of small scale eddies, and \textbf{(3)} there is a relative density contrast decrease no higher than 20\%: $|C_{\sigma}(t_f)-C_{\sigma}(t_i)|/C_{\sigma}(t_i)\leq0.2,$ where $t_i=50\,t_0$ and $t_f=300\,t_0$.
A total of 45 different runs were carried out and provide a set of dots characterising the end-of-run state of the vortex in the ($\chi, \delta, q_0$) space.
Raw results are reported in Figure \ref{app: vortices stability dots}.
They are presented another way in Figure \ref{fig: vortices stability surface}, in the form of a stability map in the left panel, and are compared to our theoretical estimates in the right panel (see Sect. \ref{subsec:theoretical estimate}).
When the aspect ratio is constant: (i) if $\chi\sim5$, vortex stability is independent of $\delta$; the vortex is stable if $q_0 > 0.375$ and unstable otherwise and (ii) if $\chi \sim 14$, vortex stability depends on $\delta$; the critical value of $q_0$ increases with $\delta$ (from $0.375$ to $0.65$), and therefore large vortices are more easily destabilised than small vortices (even at lower surface-densities).
When the radial-extent is constant: (i) if $\delta \sim 1$, vortex stability is independent of the aspect ratio; the vortex is stable if $q_0 > 0.375$ and unstable otherwise and (ii) if $\delta \sim 2.5$, vortex stability depends on the aspect ratio; the critical value of $q_0$ increases with $\chi$ (from $0.375$ to $0.65$); in this case, elongated vortices are more easily destabilised than more compact vortices (even at lower surface-densities).

In summary, SG is able to significantly change the stability of gaseous vortices. Assuming standard disc configurations with $h(r_0)=0.05$ and $\beta_\sigma=-1.1,$ we find that SG destabilises all vortices if $q_0 < 0.375$ (or again $Q_0 < 7.5$) and all vortices remain stable for moderate SG when $q_0 > 0.65$.
In a general way, small and compact vortices are more stable against SG splitting than large and elongated ones.
Finally, a destructive splitting of the vortices can occur if $Q_0$ and $Q$ are approximately of the order of $\sim 8$.

\subsection{Theoretical estimate}\label{subsec:theoretical estimate}

Gravitational stability of circumstellar discs is constrained by the famous Toomre criterion $c_s \kappa / ( \pi G \sigma )>1$, where $\kappa$ is the epicyclic frequency \citep{1964ApJ...139.1217T};
in contrast, little is known on the gravitational stability of massive gaseous vortices.
Here, in an initial attempt to resolve the problem, we extrapolate this criterion by replacing the local epicyclic frequency with the vortex vorticity $\omega=2 \, |\overline{Ro}| \, \Omega_K$; 
assuming that $|\overline{Ro}|\sim |Ro|_{max}$, we get:
\begin{equation}
\hspace{3cm} 2\frac{c_s \abs{Ro}_{max}\Omega_K}{\pi G \sigma} \gtrsim 1.
\label{SG-stab}
\end{equation}
If this condition is satisfied, we say that the vortex is stable against SG (hereafter `SG stable').
Of course such a definition is somewhat inappropriate since, stricto sensu, the criterion refers to the stability of a Keplerian disc in which the physical conditions would be the same as in the vortex core. Nonetheless, we use it below to make the discussions easier.
Condition \ref{SG-stab} can be reformulated using the fact that vortices are also pressure and density bumps, namely $P = P_{0} \left( 1 + C_P \right)$ and $\sigma = \sigma_{0} \left( 1+C_\sigma \right)$\footnote{Using definitions exposed in Table \ref{table:parameters definition} we obtain the sound speed in the vortex core, $c_s=c_{s,0} \sqrt{{(1+C_P)}/{(1+C_\sigma)}}$, where $c_{s,0}$ is the unperturbed sound speed.}.
Accounting for a scaling factor of $\sim1.65$ to better fit the numerical data, we propose an empirical condition for SG-stable vortices:
\begin{equation}\label{Eq:vortex stability final1}
\begin{array}{cr}
q_0 \gtrsim q_{0,c}=\displaystyle\frac{0.83 \, h}{\abs{Ro}_{max}}\frac{(1+C_\sigma)^{3/2}}{(1+C_P )^{1/2}} & \mbox{(hosting disc criterion),}
\end{array}
\end{equation}
or again,
\begin{equation}\label{Eq:vortex stability final2}
\begin{array}{cr}
 1.2 \abs{Ro}_{max} \, Q \gtrsim 1  & \qquad\qquad\qquad\mbox{(vortex core criterion),}
\end{array}
\end{equation}
where $q_{0,c}$ is the value above which vortices are SG stable and $Q$ is the Toomre parameter estimated in the vortex core.
Consistency of this condition with our numerical simulations was tested by computing $q_{0,c}$ at the end of the first relaxation phase (at $t=50\, t_0$).
The values of $|Ro|_{max}$, $C_\sigma$, and $C_P$ necessary for the computations are gathered in Table \ref{tab:triplets Ro, delta_sigma, delta_P}, and the values of $q_{0,c}$ are presented in a map plotted in Figure \ref{fig: vortices stability surface} (right panel).
It is striking that the two maps (right and left panels) are nearly parallel to one another in the ($\chi$, $\delta$, $q_0$) space.
Applying the above criteria to the vortex of Section \ref{sec:evolution}, we find that such a vortex is SG stable if $Q_0 \gtrsim 10.9$, or $Q \gtrsim 6.9$, a condition that is consistent with our conclusion in the aforementioned section, stating that vortices begin to split even if the hosting disc and the vortex core are stable following the standard Toomre criterion.

\begin{table}
\caption{Triplets ($|Ro|_{max}$, $C_\sigma$, $C_P$) for simulated vortices at $t=50\,t_0$.}                                  
\label{tab:triplets Ro, delta_sigma, delta_P}  
\centering                                     
\begin{tabular}{@{}l| c c c}                   
\hline                                         

\diagbox[height=2.em,width=2.em]{$\chi$}{$\delta$} &       1.0           &        1.5          &          2.5         \\ 
\hline                                         
5  &  (0.14, 0.20, 0.31) &  (0.17, 0.50, 0.78) &  (0.21, 1.73, 3.27)  \\  
8  &  (0.13, 0.32, 0.48) &  (0.15, 0.72, 1.19) &  (0.19, 2.49, 5.12)  \\  
14 &  (0.12, 0.38, 0.57) &  (0.13, 0.82, 1.37) &  (0.16, 2.72, 5.72)  \\  
\hline
\end{tabular}
\end{table}

\subsection{Summary}

The simple theoretical estimate we used to study the stability of the vortex core against SG allowed us to define a simple criterion based on a modified form of Toomre's criterion. 
The new criterion was formulated in two different ways (see Equations \ref{Eq:vortex stability final1} and \ref{Eq:vortex stability final2}) to make the comparison easier. 
The widened numerical experiment we carried out also allowed us to build up a mapping of the SG-parameter, $q_0$, in the ($\chi$, $\delta$) space.
Vortex stability is thus presented in the form of a simple map of the threshold value $q_{0,c}$ as a function of $\delta$ and $\chi$, selected in the range $0.8\leq\delta\leq3$ and $5.7\leq\chi\leq13.6$, respectively.
In light of our numerical data, the stability criterion appears satisfactory but remains a somewhat empirical relation that requires further analysis and comparison.
Our simulations showed that, when vortices are unstable with respect to SG, their destruction is preceded by a strong stretching that is likely explained by the gravitational torque exerted by the vortex itself \citep{2017MNRAS.471.2204R}. 
Finally, this stability criterion could enable to identify upper density limits in hosting discs where a vortex presence is confirmed.
This will be the subject of a discussion in Sect. \ref{subsec:observational implications}.

Thanks to the threshold parameter $q_{0,c}$ obtained in this section, we were able to choose the suited parameters for studying self-gravitating vortices in greater detail at HR.

\section{{HR simulations}}\label{sec:high resolution simulations}

%
\begin{figure*}
\centering
\resizebox{\hsize}{!}
          {\includegraphics[trim = 2.0cm 1cm 4.0cm 2.5cm, clip, width=\hsize]{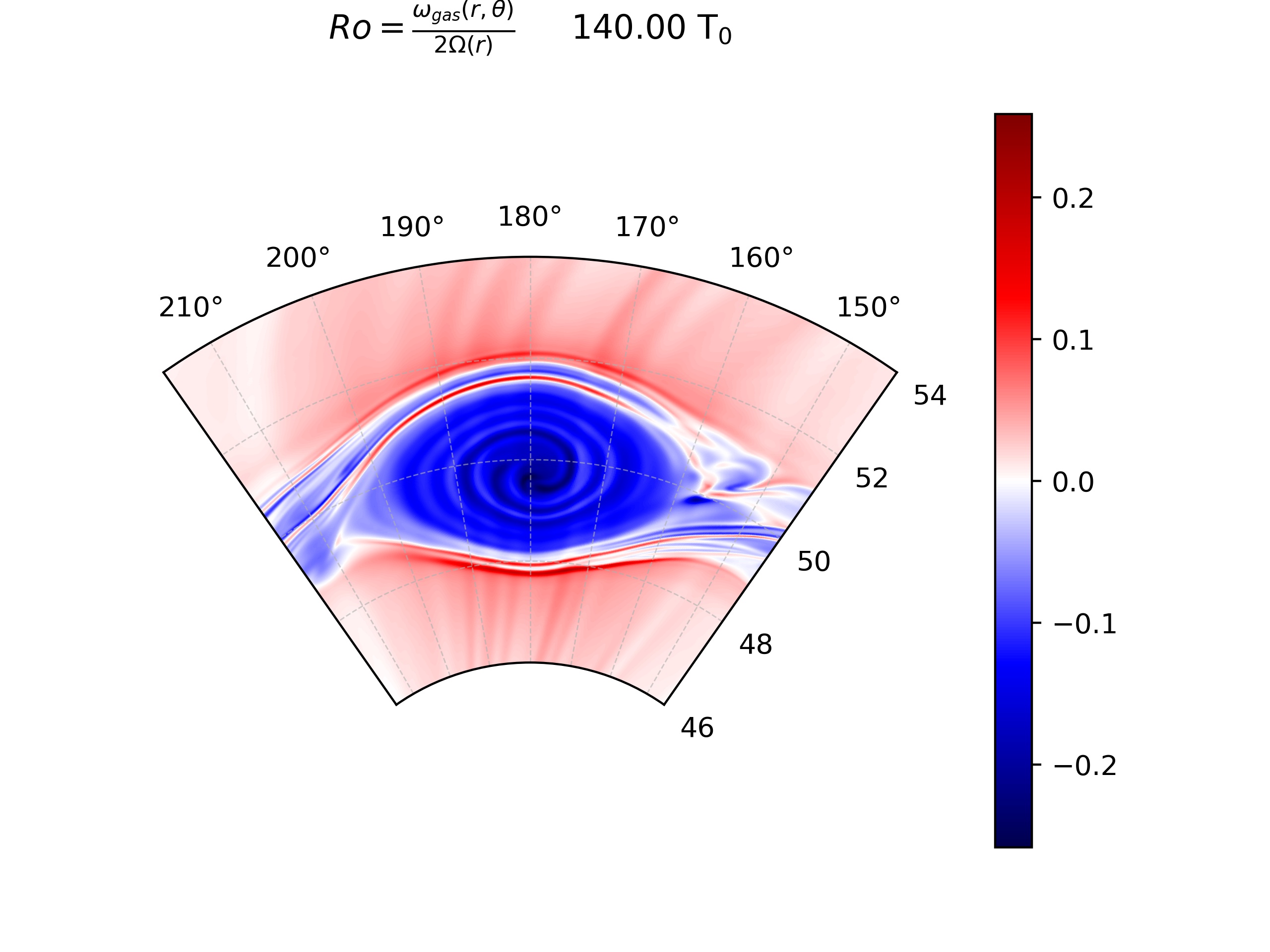}
          \includegraphics[trim = 0.8cm 0cm 1.7cm 0.75cm, clip, width=\hsize]{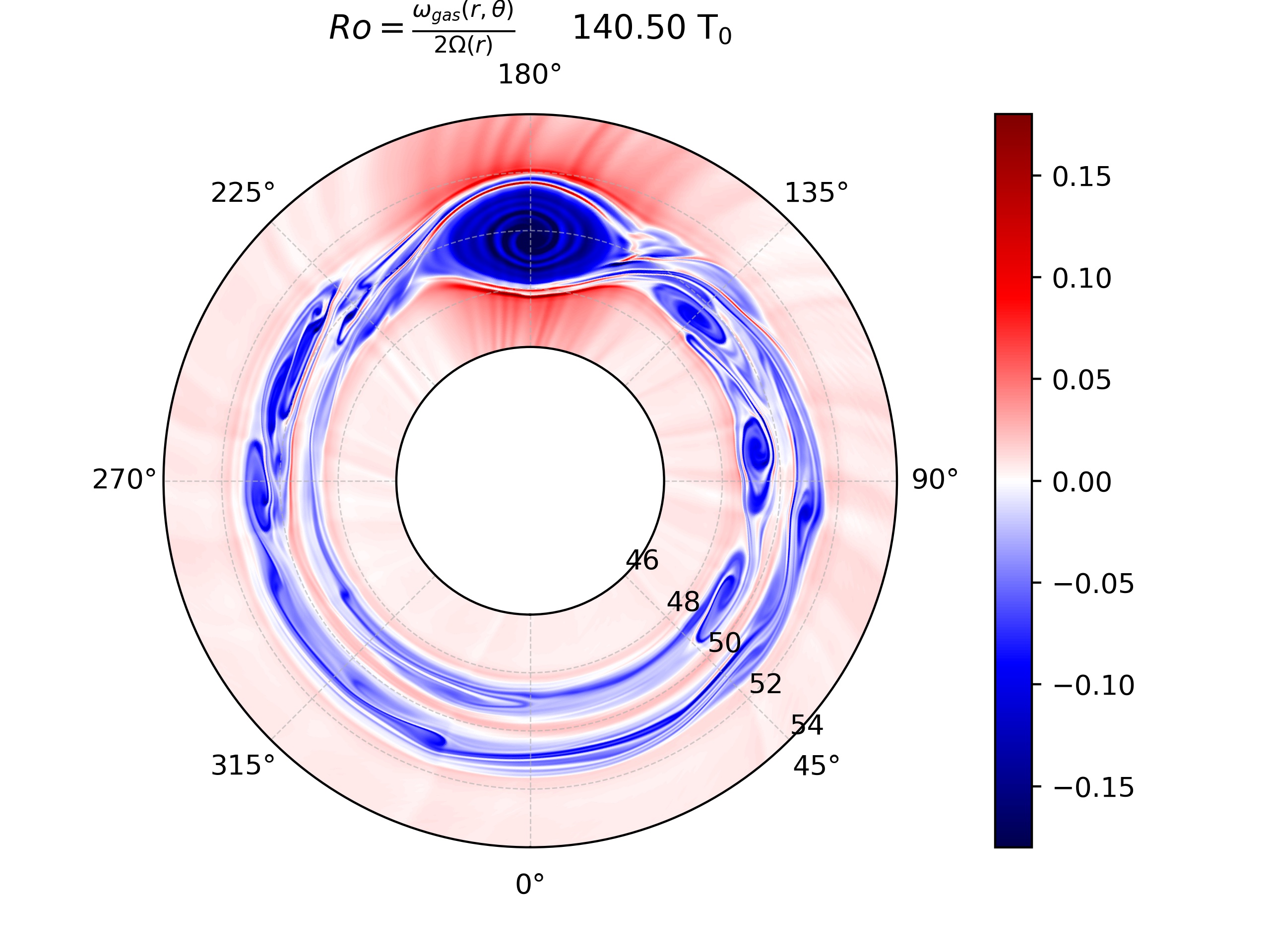}
          }
\resizebox{\hsize}{!}
          {\includegraphics[trim = 2.0cm 1cm 4.0cm 2.5cm, clip, width=\hsize]{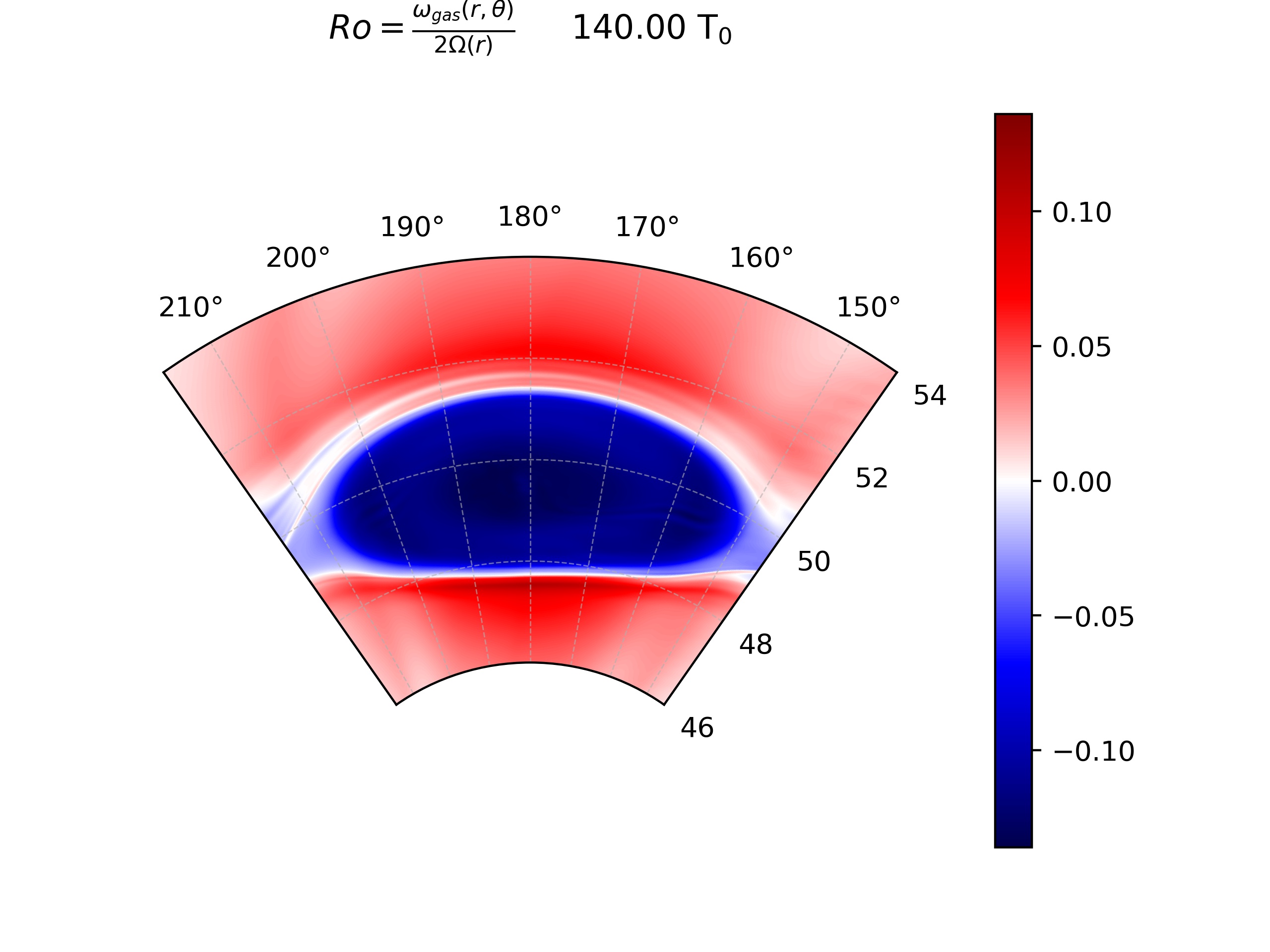}
          \includegraphics[trim = 0.8cm 0cm 1.7cm 0.75cm, clip, width=\hsize]{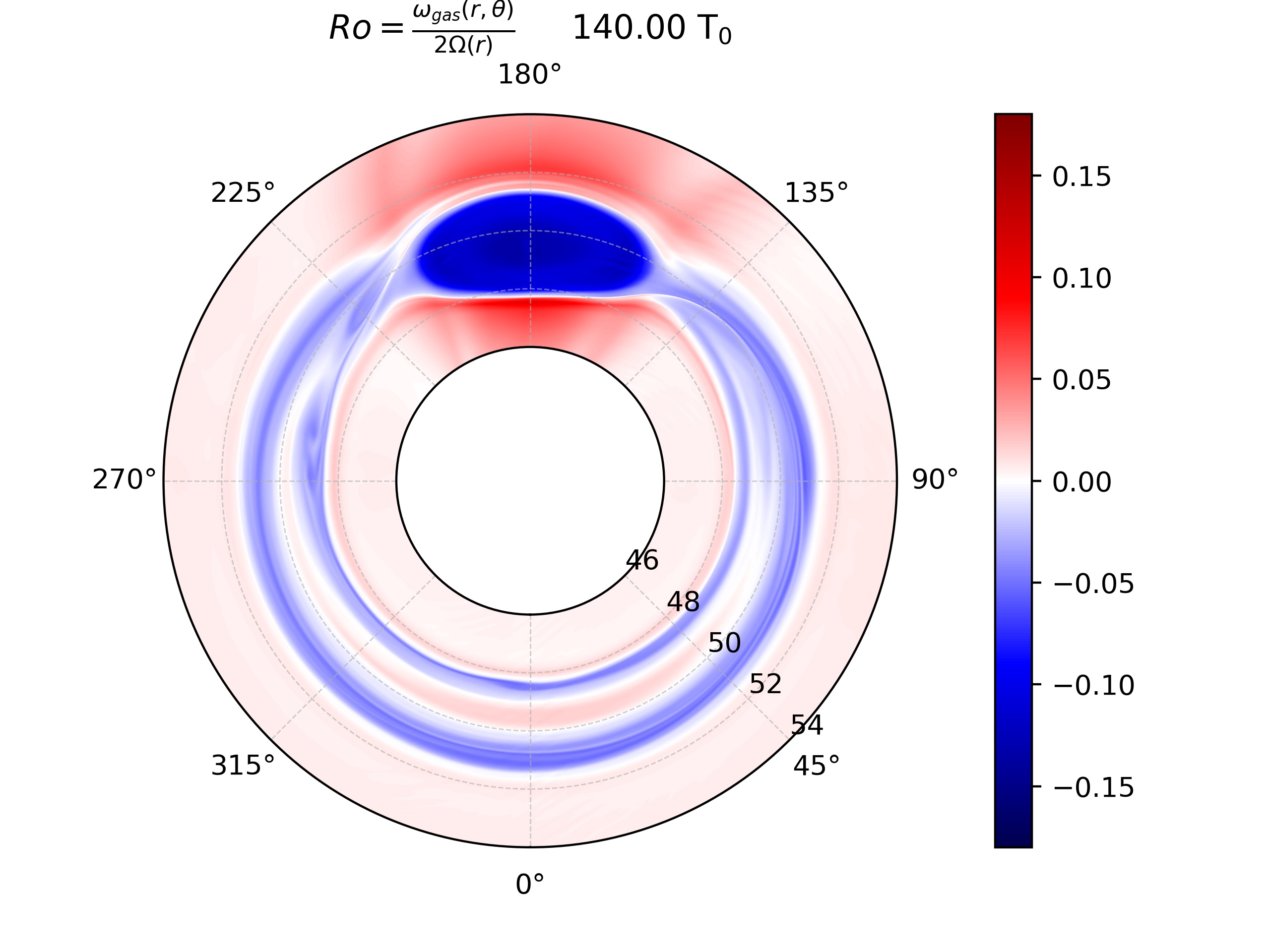}
          }
\resizebox{\hsize}{!}
          {\includegraphics[trim = 2.0cm 1cm 4.0cm 2.5cm, clip, width=\hsize]{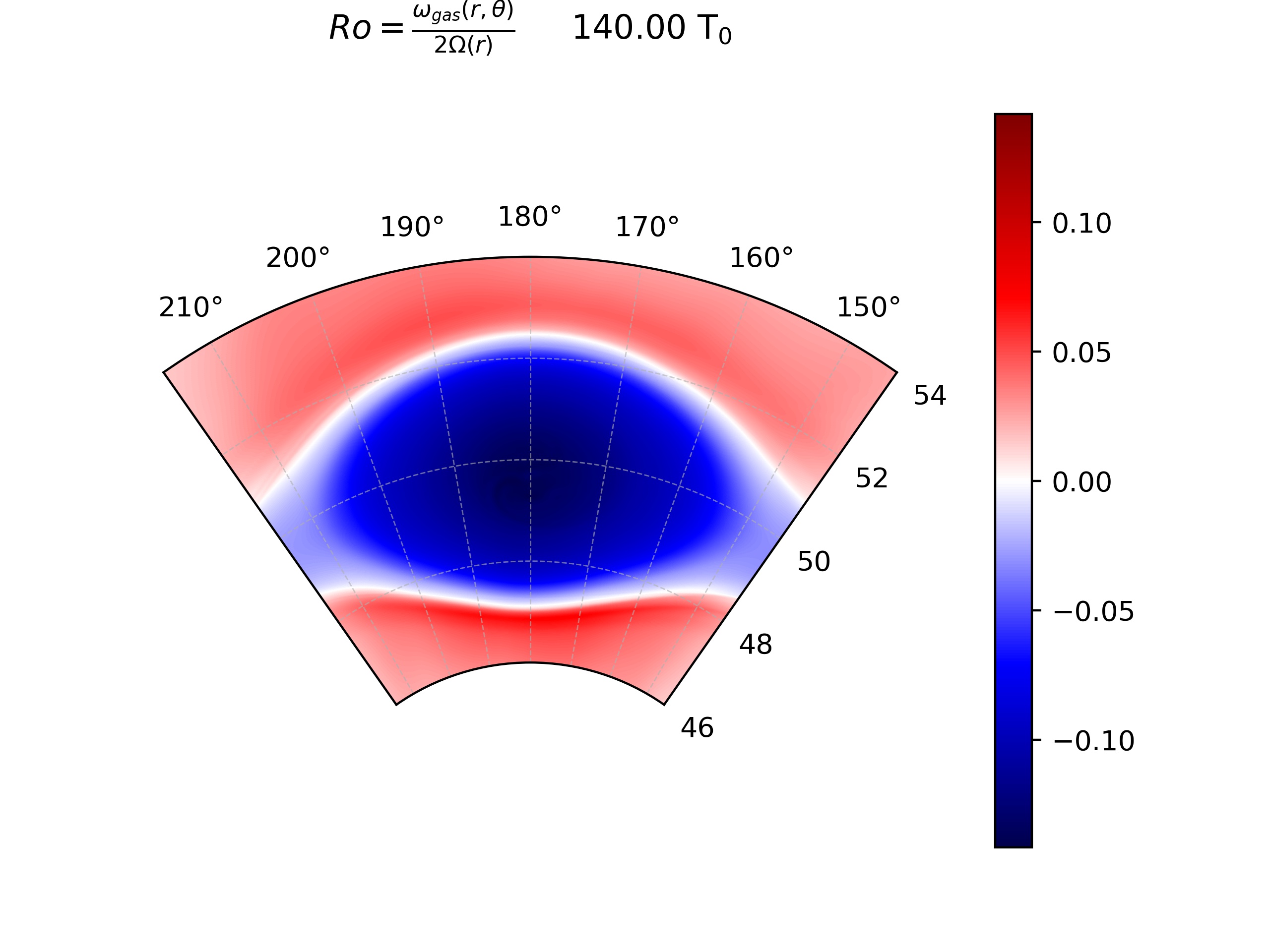}
          \includegraphics[trim = 0.8cm 0cm 1.7cm 0.75cm, clip, width=\hsize]{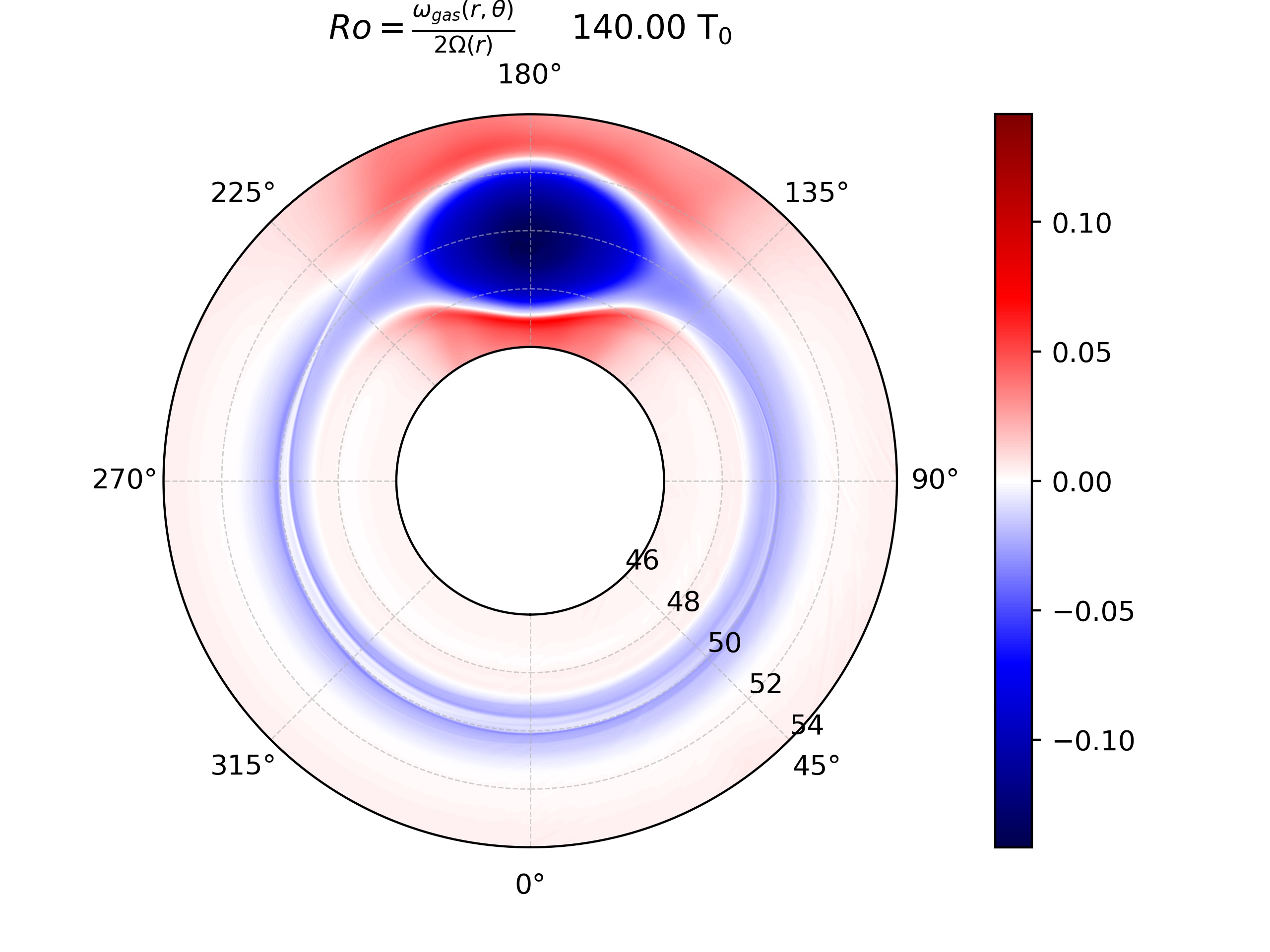}
          }
\caption{\textbf{Rossby number at $t=140\,t_0$ for an isothermal and a non-isothermal disc with SG.} Radial zoom with $r\in[46, 54]$ AU. \\
\emph{Top:} Self-gravitating vortex in an isothermal disc. \emph{\\ Middle:} Self-gravitating vortex in a non-isothermal disc. \\
\emph{Bottom:} Reference vortex, without SG and in a non-isothermal disc.}
\label{fig: comparison rossby isothermal and no isothermal}
\end{figure*}
%

%
\begin{figure*}
\centering
\resizebox{\hsize}{!}
          {\includegraphics[trim = 0.8cm 0cm 1.7cm 0.75cm, clip, width=\hsize]{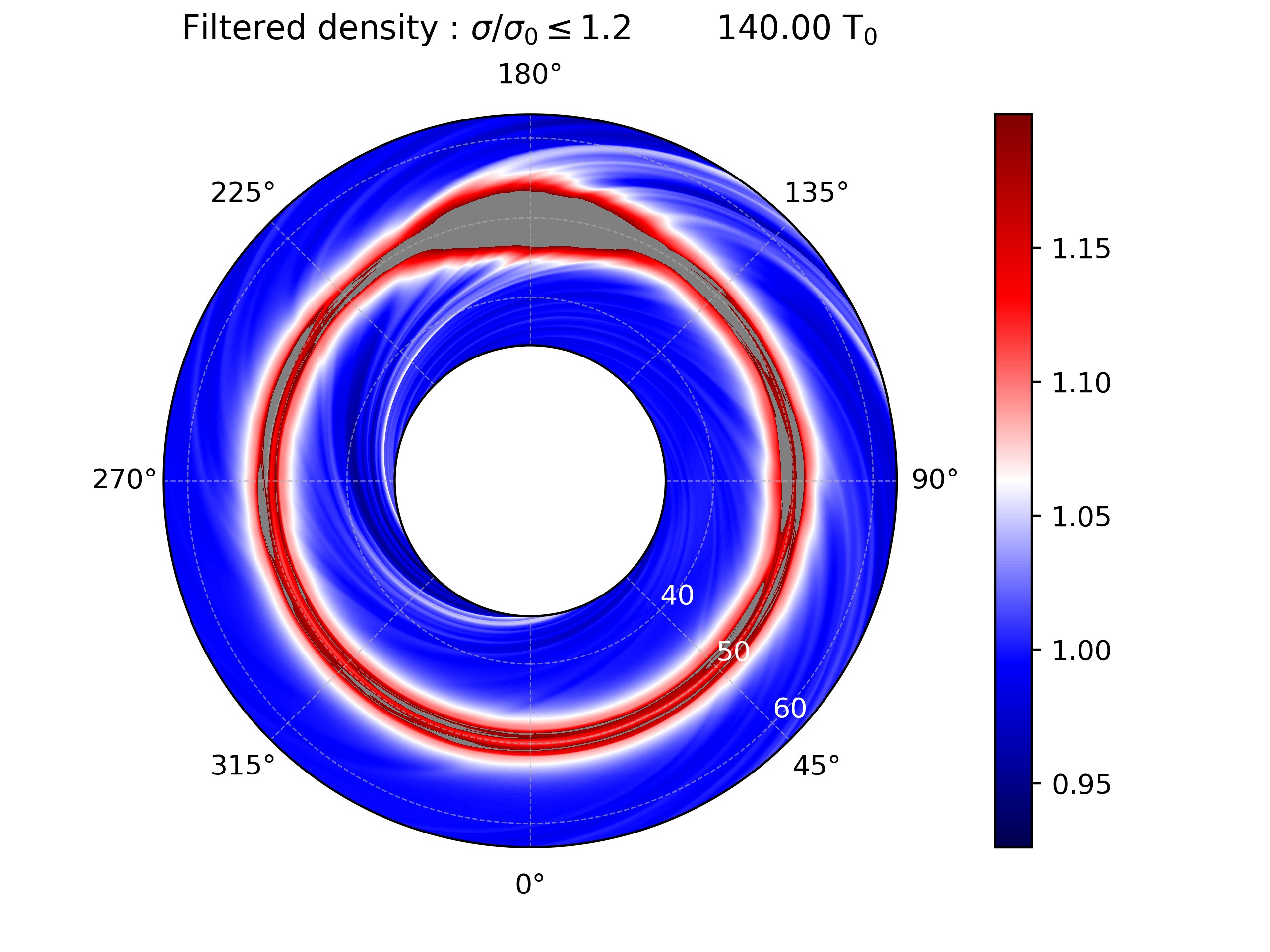}
          \includegraphics[trim = 0.8cm 0cm 1.7cm 0.75cm, clip, width=\hsize]{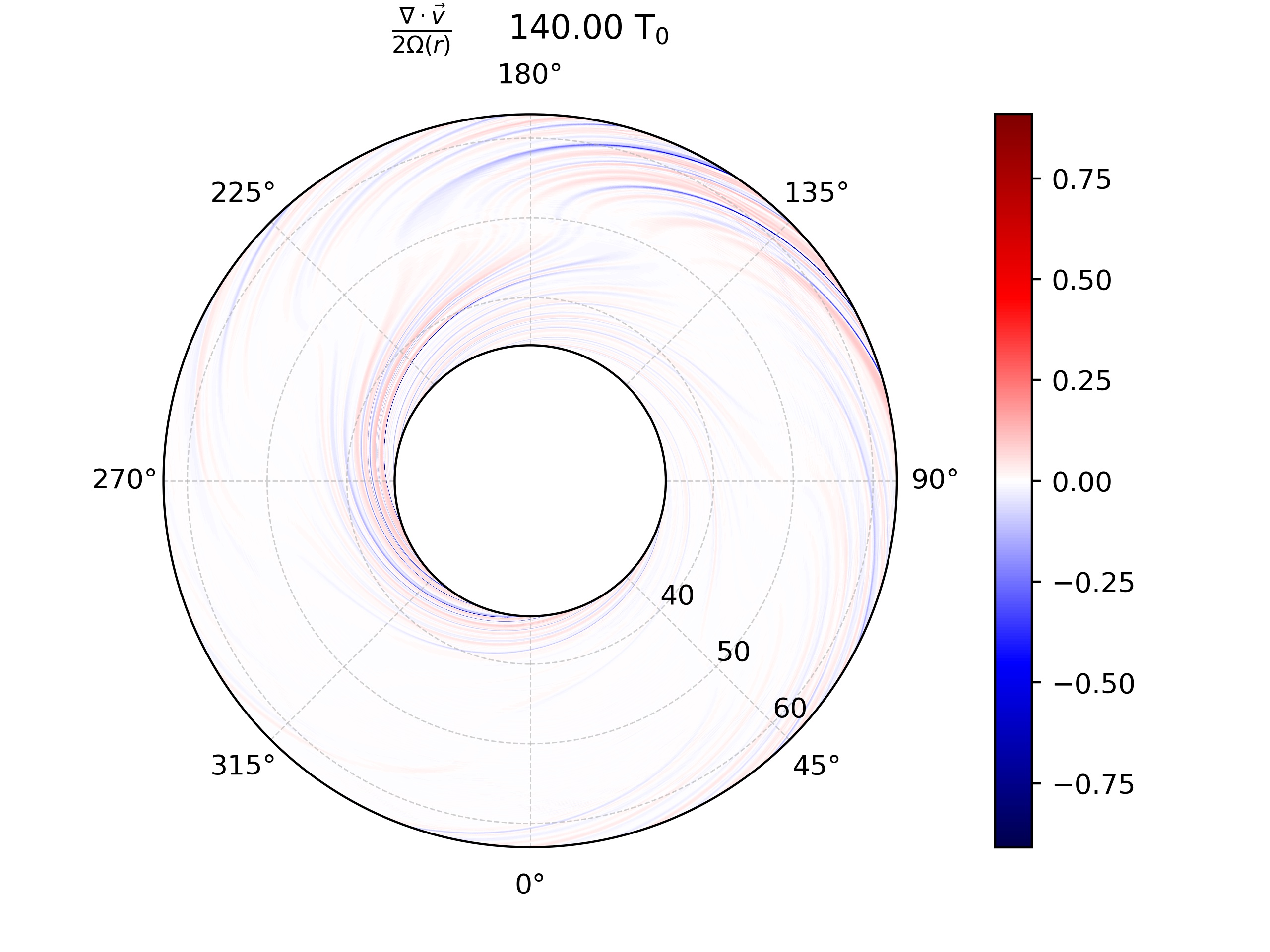}
          }
\resizebox{\hsize}{!}
          {\includegraphics[trim = 0.8cm 0cm 1.7cm 0.75cm, clip, width=\hsize]{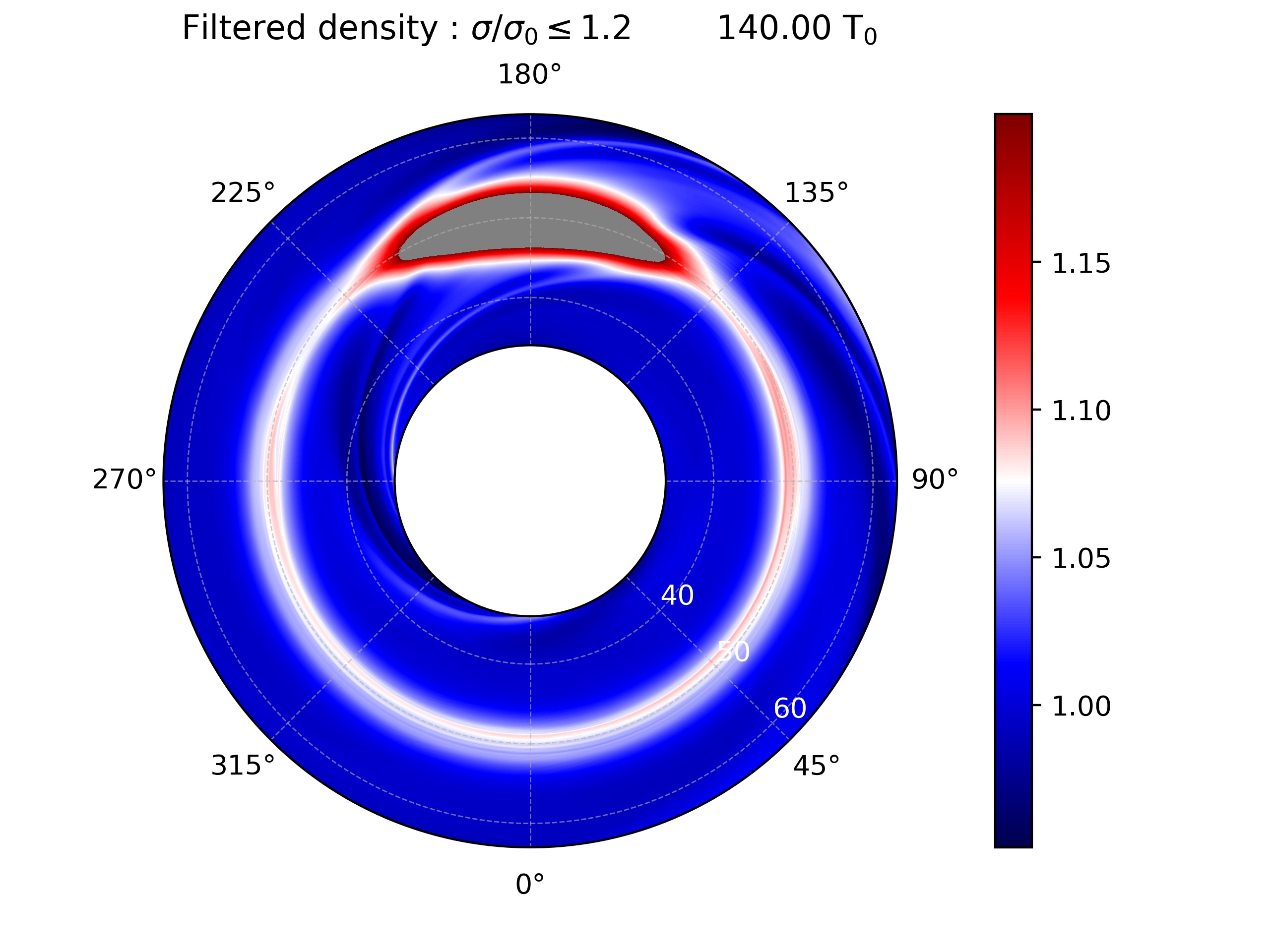}
          \includegraphics[trim = 0.8cm 0cm 1.7cm 0.75cm, clip, width=\hsize]{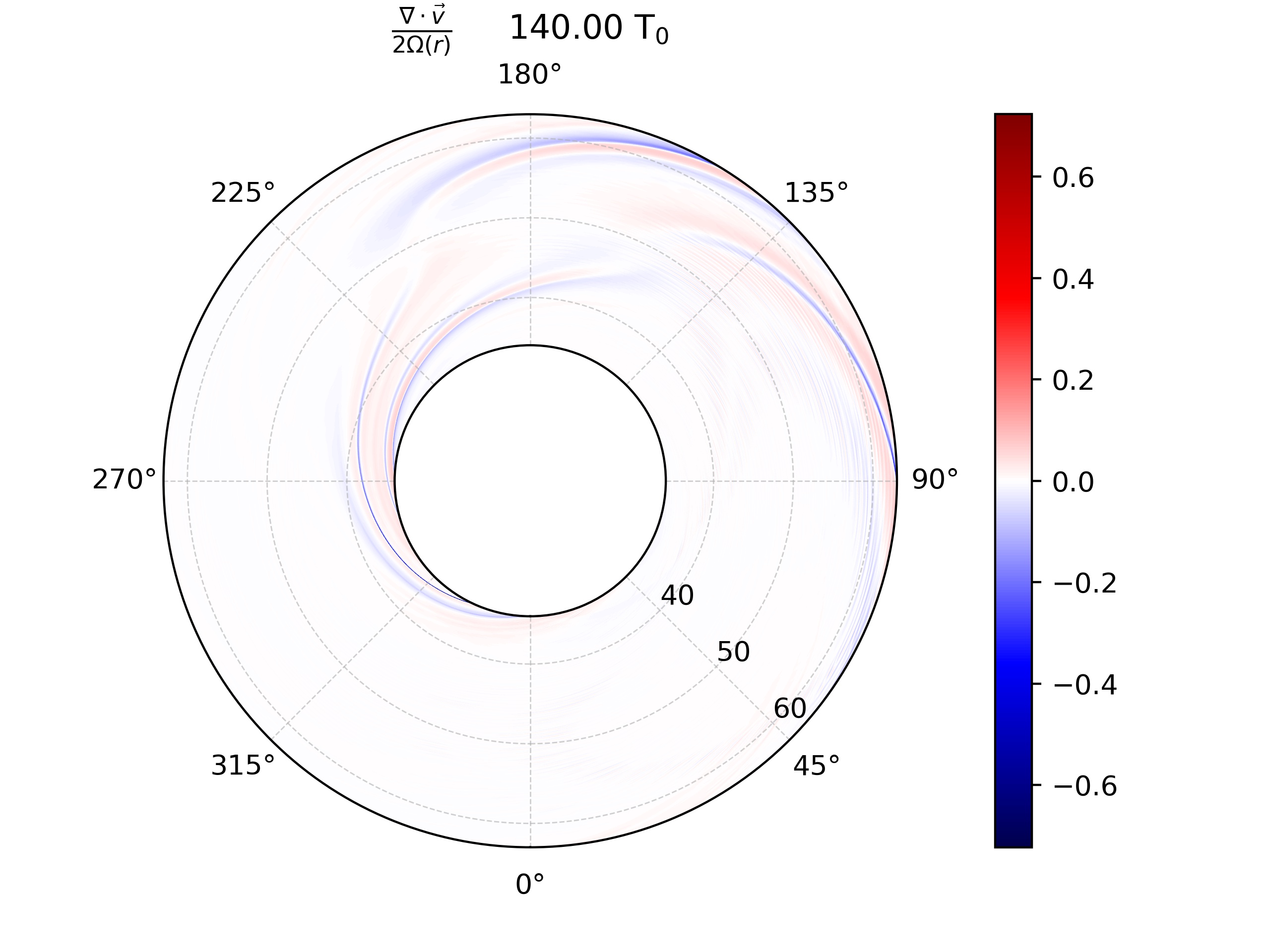}
          }
\resizebox{\hsize}{!}
          {\includegraphics[trim = 0.8cm 0cm 1.7cm 0.75cm, clip, width=\hsize]{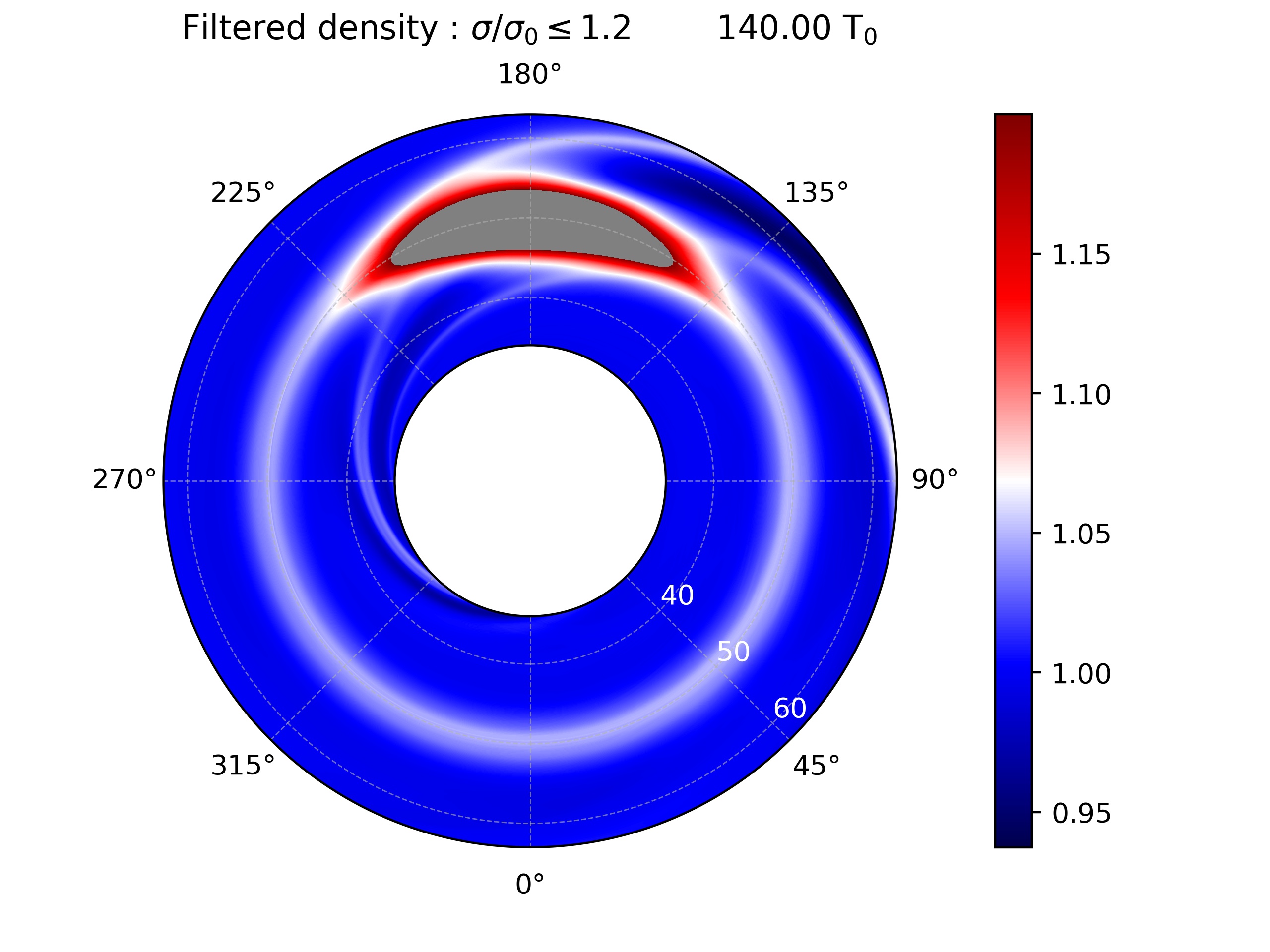}
          \includegraphics[trim = 0.8cm 0cm 1.7cm 0.75cm, clip, width=\hsize]{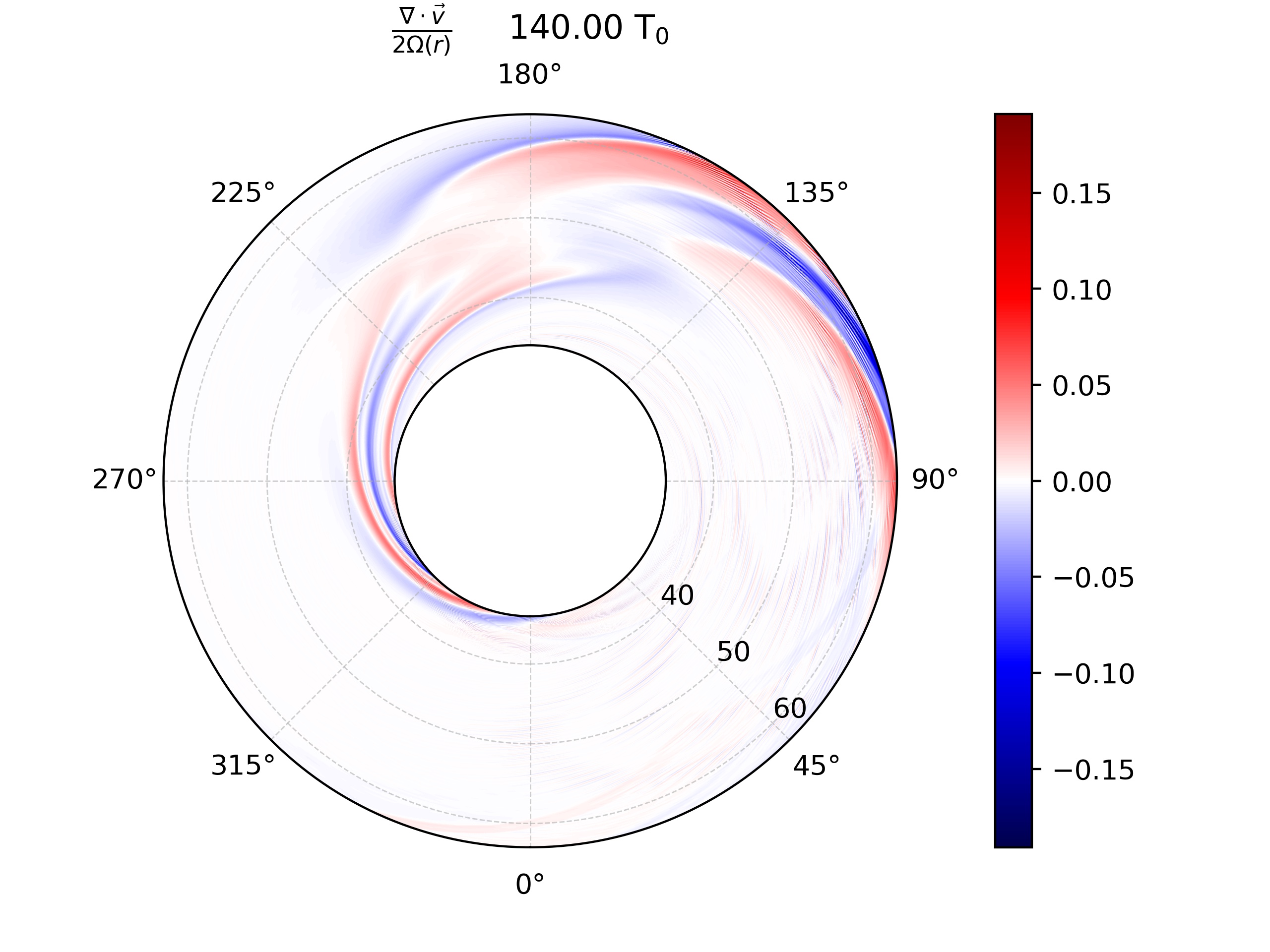}
          }
\caption{\textbf{Density and velocity-divergence at $t=140\,t_0$ for an isothermal and a non-isothermal disc (with and without SG).} \\
\emph{Left column:} Relative density ($\sigma/\sigma_0$). The main vortex is masked to distinguish the annular over-density and the spiral waves.\\
\emph{Right column:} Normalised relative velocity field divergence ($\nabla \cdot \vec{v'} / (2 \Omega_K)$). \\
\emph{From top to bottom:} Self-gravitating vortex in an isothermal disc, self-gravitating vortex in a non-isothermal disc, and the reference vortex (without SG and in a non-isothermal disc), respectively.}
\label{fig:density, div isothermal and no isothermal}
\end{figure*}
%

%
\begin{figure*}
\centering
\resizebox{\hsize}{!}
          {\includegraphics[width=\hsize]{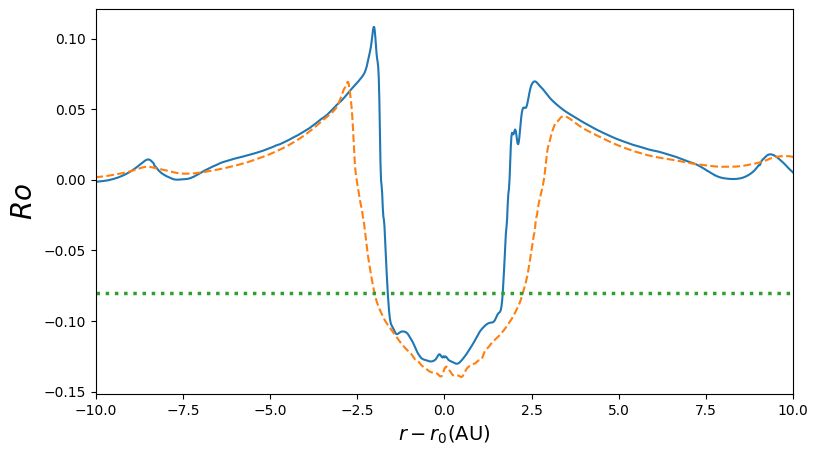}
           \includegraphics[width=\hsize]{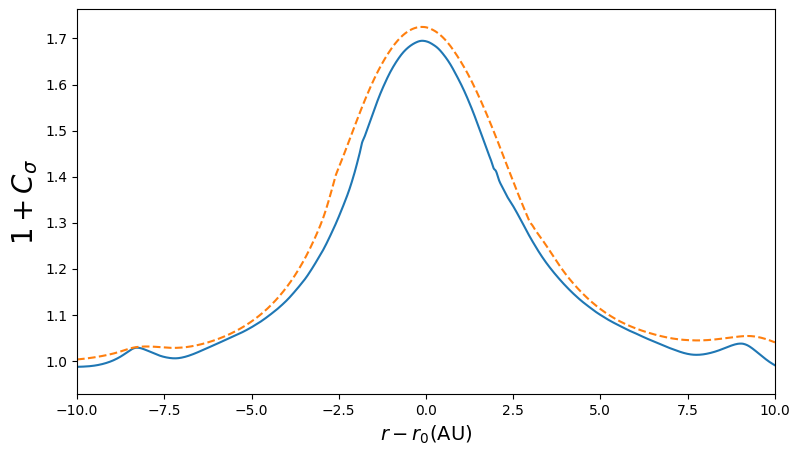}
          }
\resizebox{\hsize}{!}
          {\includegraphics[width=\hsize]{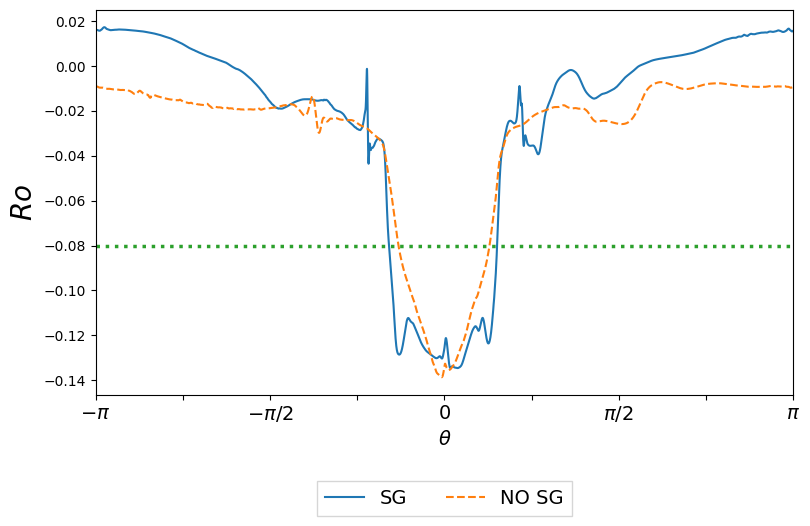}
           \includegraphics[width=\hsize]{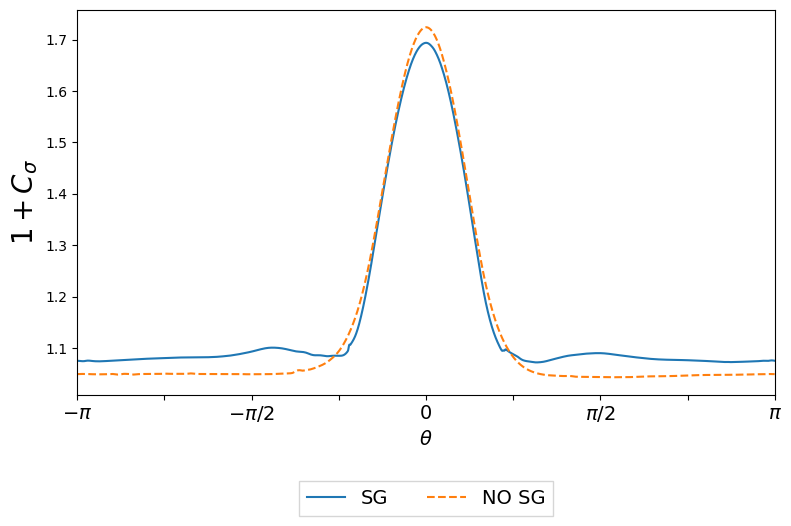}
          }
          
\caption{\textbf{Radial and azimuthal profiles of the Rossby number and the relative density at $t=140 t_0$ for a non-isothermal disc.} \\
\emph{Top:} Radial profile of the Rossby number and the relative density ($\sigma/\sigma_0$), from left to right, respectively.\\ 
\emph{Bottom:} Azimuth profile of the Rossby number and the relative density ($\sigma/\sigma_0$), from left to right, respectively. \\
The dotted green line corresponds to the level $Ro=-0.08$ 
used to compute the vortex geometrical parameters.
}
\label{fig:rho and rossby radial and azimuthal profiles}
\end{figure*}
%

%
\begin{figure*}
\centering
\resizebox{\hsize}{!}
{\includegraphics[trim = 0.5cm 0.0cm 1.0cm 1.0cm, clip, width=\hsize]{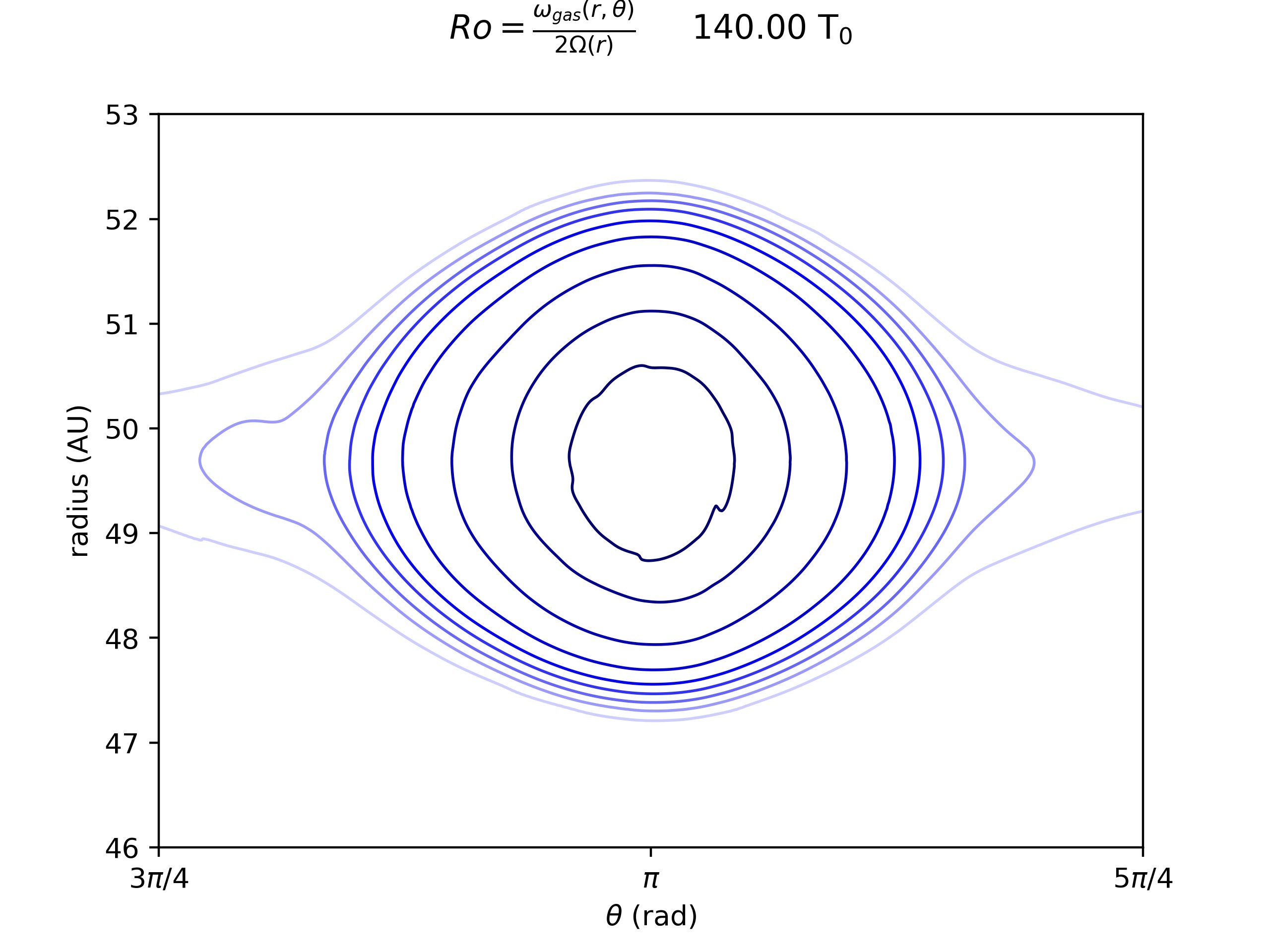}
 \includegraphics[trim = 0.5cm 0.0cm 1.0cm 1.0cm, clip, width=\hsize]{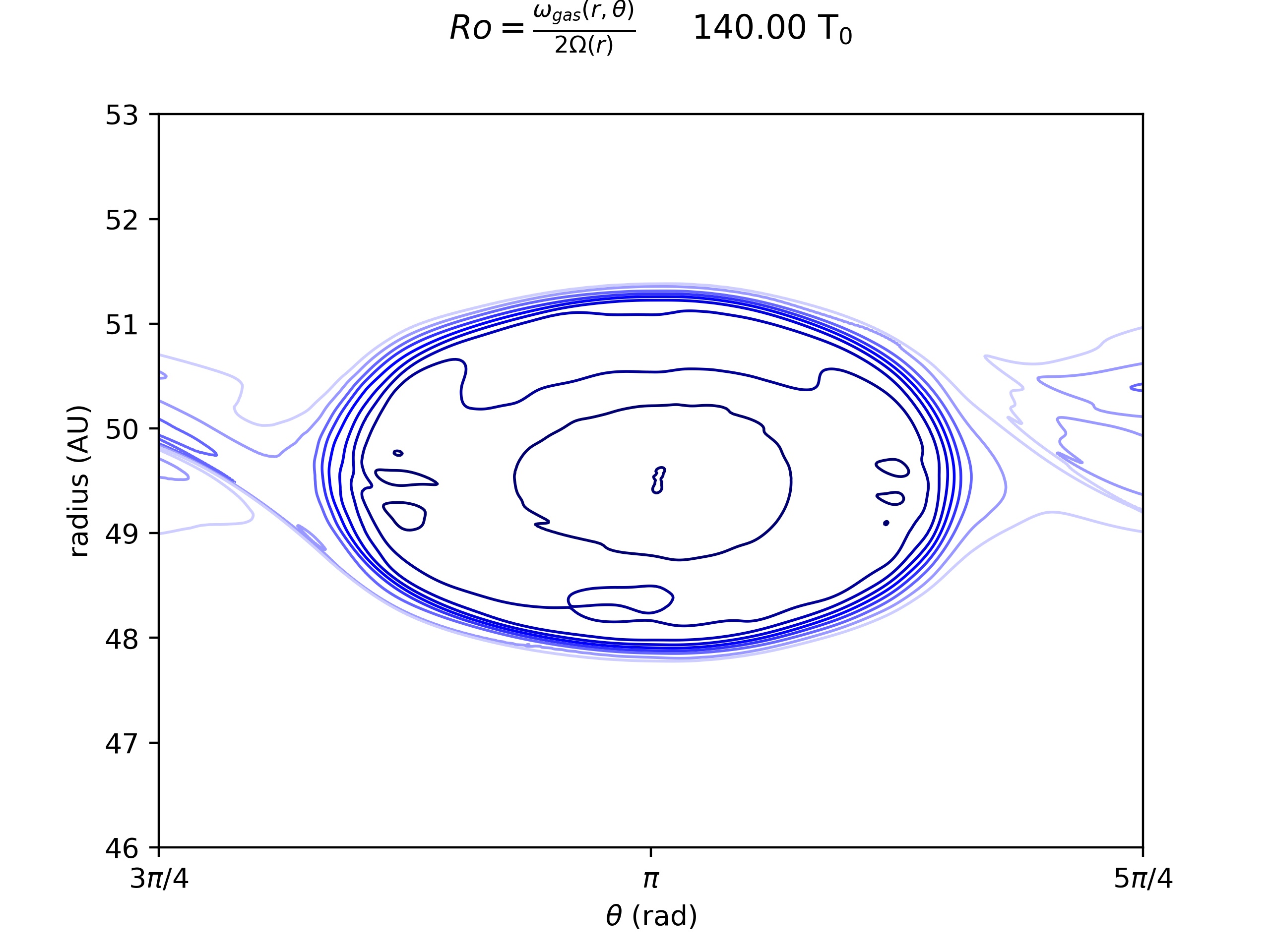}
          }
\caption{\textbf{Iso-contours of the Rossby number in a non-isothermal disc at $t=140\,t_0$.}
\emph{\\ Left:} Without SG.\\  \emph{Right:} With SG.}
\label{fig:rossby contour plots}
\end{figure*}
%

We performed HR simulations to better characterise the internal structure of the self-gravitating vortices identified in Section \ref{sec:vortex stability}.
We also studied two different cases, one in which the disc at equilibrium was isothermal and one in which it was not.
In Figures \ref{fig: comparison rossby isothermal and no isothermal} and \ref{fig:density, div isothermal and no isothermal}, the Rossby number, density, and compressibility at $t=140\,t_0$ are plotted.
For the density plots (right panel of Figure \ref{fig:density, div isothermal and no isothermal}), the main vortex was masked to highlight the weak contribution of other substructures associated with a vortex.
Besides the vortex internal structure, other features of interest are the spiral waves and the annular overdensity, which are intrinsically associated with vortices.
Indeed, such features are possible clues for theoretical models and observational studies as well. 
For reference comparison, the case of a vortex in a non-SG and non-isothermal case has been added to these figures. 
In Figure \ref{fig:rho and rossby radial and azimuthal profiles} the radial and azimuthal profiles of the density and the Rossby number are plotted for a non-isothermal disc in order to show the differences between quasi-steady vortices where SG is present and where it is not.
The analysis of local spin inside the vortex is completed by Figure \ref{fig:rossby contour plots}, where the iso-contours of the Rossby number have been plotted for a non-isothermal disc.
Finally, for completeness, the SG potential of the vortex and the induced forces in the radial and azimuthal directions are reported in Figure \ref{fig:SG potential and forces} in the case of the non-isothermal disc.

\subsection{Numerical resolution in the vortex core}

Numerical resolution is important for relevant descriptions of SG and gas evolution in the inner region of vortices.
In a circular vortex patch around a constant-density core, the enclosed mass is $m\sim\pi \displaystyle\left(\Delta r\right)^2 \sigma$.
Thus, the gravitational influence of this region of the vortex extends to a mean distance given by Hill's radius, $R_H=r_0 \displaystyle\left({m}/{3 M_{\odot}} \right)^{1/3}$. 
Then, if the radius of the circular patch reaches $\sim R_H$, the enclosed mass is $m\sim\pi R_H^2 \sigma$ and we get:
\begin{equation}\label{eq:hill vortex}
\hspace{3.25cm} R_H = \frac{H}{3 Q}.
\end{equation}
For instance, if $Q=10$ , we need to resolve (at least) 30 times the pressure scale height to correctly describe the vortex core.
In the foregoing results (cf. Section \ref{sec:evolution}), this condition is better satisfied in the radial than in the azimuthal direction.
This could contribute to the stronger contraction of the vortex observed in the radial than in the azimuthal direction.
Motivated by Equation \ref{eq:hill vortex}, we undertook HR simulations, the steps for which are described in next subsection.

\subsection{Activating SG for HR simulations}\label{subsec:Numerical resolution and SG continuous activation}

From a number of preliminary tests, we learned that the procedure to smoothly activate SG must be adapted to the numerical resolution.
Indeed, when activating SG, the vortex evolution is two fold: higher resolution may lead to undesired splitting, while lower resolution, at long timescales, leads to enhanced decay. 
In fact, an appropriate choice of the time steps and the resolution, at which numerical convergence is reached (see Appendices \ref{subsec:convergency1} and \ref{subsec:convergency2}), can avoid spurious evolutions.
Thus, a new sequence for SG activation at HR has been defined with the following steps:
\begin{equation*}
\begin{array}{lcl}
t=0                 & : & \mbox{Gaussian vortex at LR}    \\
                    &   & \mbox{with } (N_r, N_\theta)=(500, 640),     \\
t=60\,t_0           & : & \mbox{SG is smoothly activated,}             \\
t=120\,t_0          & : & \mbox{Resolution increased to (1800,16000),} \\
120\,t_0<t<140\,t_0 & : & \mbox{HR simulation.}        
\end{array}
\end{equation*} 
The final resolution is equivalent to 146 cells/H and 127 cells/H in the radial and azimuthal directions, respectively.
Such a resolution was chosen so that both directions are almost equally resolved, which is important to correctly account for the SG.
For obvious reasons of computational costs, the last HR-stage was stopped after only 20 orbits; despite this drawback, results are sufficient to distinguish substructures and to guess a possible steady state in the evolution of the main vortex.
The outline for our HR simulations are the following: $r \in [34, 63]$ AU, $h_0$=0.05, $r_0$=50 AU, and $(\chi,\delta)=(14, 0.92)$ at $t=120\,t_0$. 
Furthermore, in order to choose the Toomre parameter  most suited to obtaining self-gravitating vortices, several values of $q_0$ were tested close to the critical limit provided by our stability Criterion \ref{Eq:vortex stability final1}.
This relation leads to values of $q_{0,c}$ equal to $0.74$ and $0.44$ for the isothermal and the non-isothermal cases, respectively.
Finally, we found that the best critical values of the parameters are ($q_{0,c} = 0.7$ , $\sigma_0=3.35$ g.cm$^{-2}$) and ($q_{0,c}=0.425$ , $\sigma_0=5.52$ g.cm$^{-2}$) in the isothermal and the non-isothermal cases, respectively.
For both simulations we get $Q\sim6,$ which means that, in our simulations, the Hill radius of the vortex is resolved at least eight times.

We will now use our HR results to discriminate between vortices evolving in an isothermal or a non-isothermal disc.
We will also use the results to better describe both the internal structure of a SG vortex and its co-orbital environment.

\subsection{Isothermal versus non-isothermal}\label{subsec:Isothermal versus}

In both these two cases, SG is present.
Comparisons are based on Figure \ref{fig: comparison rossby isothermal and no isothermal} (the two upper panels) and on Figure \ref{fig:density, div isothermal and no isothermal}.

\subsubsection{Vortex core}\label{subsubsec:Core region}

In the vortex core, the most striking difference between the isothermal and the non-isothermal cases is the presence of a rotating spiral in the Rossby number and the density distributions, which both evoke Taylor-Green vortices.
This spiral pattern, which barely appears during the LR-stage, tends to strengthen during the HR-stage. 
It is also associated with the emission of weak compression waves from the core that are spiralling outwards.
On the other hand, the outer region that directly surrounds the vortex-core seems to keep a stable elliptical shape.
In contrast, in the non-isothermal case, the vortex keeps a smooth and steady elliptical shape with a nearly constant vorticity (or  $\overline{Ro}\simeq-0.12$).
The same spiral structure is present in this case but is very weak and barely visible, likely due to the near incompressibility of the flow in the vortex core.
These results suggest that stationary SG vortices have to be found in non-isothermal discs.

\subsubsection{Annular bump and waves}\label{subsubsec:spiral waves and annular over-density}

Gas flowing in the horseshoe region of the main vortex is one of the peculiarities of self-gravitating vortices.
This is, indeed, obvious in the two upper panels of Figure \ref{fig: comparison rossby isothermal and no isothermal}.
The flow appears rather laminar in the non-isothermal case, but a more complex evolution is possible in the isothermal case, such as observed in the top panel of Figure \ref{fig: comparison rossby isothermal and no isothermal}.
This snapshot at $t=140\,t_0$ shows six small vortices in the horseshoe region, located at $[50^{\circ}, 90^{\circ}, 95^{\circ}, 135^{\circ}, 250^{\circ},   270^{\circ}]$.
These vortices are emitting small amplitude spiral waves with density contrasts lower than 5\%, which radially  damp in less than 10 AU.
The corresponding evolution sequence, available in the \textbf{online movie 3}, shows that these small vortices are not ejected by the main vortex but are instead formed in the shear of the horseshoe region (this effect amplifies with increasing resolution).
In the non-isothermal case, the flow structure at $r_0=50$ and near $\theta=[95^{\circ}, 265^{\circ}]$ could be misinterpreted as eddies (see Figure \ref{fig:density, div isothermal and no isothermal}).
Indeed, because of the presence of the massive vortex, the trajectories of small scale eddies are expected to be horseshoe.
However, a detailed examination of our data shows that these weak overdensities are stationary in the frame centred on the vortex and don't correspond to vortical structures (see Figure \ref{fig: comparison rossby isothermal and no isothermal}).
The above arguments suggest that these two structures correspond instead to transition regions between the annular bump and the vortex itself.

\subsection{With and without SG} 

This section focuses on vortices in non-isothermal discs and stresses the difference between self-gravitating and gravity-free vortices. Comparisons are based on Figures \ref{fig: comparison rossby isothermal and no isothermal} and \ref{fig:density, div isothermal and no isothermal} (the two lower panels), and on Figures \ref{fig:rho and rossby radial and azimuthal profiles} \& \ref{fig:rossby contour plots}. 

\subsubsection{Vortex}

Vortex shape can be estimated from the distribution in space of the vorticity or the density.
The mapping and the iso-contours of the Rossby number in Figures \ref{fig: comparison rossby isothermal and no isothermal} and \ref{fig:rossby contour plots}, respectively, clearly show that the vortex shape significantly differs when SG is or isn't taken into account. 
Under SG, the vortex is found to contract in the radial direction (by a factor of $\sim 1.5$) and to remain nearly unchanged in the azimuthal direction. 
Self-gravitating vortices appear flatter than the gravity-free vortices, which are only constrained by the shear.
All these findings are consistent with previous works and particularly with the stretching mechanism revealed by \citet{2017MNRAS.471.2204R}.
When comparing the radial and azimuthal profiles of the Rossby number and the density in Figure \ref{fig:rho and rossby radial and azimuthal profiles}, we find, in addition to the abovementioned vortex flattening, a much steeper increase in the vorticity and a lower density peak correlated with a higher annular bump.
When SG is at work, the uniform distribution of vorticity and the strong tightening in the Rossby iso-contours for the external vortex regions suggest that gas compression is more important for self-gravitating vortices, as will be discussed in the next subsection.

Finally, it must be noted that a non-stationary spiral pattern is always present in the density distribution, near the vortex centre.
However, this is hardly perceptible.

\subsubsection{Annular bump and waves}
In the gravity-free case, the vortex is commonly associated with an annular bump in the density and the vorticity.
On the other hand, when SG is taken into account, the vorticity bump splits into the two branches of a horseshoe.
Thus, in this annular region, the flow is either a simple shear with systematic interactions with the vortex, or a complete horseshoe oscillation between a U-turn on each side of the vortex.
This is a situation already observed in the LR simulations of Section \ref{sec:evolution}, where secondary vortices coexist with the massive primary, which is an obvious consequence of three-body interactions. 
Finally, both vortices emit spiral waves with low density contrast ($\lesssim 7\%$).
Interestingly, when SG is absent, the spiral pattern is made of a rarefaction region located between two compression zones, while when SG is present, it is dominated by compression \citep{2001ApJ...551..874L}.
The compression of the gas flow in the spiral waves (right panel of Figure \ref{fig:density, div isothermal and no isothermal}) is five times stronger with SG than without SG, likely due to a stiffer connection between the vortex and the background.

\subsection{Summary on HR results}

Our HR simulations enabled detailed comparisons between SG vortices in the isothermal and the non-isothermal cases, but also between gravity-free and self-gravitating vortices in the non-isothermal case.
Important conclusions are reached concerning the shape, the evolution, and the co-orbital environment of the vortices.
When SG is taken into account: 
(i) vortices in an isothermal-disc tend to distort and may be easily destabilised in a chain of eddies moving along a horseshoe (a lot of waves are spiralling outwards), and the vorticity distribution of the initial vortex presents a rotating spiral pattern in its core;\
(ii) vortices in a non-isothermal disc keep a stable elliptical shape until the end of run, and outside the vortex the flow of gas in the co-orbital region is horseshoe shaped.
In the case of non-isothermal discs:\ 
(i) self-gravitating vortices are flatter than gravity-free vortices (with only a radial contraction), gas compression in spiral waves is stronger (in relation with the vortex distortions), and the vorticity bump splits into the two branches of a horseshoe; 
(ii) if gravity is neglected, the vortex remains in a steady state and is associated with an annular bump in the density and vorticity distributions. 
We want to stress that the flow of gas in this annular region can be part of the vortex solution at steady state (cf. Section \ref{Theoretical}).
The gas compression in the spiral waves is weak.\\

HR simulations have shown that large-scale vortices reach a quasi-steady state in non-isothermal discs, which is not the case in isothermal discs.
The disc temperature seems to play a central role in the dynamical evolution of the vortices, particularly when SG is taken into account.
Finally, this study shows that higher resolutions are needed for correctly and accurately studying SG in vortices, but also to describe small-scale structures, such as the ones observed in the isothermal run of Figure \ref{fig: comparison rossby isothermal and no isothermal}, or in future dust trapping simulations.

\section{Discussion}\label{sec:discussion}

This section comes back to the strategy used in this paper to simulate the evolution of self-gravitating vortices.
We discuss the interest and drawbacks of the method, but also its limitations and the possible improvements. 
Our results are compared to those obtained by \citet{2016MNRAS.458.3918Z}, \citet{2021MNRAS.tmp.3261T}, and \citet{2018MNRAS.478..575L}, particularly where they concern vortex stability.
Finally, we discuss possible observational implications.

\subsection{Single-vortex strategy}

In the absence of SG, previous studies showed the possibility of finding approximate vortex solutions of the stationary Euler equations. When injected in the numerical simulations, these models relax into long-lived and quasi-steady vortices. 
This strategy was successful in the case of non-SG discs \citep{Surville2015} and is also appealing when SG is taken into account.
However, this generalisation requires analytical vortex solutions of equations that include SG.
We didn't find analytical solutions in this case, even in an approximate form, but nonetheless decided to keep the same strategy, with the idea of reaching the solution numerically following a succession of steady states.
This is the reason for the construction of numerical procedures in Sections \ref{sec:evolution} and \ref{sec:high resolution simulations} to get a gradual activation of SG starting from a model of a gravity-free vortex.

\subsubsection{Learning from HR runs}

In fact, the numerical procedures used in Sections \ref{sec:evolution} and \ref{sec:high resolution simulations} manage two different effects: a slow variation of SG as a function of time (which does not correspond to a true physical situation), and an increase in the numerical resolution, which necessarily introduces noise in the computations.
Of course an increase in the numerical resolution also corresponds to a decrease in the numerical viscosity, which at the small sizes scales as $\sim \delta_*^2 \Omega,$ where $\delta_*$ is the mesh size.
Therefore, in the HR case, the relaxation process is longer and often more complex than in the LR case.
For example, in a number of HR runs, vortices are observed splitting into smaller vortices. This clearly stresses that numerical resolution is key in our SG-activation procedure.

\subsubsection{Required time steps for SG activation}

Moreover, following our strategy, HR should not be increased before a quasi-steady state is reached. 
Thus, so that our procedure does not lead to spurious vortex evolutions, a nearly steady state must be reached before any increase in the numerical resolution. In other words, the procedure should satisfy the following sequence:\ (i) activate smoothly SG at LR, (ii) proceed until a quasi-steady state is reached and (iii) increase resolution.

\subsection{Instability-generated vortices}

Another way to study the evolution of self-gravitating vortices in PPDs is to assume that they are first formed following one of the numerous instabilities known to produce large-scale vortices, such as RWI \citep{2013MNRAS.429..529L, 2018MNRAS.478..575L} or the baroclinic instabilities \citep{2003ApJ...582..869K,2004ApJ...606.1070K}.
The constraint, in this case, is that the outline of the computations is imposed by the specific conditions necessary for the growth of the instability. 
Thus, the vortex evolution is model-dependant, which is not the case in the single-vortex approach developed above.

\subsection{Comparisons with other works}

We now compare our results with those obtained in three different papers.
These are \citep{2021MNRAS.tmp.3261T}, \citep{2018MNRAS.479.4878P} and \citet{2016MNRAS.458.3918Z}.

\subsubsection{Comparison with \citet{2021MNRAS.tmp.3261T}}

The authors performed a similar study to our own using a wider sample (1980 different runs).
In contrast to our work, their outline is different since vortices are generated by RWI at the edge of a magnetically dead zone in a viscous $\alpha$-disc model with $\alpha=(10^{-4}, 10^{-5}),$ and their simulations had LR with $(N_r,N_{\theta})=(256, 512)$.
In spite of the differences, mainly due to a stronger viscosity (turbulent and numerical, both), the results they found are quite similar to ours with the discrimination of various classes: vortex survival, vortex splitting at various scales, and ring formation.

Following Figure 5 of their paper, we notice that vortex splitting occurs for $Q\simeq 5$; however, to check the relevance or not of the stability criterion reported in our Equation \ref{Eq:vortex stability final2}, the corresponding value of $Ro$ is also required.
This value is not provided by the authors, but it can be estimated in a simple way using the approximate dependence between $Ro$ and the aspect ratio \citep{survi2013}.
For elongations between five and 20, we have $0.1 \lesssim |Ro| \lesssim 0.2$ and we get $0.6 \lesssim 1.2\, |\overline{Ro}| \,Q \lesssim 1.2,$ which is consistent with our criterion in Equation \ref{Eq:vortex stability final1}.

\subsubsection{Comparison with \citet{2018MNRAS.479.4878P}}
\begin{table}
\caption{Simulation parameters for isothermal runs in \citet{2018MNRAS.478..575L} at the 500$^{th}$ orbit. \\
Here isothermal refers to the convention used in the aforementioned article.} 
\label{tab:pierens}                            
\centering                                     
\begin{tabular}{c c c | c}                     
\hline                                         
$f$ & $Q$ & $|Ro|$ & $1.2 \,|Ro| \, Q$ \\  
\hline                                         
1   & 8   & 0.12 & 1.15  \\  
2   & 4   & 0.10 & 0.48  \\
\hline
\end{tabular}
\end{table}
As in the previous paper, the authors study Rossby vortices in a disc with SG. 
These vortices are generated with or without $\beta$-cooling at the frontier of the dead-zone, where the alpha-viscosity jumps from $10^{-4}$ inside the dead zone to $10^{-2}$ in the magnetically active region.
Their simulations were performed with a radial and an azimuthal resolution at the vortex position equal to 16 cells/H and 10 cells/H, respectively. The obtained vortices are either similar to our SG vortices, $-0.13\leq Ro \leq -0.10$ and $5\leq\chi\leq7$, or very different with $Ro=-0.05$ and $\chi\geq25$. 
These very weak and elongated structures are likely due to the artificial $\alpha$-viscosity and the low numerical resolution.
Care should be paid to the fact that the authors named theirs studies without and with $\beta$-cooling and thermal radiation as isothermal and non-isothermal,  respectively.
For meaningful comparisons and possible testing of Criterion \ref{Eq:vortex stability final2}, we focus on their isothermal case and excluded the cases $f=[4,8]$ because in our simulations, we never encountered vortices with such low Rossby numbers and such high elongations ($Ro=-0.05$, $\chi\geq25$).

The estimated vortex parameters and the values of $|Ro|$ and $Q$ are gathered in Table \ref{tab:pierens} at the 500$^{th}$ orbit, which is the time at which RWI saturated and formed a unique vortex.
In this table $f$ is the density scaling factor used by the authors, and the last column shows that our criterion is also comforted by this work, since the two vortices satisfying $1.2 \, |Ro| \, Q \lesssim 1$ are decaying.

\subsubsection{Comparison with \citet{2016MNRAS.458.3918Z}}
\begin{table}
\caption{Our SG parameter, $q_0$, corresponding to the density factor, $\Sigma_0$, of \citet{2016MNRAS.458.3918Z}} 
\label{tab:zhu}                            
\centering                                     
\begin{tabular}{c c}                     
\hline                                         
$\Sigma_0$ & $q_0$\\  
\hline                                         
0.0002  & 15.7   \\  
0.002   & 1.6    \\
0.005   & 0.63   \\
0.01    & 0.31   \\
\hline
\end{tabular}
\end{table}

The authors studied the effect of SG on Rossby vortices but also considered the effect of the indirect gravitational potential of the $\{$star+disc$\}$ system. 
While SG alone tends to delay RWI and inhibit low $m$ modes, they found that the indirect potential tends to strengthen vortices. 
Such conclusions were also found by \citet{2017MNRAS.471.2204R}.

In their Figure 3 (bottom panels), they report the results of various runs (g5gi, g0p2gi, g2gi and g10gi), which can be fruitfully compared to ours.
In all these runs, no splitting or vortex destruction is observed; this suggests, following our criterion, that $q_0$ lies above the threshold value $q_{0,c}$.
The morphological and SG parameters necessary to test the criterion can be found in Figure 3 of their paper and in our Table \ref{tab:zhu};
thus, we get $(\chi,\delta)=(6,0.85)$ and with our Figure \ref{fig: vortices stability surface}, we estimated that $q_{0,c} \leq 0.375$\footnote{The map of Figure \ref{fig: vortices stability surface} is valid only in the range $0.8\lesssim \delta\lesssim 3$}.
The results gathered in our Table \ref{tab:zhu} are found to be consistent with our criterion ($q_0 \geq 0.375 \geq q_{0,c}$), except in the last line where the inequality is indefinite.

\subsubsection{Comparison summary}

Equations \ref{Eq:vortex stability final1} and \ref{Eq:vortex stability final2} were compared to the three independent works and exhibit satisfactory results.
It is worth noticing that when more physical parameters are taken into account, there should exist a more general stability criterion under the form: $f(Q,|Ro|,h, Re,\tau_c)\geq1$, where $Re$ is the Reynolds number and $\tau_c$ is the beta-cooling time normalised with respect to the Keplerian period.
Finding the general function $f$ could be a very difficult task, if not impossible, but as a first step, it could be really interesting if researchers possessing a large simulation sample and more physical ingredients, such as \citet{2021MNRAS.tmp.3261T}, checked if a simple stability criterion involving power laws, $Q^a |Ro|^b Re^c \tau_c^d h^k \geq 1$, can succeed.

\subsection{Theoretical models for steady vortices in SG-free discs} \label{Theoretical}

Simple vortex solutions of the incompressible Euler equations at steady state have been studied by a number of authors \citep[hereafter GNG]{1987MNRAS.225..695G}, \citep{Kida1981, Chavanis2000}. 
GNG's model provides a rough solution of the Euler and continuity equations for a polytropic\footnote{$P=A\sigma^{1+{1}/{n}}$} sheared flow.
It also indicates that, for a vortex patch of constant vorticity, the pressure and density contours should be ellipses with equal aspect ratios.
More generally, a stationary, incompressible, barotropic\footnote{$P=P(\sigma)$} vortex solution in a sheared flow should satisfy the requirement that the surface density, pressure, and vorticity are functions of the stream function, which indirectly implies that these three quantities must have the same aspect ratios.

Nonetheless, the nearly steady state reached after the relaxation of the Gaussian model in the present work (non-isothermal and gravity-free run studied in Section \ref{sec:high resolution simulations}) and in \citet{Surville2015} suggests the existence of an analytical solution of the steady, compressible Euler equations in the case of non-barotropic fluids.
Such a solution should satisfy the following requirements:
(i) Distributions of pressure, density, and vorticity, which can differ from one another. For instance, we found different aspect ratios for these quantities in the gravity-free runs (i.e. $\chi_{Ro}=9$, $\chi_\sigma=7.7,$ and $\chi_P=7.3$, respectively), 
(ii) an analytical way to connect the vortex and the annular bump, 
(iii) a steady spiral pattern that can account for the vortex wake and 
(iv) compressibility in the vortex core.
This is a difficult task and the necessary developments to improve GNG's model are out of the scope of the present paper.

\subsection{Possible observational consequences}

Although the capture of dust and its concentration in vortices has not been addressed in the present paper, we think that some of our results could already provide interesting clues to test the presence of vortices in PPDs or infer disc properties from the survival of SG vortices.
More details on observational signatures of vortices can be found in \citep{2012MNRAS.419.1701R,2020A&A...641A.128R} and the references therein.

\subsubsection{An estimation for HD163296 surface density}\label{subsec:observational implications}

Around this Herbig Ae star of mass $2.3 M_{\odot}$, a dust clump orbiting at a distance of $\sim 0.3$ AU (period $\sim 40$ d) was recently detected by the VLTI/MATISSE and GRAVITY instruments \citep{2014Msngr.157....5L,2017A&A...602A..94G, 2021A&A...647A..56V}.
The authors tested different possible observational scenarios that could account for this feature, such as a stellar or a sub-stellar companion, and an inner disc misalignment with the line of sight.
Following their conclusions, the most likely mechanism to explain such a dynamic asymmetry is growing dust grains being captured by a gaseous vortex. 
Indeed, the proximity of the star could produce axisymmetrical step jumps in the density and the pressure that induce a local extrema in the generalised potential vortensity distribution 
\citep{2000ApJ...533.1023L};
this is known to be favourable to RWI and the generation of a large-scale vortex.
The presence of a vortex seems to be supported by the shadow variations revealed by the Hubble Space Telescope \citep{2020ApJ...902....4R}, since vortices are also pressure bumps whose vertical extent easily exceeds the mean scale height.\\

The results we get in this paper have potential implications on the disc density that  observational predictions could infer.
If a large-scale vortex is definitely present in the HD163296 disc and at $r_0=0.3$ AU from the star, it should also be stable against SG forces;
thus, following our criterion, we should have $q_0 \geq q_{0,c}$.
Of course, we are aware that, in the case of HD163296, the physical context is at the limit of validity of the method used in the present paper, but we have found it interesting to test this idea nonetheless.
For this test the numerical outline is the following: an isothermal  gas at $T\sim 1400 K$ with a similar molecular weight as H$_\mathrm{2}$ and a vortex located at $r_0=0.3$ AU from the star with $\chi\sim 7$ and $\delta \sim 1$.
Thanks to previous assumptions, we were able to estimate the disc flaring equal to $h=0.04$ and $Q_c=25$.
Then, Figure \ref{fig: vortices stability surface} provides the threshold value $q_{0,c}=0.375,$ which fixes an upper limit to the density:
\begin{equation*}
\sigma_{0, \mathrm{max}}(0.3 \mbox{AU})=\frac{c_s \Omega_K}{0.375 \cdot 4 \pi G Q_c} = 62 \cdot 10^{3} \mbox{ g.cm}^{-2}.
\end{equation*}  
Finally, if a vortex is really present, the surface density of the disc should not be larger than $62 \cdot 10^{3} \mbox{ g.cm}^{-2}$ at 0.3 AU to avoid being destroyed by SG splitting.
This is, of course, a very crude approach to the problem, but our goal was only to stress that new constraints can emerge when SG is taken into account.

\subsubsection{Vortex-induced spiral waves}

In the absence of SG, vortices are known to excite spiral waves that cause them to migrate inwards towards the star.
These waves, in contrast to the spiral density waves produced in disc or planet interactions, originate from the compression and expansion of the gas layers sandwiched between the vortex core and the background flow.
Following \citet{2019ApJ...883L..39H}, the density contrast of these waves (lower than 20\%) is much too weak to be observed in scattered light.

On the other hand, when SG is taken into account, the mass of the vortex also contributes to the wave emission with the excitation of spiral density waves.
Of course, the more massive the vortex, the higher the density contrast of the waves.
In fact, the maximum mass a vortex can reach is limited by its stability against SG, which is constrained according to Criteria \ref{Eq:vortex stability final1} or \ref{Eq:vortex stability final2}. 
Anyway, it can be concluded that spiral waves are always much weaker if excited by a gaseous vortex than if excited by a planet.
The density contrast of the waves associated with the vortex remains very weak (see Figures \ref{fig:density, div isothermal and no isothermal} and \ref{fig:extent box}), and the only way to obtain stronger waves is to have large amount of solid material trapped in the core of the vortex, mimicking the effect of a planet.

\subsubsection{Annular substructures}

Vortices in PPDs are known to easily capture solid particles with optimal Stokes number ($St \sim 1$) and to confine them in the core \citep{Barge1994,1995A&A...295L...1B,1996Icar..121..158T,1999PhFl...11.2280B}.
As the vortices are always associated with an annular bump (see Figure \ref{fig:density, div isothermal and no isothermal}), we have found it interesting to explore the possible evolution of the solid particles in this annular region during the trapping-in-vortex stage.
Indeed, dust concentrations in this region could be observational markers of the vortices.
As noted in Section \ref{sec:high resolution simulations}, this annular region could be a simple annular pressure and density bump in a nearly Keplerian shear, for weak SG, or a more complex region with a horseshoe flow and possible substructures, for  intermediate SG. 
Since pressure bumps are possible sinks and vortical substructures are potential traps, we expect that solid particles with $St\leq 1$ should form a ring strewed with various clumps. These clumps may have horseshoe motions.
Finally, these annular bumps, which can trap a non-negligible amount of solid particles, can be considered as a tank of intermediate-sized particles, which possibly `feed' the vortex during its evolution.

More precisely, since the density amplitudes between the vortex core and the ring differ, we expect that a segregation with respect to dust size should occur, which implies that dust asymmetries and rings located at the same radial position should be observable at different wavelengths in scattered light.
Here, we conjecture that this mechanism of gap cleaning and the ring distribution of dust induced by vortices are due to the small amplitude overdensity ring, which always goes along a vortex.
This is out of the scope of current paper, but would be an interesting work to explore and to be compared with observations.

\section{Conclusions and perspectives}

In this paper we reported the evolution, in an inviscid disc, of vortices under the effect of SG thanks to HR numerical simulations.
As an initial state, vortices were injected considering a Gaussian model \citep{Surville2015} and relaxed by numerical viscosity.
Our study shows that, in the presence of SG, vortices evolve following three possible regimes:
\begin{itemize}
  \item \textbf{Shear dominated:}\\
vortices are modelled by the shear but stay nearly unchanged with respect to the non-SG case,
  \item\textbf{Self-gravitating vortices:}\\
vortices stretch out, contract radially, and have a core spinning slightly faster. The main vortex eventually releases small secondary vortices.
  \item\textbf{Splitting cascade:}\\
the initial vortex is gradually destroyed and, after a few hundred orbits, spreads into a residual annular bump that eventually contains small-scale vortex remnants; this can occur even if Toomre's parameter $Q_0 \sim 10$.
\end{itemize}

From these simulations it seems that for self-gravitating vortices, vortex strength is not necessarily proportional to the aspect ratio inverse. 
This should be confirmed by other independent works.
The other conclusions we get concern vortex stability, particularly in regard to numerical resolution and disc temperature.
The most interesting points are the following:
\begin{itemize}
\item \textbf{Approximate stability criterion:}\\
from an extended numerical study of vortices submitted to their own gravity, and after testing with simple analytical arguments, we propose a new criterion, which states that in inviscid discs, vortices resist SG destabilising effects if $1.2 \,|Ro|_{max} Q \gtrsim 1,$ or again if $q_0 \gtrsim q_{0c}$. 
This criterion is supported by comparison with the results of three independent studies \citep{2021MNRAS.tmp.3261T, 2016MNRAS.458.3918Z, 2018MNRAS.479.4878P}.
In addition, the aforementioned criteria were also applied to a possible vortex candidate in HD163296 disc \citep{2021A&A...647A..56V} which allowed an estimation to be made of the upper limit for the gas density.
\item\textbf{Numerical resolution}\\
In the absence of viscosity, we found that simulations with at least $(71,24)$ cells/$H$ in the radial and azimuthal directions, respectively, are needed to accurately describe vortices.
If such a requirement is not fulfilled, vortices lose 15\% of their mass during 250\ orbits (see Appendix \ref{subsec:convergency1}) and this trend is accentuated in the presence of SG.
On the other hand, simulations with $(N_r, N_\theta)=(1800, 16000)$, have shown that numerical resolution is key to correctly and finely describe the shape and evolution of SG vortices. 
Nearly uniform resolution, in $r$ and $\theta$, is necessary and simple estimates show that simulations should resolve distances of the order of $\sim{H}/{(3Q)}$.
\item\textbf{Disc temperature}\\ 
Comparisons between the HR simulations of Sections \ref{sec:high resolution simulations} clearly show that SG vortices in non-isothermal discs: (i) evolve more slowly than vortices in isothermal discs, and (ii) keep a smooth elliptical shape throughout the run.
On the other hand, vortices in isothermal discs are distorted by the SG forces and vortical substructures are generated in the horseshoe region; they also excite more  compressive waves than in the case of non-isothermal discs.
The key point here is that a temperature gradient of the disc could be necessary for SG vortices to reach a nearly steady-state.
\end{itemize}

\citet{2016MNRAS.458.3918Z}, \citet{2017MNRAS.471.2204R} and \citet{2021MNRAS.tmp.3261T} found that vortices cannot form in massive discs since lower mode numbers are inhibited by SG.
Although the strategy in this paper was different, our results are consistent with their findings: not only can vortices not form in massive discs, but even if they are well established (as the initial state in our simulations), they cannot be sustained.
Thus, if observed crescent-shaped asymmetries are indeed vortex presence tracks, these vortices should be hosted in non-massive discs.
However, in light of self-gravitating vortices stability (as defined in this work), the definition of non-massive disc should be clarified with respect to the other physical ingredients governing the dynamics, such as viscosity or thermal cooling.
In a roundabout way, we provided an attempt of this definition for inviscid discs through the unique Equation \ref{Eq:vortex stability final1} (or Equation \ref{Eq:vortex stability final2}).

Finally, the results of this paper also provide useful input to prepare dust-trapping simulations in permanent self-gravitating vortices. 
For example, these simulations should be carried out in non-isothermal discs in which $q_0 \geq 1.2 \, q_{0,c}$ minimises the destabilising effects of SG, in anticipation of their possible enhancement by the dust back-reaction;
but also a $q_0$ value relatively close to $q_{0,c}$ would imply non-negligible effects from SG in the gas and dusty phases.

\begin{acknowledgements}
We thank the referee, Zsolt Reg{\'a}ly, for his helpful comments which improved the quality of this work.
We would like to warmly acknowledge Stéphane Le Dizès for fruitfull discussions during the preparation of the paper and also for his help to find funding of part of the project.
This research has made use of computing facilities operated by CeSAM data center at LAM, Marseille, France.
The Centre de Calcul Intensif d’Aix-Marseille is acknowledged for granting access to its high performance computing resources. 
This work was granted access to the HPC resources of CINES/IDRIS under the allocation 2020-A0090411982 made by GENCI.
This work was granted access to the HPC/AI resources of CINES under the allocation 2018-A0050407407 made by GENCI.

\end{acknowledgements}

\bibliographystyle{aa}
\bibliography{bibliography}

\begin{appendix}

\section{Full simulation box and emitted spiral}

\begin{figure}[h]
\centering
\includegraphics[trim = 0.8cm 0cm 1.5cm 0.75cm, clip, width=0.98\hsize]{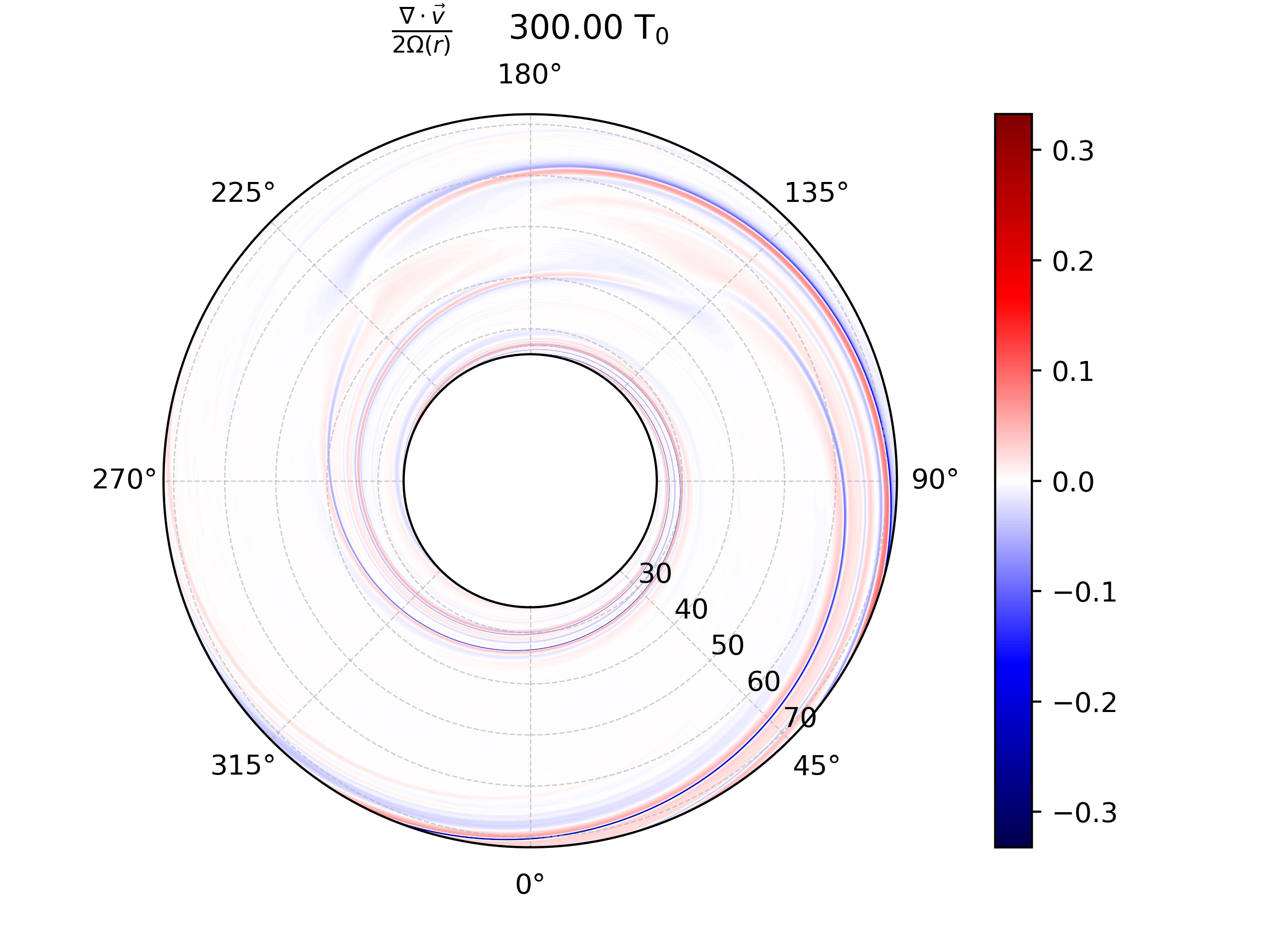}
\caption{\textbf{Compressibility waves, $\nabla \cdot \vec{v}' / \left(2 \Omega_K\right)$, emitted by a quasi-steady vortex in its full simulation box, $r\in[25,72]$ AU, as used in Section \ref{sec:evolution}.}
}
\label{fig:extent box}
\end{figure}

Figure \ref{fig:extent box} presents the divergence of the velocity field normalised with respect to the gas rotation rate in order to show the full extent of the simulation box used in Section \ref{sec:evolution}.

%
%

\section{Stable and unstable vortex structures with respect to Toomre's parameter}\label{app:stable and unstable vortex structures}

\begin{figure}[h]
\centering
\includegraphics[trim = 4.0cm 0.1cm 0.5cm 3.0cm, clip, width=8cm]{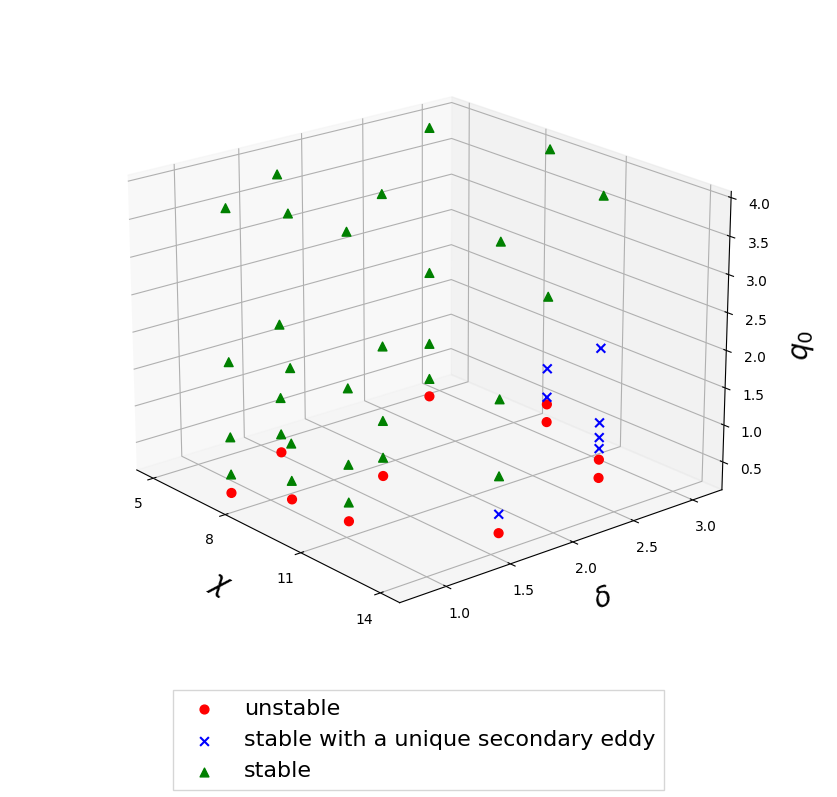}

\caption{\textbf{Vortex stability in the ($\chi$, $\delta$, $q_0$) space.} Each point is an intermediate resolution simulation : $(N_r, N_\theta)=(1500, 3000)$.
$\chi$ and $\delta$ are, respectively, the vortex aspect ratio and the radial extent at $t=50\,t_0$.}    
\label{app: vortices stability dots}
\end{figure}

This section refers to the definition of stability provided in Sect. \ref{subsec:numerical study}.
Figure \ref{app: vortices stability dots} illustrates for each triplet ($\chi$, $\delta$, $q_0$) the vortex stability with respect to SG. In order to improve results,
three additional simulations were performed for the following triplets: $(\chi,\delta,q_0)=[(12.6,2.7,0.8), (12.6,2.7,0.65), (9.00, 3.06,0.6)]$. 
The blue surface of Figure \ref{fig: vortices stability surface} (left panel) was obtained estimating the transition zone between stable and unstable vortices. Below are the prescriptions used to estimate, for a given couple ($\chi$, $\delta$), the transition, in the $q_0$ direction, from the stable to the unstable region.
\begin{enumerate}
\item[$\bullet$] No blue cross in the $q_0$ direction indicates that the transition between stable and unstable vortices is located in the mid-value between the lowest green triangle and the upper red dot.

\item[$\bullet$] A blue cross in the $q_0$ direction indicates that the transition between stable and unstable vortices is the lowest blue cross.
\end{enumerate}

\section{Computed quantities}\label{app: section - computed quantities}

This section is devoted to introducing in detail the way in which we computed different quantities, such as the
distances, the vortex mass, the mean Rossby number, and the vortex geometrical parameters.
We begin describing the way in which we estimated the vortex position.

\subsection{Main vortex position and distance to the star}\label{app:computation distance to the star}

We found that even in presence of secondary vortices the position of the main vortex is given by the pressure maximum.
For the above reason, the distance to the star and the angular position of the vortex are computed by tracking the position of the maximum of pressure contrast $C_P=\frac{P}{P_0}-1$.
We denote the vortex angular position by $\theta_{max}$.

\subsection{Mass of the vortices}\label{app:mass computation}

\begin{figure}
\centering
\includegraphics[trim = 1.8cm 1cm 2.2cm 1.5cm, clip, width=8cm]{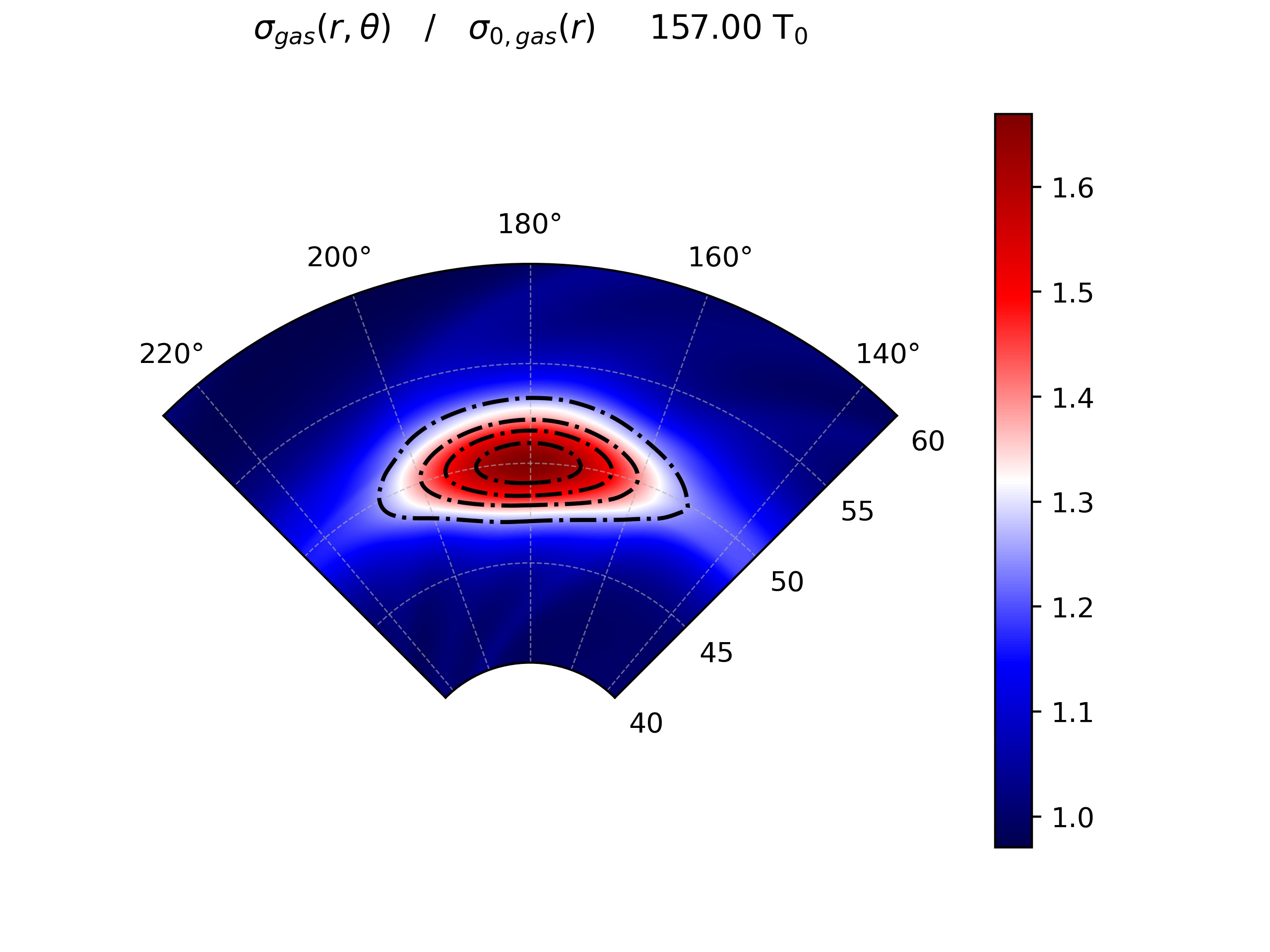}
\caption{\textbf{Mass of the vortex core with respect to the $\mu$ threshold at $t=158\,t_0$.}\\
$\mu+1=[ 1.25, 1.40, 1.50, 1.60]$ (from the exterior to the vortex core).
The respective vortex-core masses are 0.18, 0.1, 0.06, and 0.02 (in $M_J$).
The initial vortex parameters are: $(\chi,\delta)=(14,1.5)$ ($t=0$) and $q_0=0.5$.}
\label{fig: mass with respect to density threshold}
\end{figure}

For comparison of the vortices studied in Section \ref{sec:evolution}, it was decided to define the core `mass' of a vortex as the region where the density contrast is higher than a given threshold $\mu$.
In order to choose appropriately this threshold we plotted in Figure \ref{fig: mass with respect to density threshold} contours for different values of $\mu$ and provided the associated vortex masses in the caption.
Of course, the higher the threshold, the smaller the captured mass that leads to the lowest (preferred) value for $\mu$.
However, a lower threshold, such as $\mu=[0.05, 0.15]$, is not convenient since the numerical filter captures the annular overdensity and a small share of the spiral waves emitted by the vortex (see top panels of Figure \ref{fig:filter 1 and 2}).
In order to overcome this issue, we decided to set the threshold parameter at a value of $\mu=1.25-1=0.25$.
Yet even with this clip, another problem arises because of the presence of secondary vortices (see cases $q_0=[0.25, 0.5]$ in Section \ref{sec:evolution} or the vortex hosted in an isothermal disc in Section \ref{sec:high resolution simulations}).
We get rid of them by applying a second filter that extracts only the region located at a maximal angular position of 45$^{\circ}$ from the pressure maximum.
The bottom panels of Figure \ref{fig:filter 1 and 2} show how the density is filtered without (left) and with (right) this second filter. 
Despite our best efforts, the reader should keep in mind that this second filter is not optimal when secondary vortices enter the region $|\theta-\theta_{max}|\leq45^{\circ}$.
In summary, the mass of a vortex core is computed numerically in our code as:
\begin{equation*}
m_{\mathrm{core}}=\int\limits_{\tau} \sigma(r,\theta) \, r \, dr \, d\theta
\end{equation*}
where $\tau=\Big\{ (r,\theta) \in \mathbb{R}^+ \times \left[0,2\pi\right[  \, : \, C_\sigma \geq 0.25 \, \mbox{and} \, |\theta-\theta_{max}|\leq45^{\circ} \Big\}$. 

\subsection{Mean Rossby number}\label{app:rossby computation}

Similarly to the previous subsection, we applied a clip to the Rossby number in order to extract the `vortical' core region: $Ro \leq -0.8$.
The mean Rossby number is computed numerically in the extracted region:
\begin{equation*}
\overline{Ro}= \left( \int\limits_{\Gamma} Ro \, r \, dr \, d\theta \right) / S, 
\end{equation*}
where $S=\int\limits_{\Gamma}  \, r \, dr \, d\theta$ is the total surface of the extracted region and $\Gamma=\Big\{ (r,\theta) \in \mathbb{R}^+ \times \left[0,2\pi\right[  \, : \, Ro \leq -0.8  \, \mbox{and} \, |\theta-\theta_{max}|\leq45^{\circ}  \Big\}$.

\subsection{Aspect ratios and radial width}\label{app:computation aspect ratios}
\begin{figure}
\centering
\includegraphics[width=\hsize]{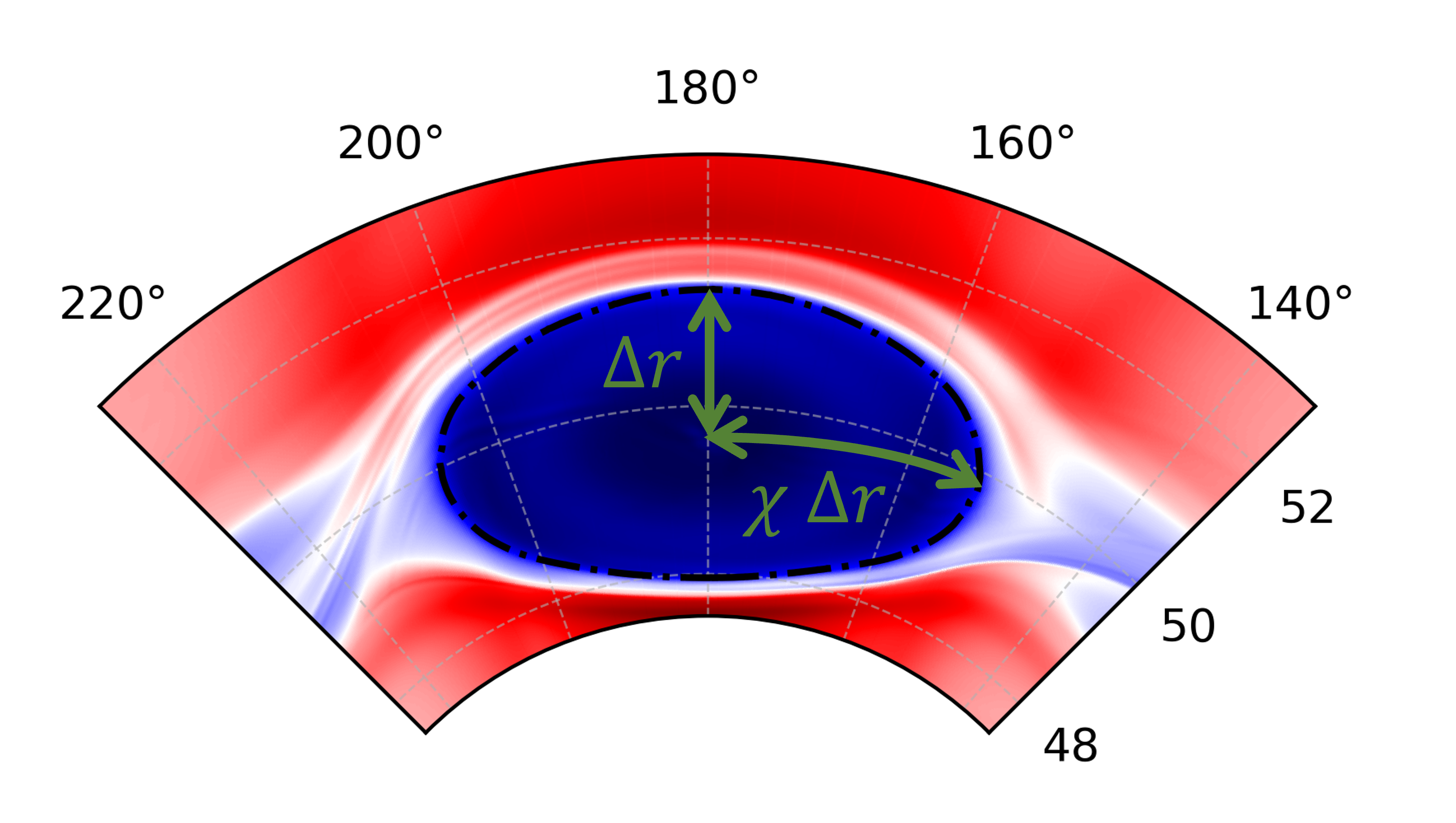}
\caption{\textbf{Definition of aspect ratio, $\chi$, and radial extent, $\Delta r$, fitting the $Ro=-0.8$ iso-contours.} \\
In  the $(r,\theta)$ coordinate system, Rossby iso-contours are ellipses, while in the Cartesian coordinate system, ellipses are curved.}
\label{fig: definiton aspect ratio and radial extent}
\end{figure}

Aspect ratios were computed thanks to the python routine Least Squares fitting of ellipses \citep{ben_hammel_2020_3723294}, which provides the ellipse parameters (semi-major and semi-minor axes, inclination) which better fits a given set of points. 
Since this technique is based on the least square method, intrinsically it cannot provide uncertainties, but an estimation was performed in Appendix \ref{app:section uncertainties} thanks to the grid resolution.
Figure \ref{fig: definiton aspect ratio and radial extent} shows how contour plots were performed for the Rossby number, how these contours were fitted with the best ellipse, and the definitions of radial width, $\Delta r$, and aspect ratio, $\chi$.

However, when vortices start to split in many substructures, it is not possible to discriminate the main vortex using Rossby number contours.
Therefore, we applied the same filter used in Sects. \ref{app:mass computation} and \ref{app:rossby computation} in order to isolate the main vortex.
We decided to use Rossby iso-contours, for tracking geometrical parameters, since the extensive study achieved in Section \ref{sec:vortex stability} showed that the only common quantity shared by all 45 vortices was the Ro=-0.08 contours, while density and the pressure maximum amplitude vary for all different structures, which makes comparison a difficult task.

\begin{figure*}
\centering
\resizebox{\hsize}{!}
          {\includegraphics[trim = 0.8cm 0cm 2cm 0.1cm, clip]{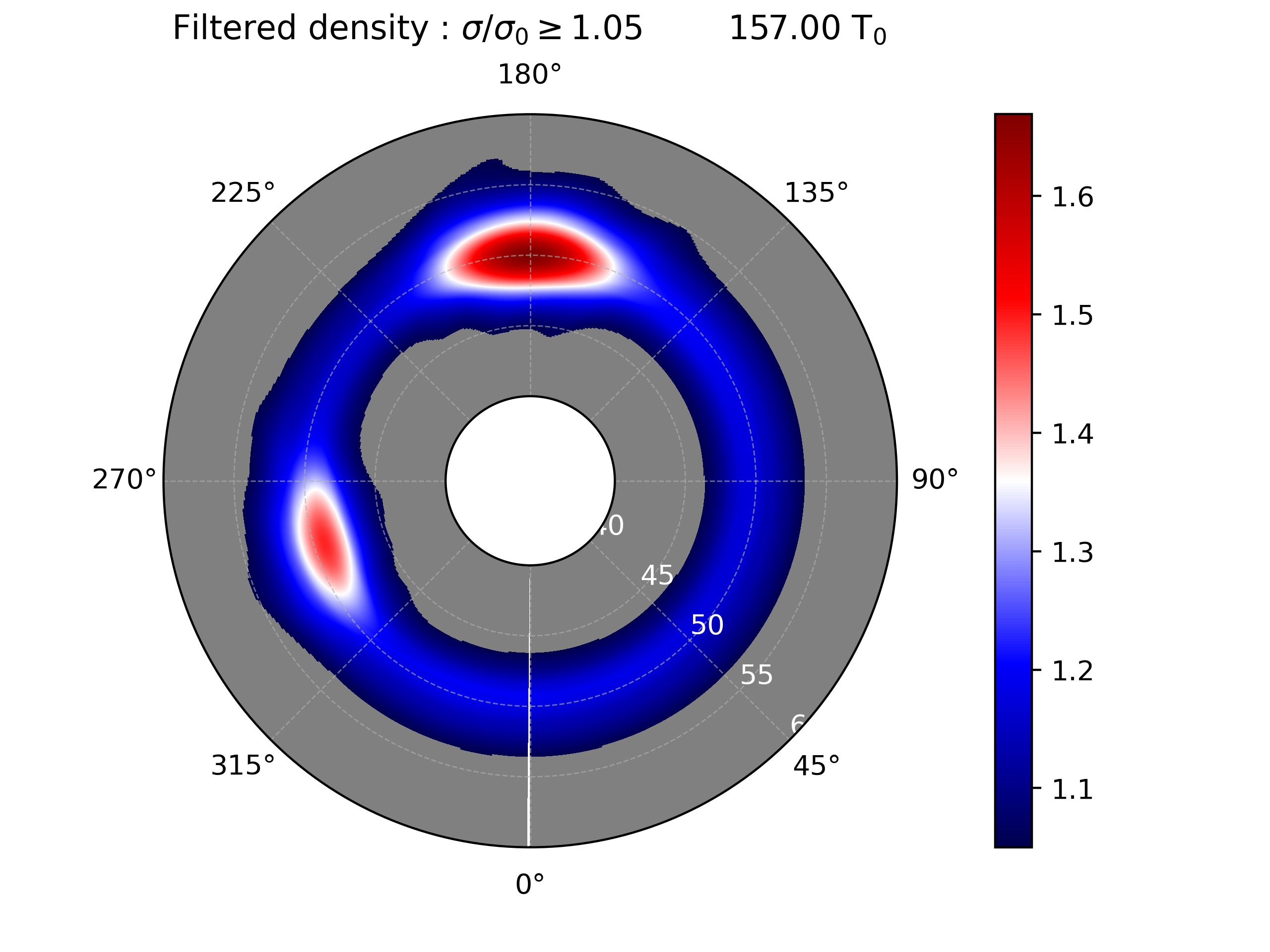}
           \includegraphics[trim = 0.8cm 0cm 2cm 0.1cm, clip]{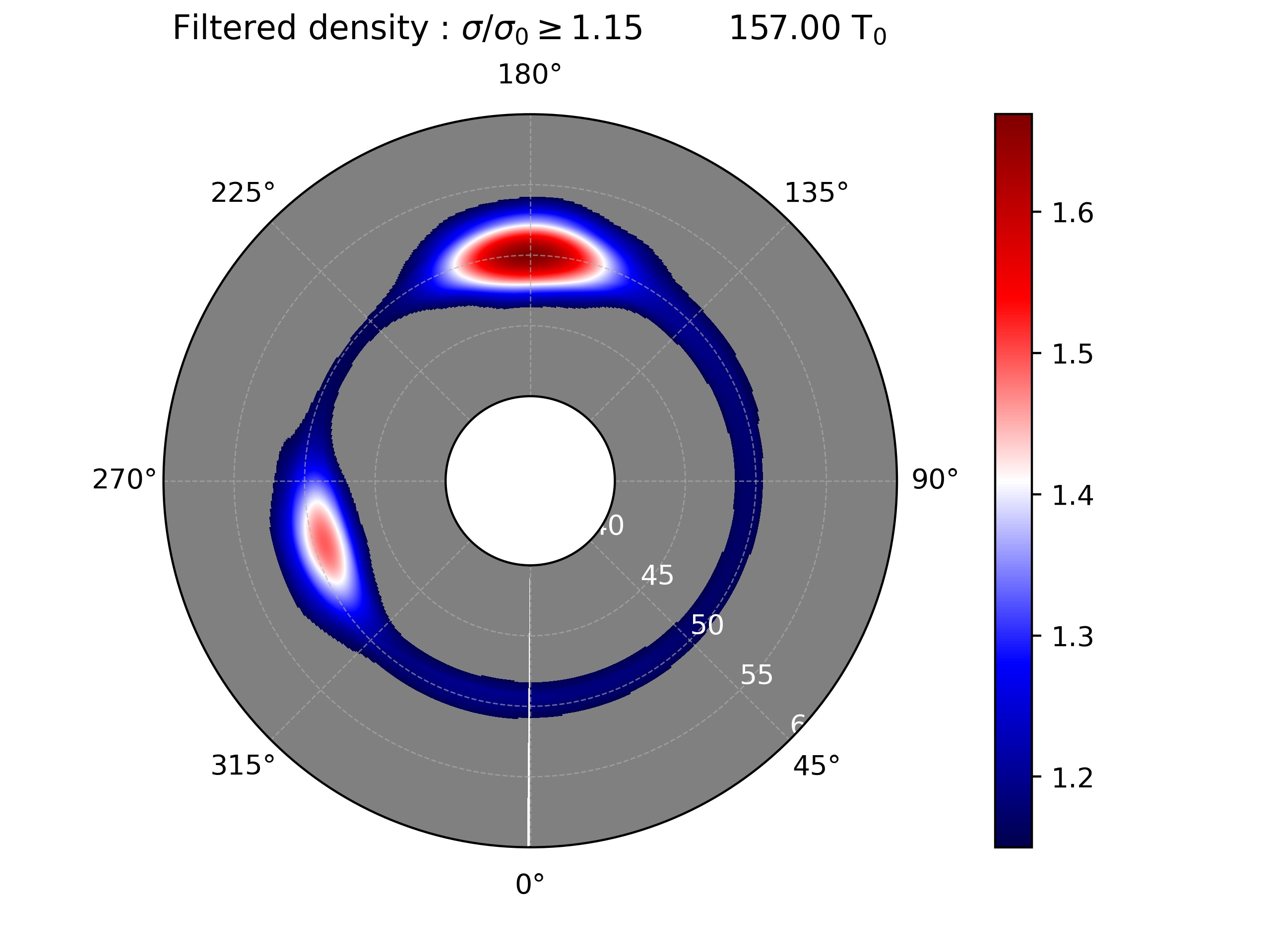}
          }
\resizebox{\hsize}{!}
          {\includegraphics[trim = 0.8cm 0cm 2cm 0.1cm, clip]{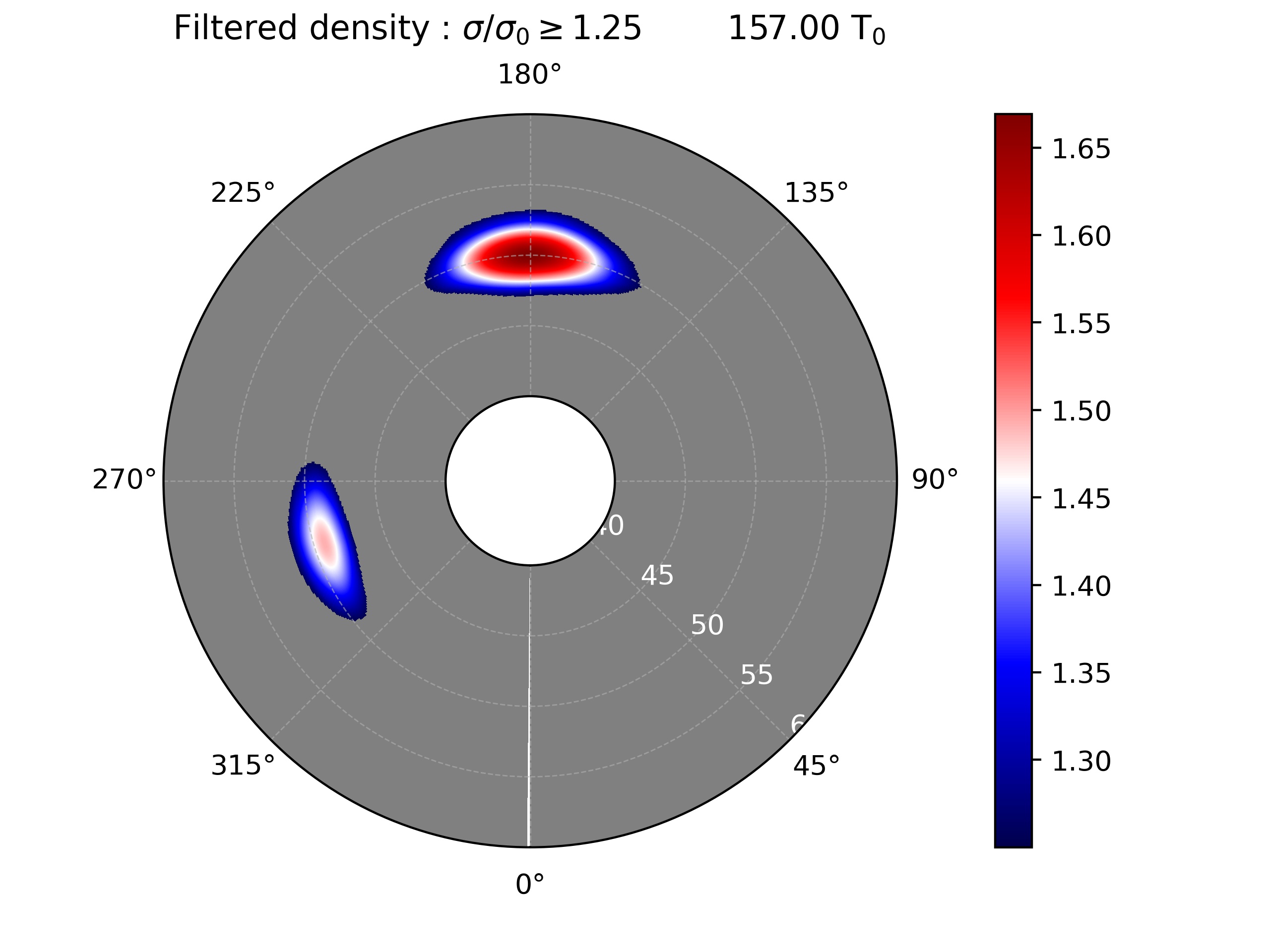}
           \includegraphics[trim = 0.8cm 0cm 2cm 0.1cm, clip]{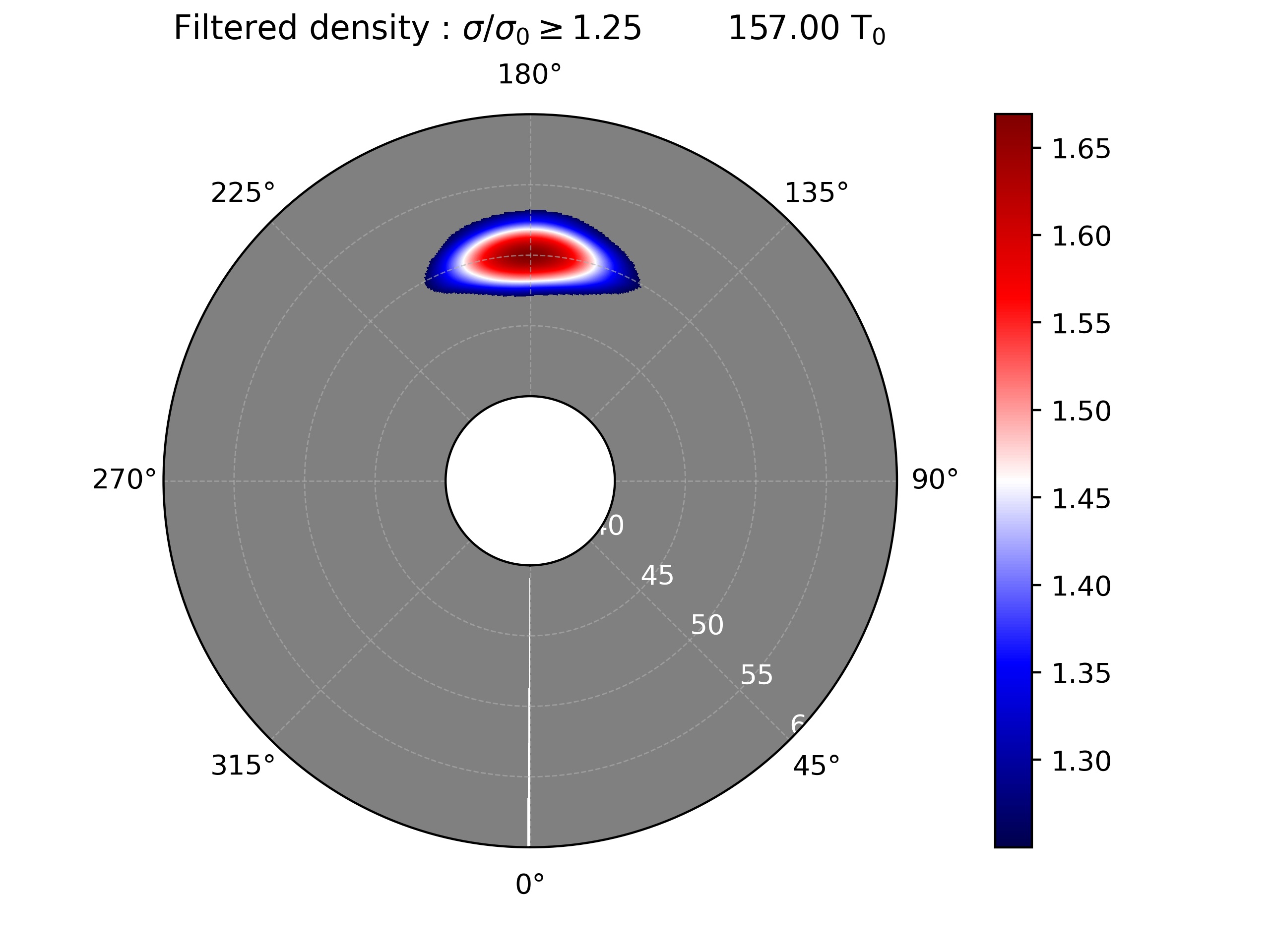}
          }
\caption{\textbf{Vortex masses with respect to different density thresholds and filters.} \\
\emph{Top} and \emph{Bottom left} panels: We applied only a clip with respect to the density threshold. \\
\emph{Bottom right} panel: We applied an additional filter that extracts only the region located at a maximal angular position of 45$^{\circ}$ from the pressure maximum (see Appendix \ref{app:mass computation}).\\
\emph{Top left: }$\mu+1=1.05$, $m_{core}=1.69 M_J$. \emph{Top right: }$\mu+1=1.15$, $m_{core}=0.78 M_J$ \\
\emph{Bottom left: }$\mu+1=1.25$, $m_{core}=0.28 M_J$. \emph{Bottom right: }$\mu+1=1.25$, $m_{core}=0.18 M_J$} 
\label{fig:filter 1 and 2}
\end{figure*}

\section{Uncertainties}\label{app:section uncertainties}

In numerical experiments the accuracy arises from the used numerical method.
In particularly, for a finite volume method, it depends on the time step and on the approach used for computing fluxes between cells.
Uncertainties on the computed quantities arising from finite volume method are difficult to quantify, which encouraged us to consider that uncertainties of computed quantities from raw data arise only from the grid resolution. 

\begin{enumerate}
\item[$\bullet$] \textbf{Distances:} Uncertainties related to distances (distance to the star and vortex radial width) are given by the maximum value between the resolution in the radial or azimuthal direction: $\delta_* = \mathrm{Max}(\delta_r, r \delta_\theta)$.
\item[$\bullet$] \textbf{Aspect ratios:} The aspect ratio of the fitted ellipse (see Appendix \ref{app:computation aspect ratios}) is defined as $\chi=\frac{a}{b}$, where a and b are, respectively, the semi-major and semi-minor axes. 
Thus, the uncertainty is : $\Delta \chi = \frac{\chi}{b} \left( \frac{\Delta a}{\chi} + \Delta b \right)=\frac{\delta_*}{b} \left( 1 + \chi \right)$.
\end{enumerate}

\section{Numerical convergency}

\begin{table}
\caption{Resolution for numerical convergency tests}    
\label{tab:numerical convergency1}                     
\centering                                              
\begin{tabular}{c || c c || c c c}  
\textbf{Name} & \multicolumn{2}{c}{Sections \ref{sec:evolution} and \ref{sec:vortex stability}}  &    \multicolumn{2}{c}{Section \ref{sec:high resolution simulations}} \\              
\hline
\hline                                                 
          & $(N_r, N_\theta)$    & cells/$H$   & $(N_r, N_\theta)$    & cells/$H$ \\                  
\hline                                         
\emph{Res1} & (240, 320)           &   (11, 2.6) & (1500, 3000)  & (122, 24)  \\  
\emph{Res2} & (500, 720)           &   (24, 5)   & (1400, 11520) & (113, 92)  \\ 
\emph{Res3} & (750, 1536)          &   (34, 12)  & (1800, 16000) & (146, 127) \\  
\emph{Res4} & (1500, 3000)         &   (71, 24)  & (2200, 19840) & (178, 158) \\  
\hline
\end{tabular}
\end{table}

\begin{figure}
\centering
\includegraphics[width=\hsize]{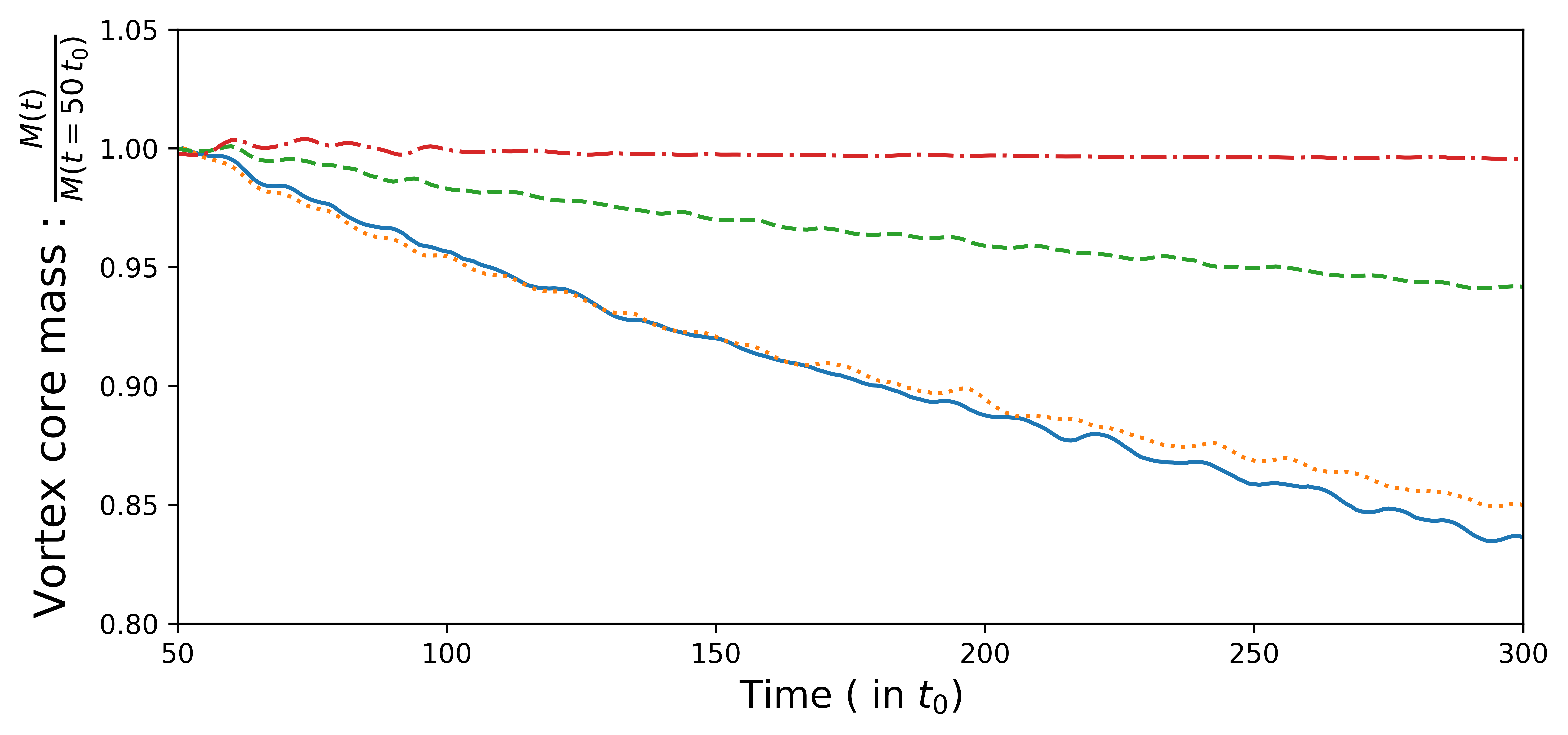}
\includegraphics[width=\hsize]{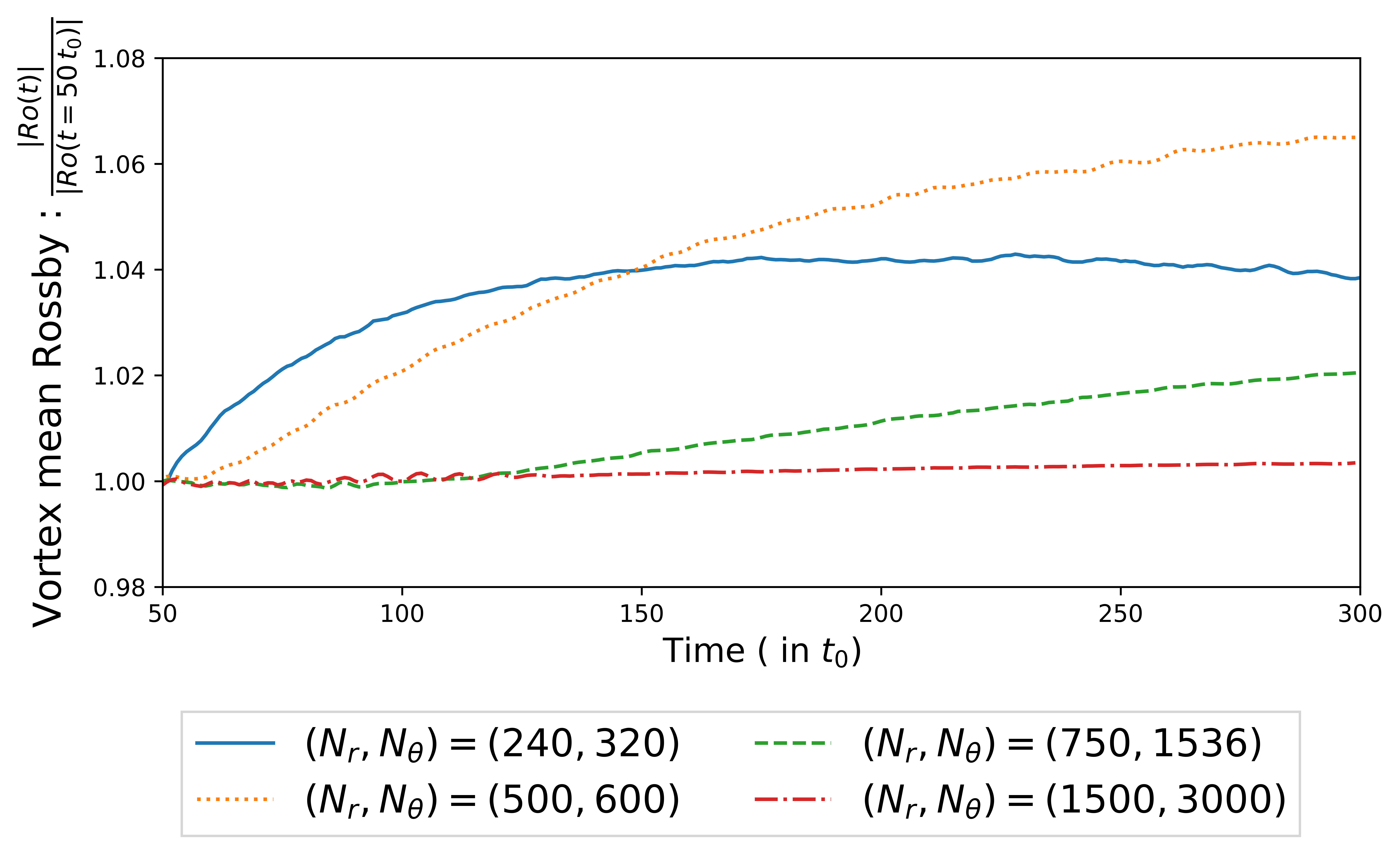}
\caption{\textbf{Numerical convergency for Sections \ref{sec:evolution} and \ref{sec:vortex stability}}\\
\emph{Top:} Convergence of the vortex mass.\\
\emph{Bottom:} Convergence of the Rossby number.\\
All quantities were normalised with their respective value at $t=50\,t_0$.}
\label{fig:numerical convergency 1}
\end{figure}

\begin{figure}
\centering
\includegraphics[width=\hsize]{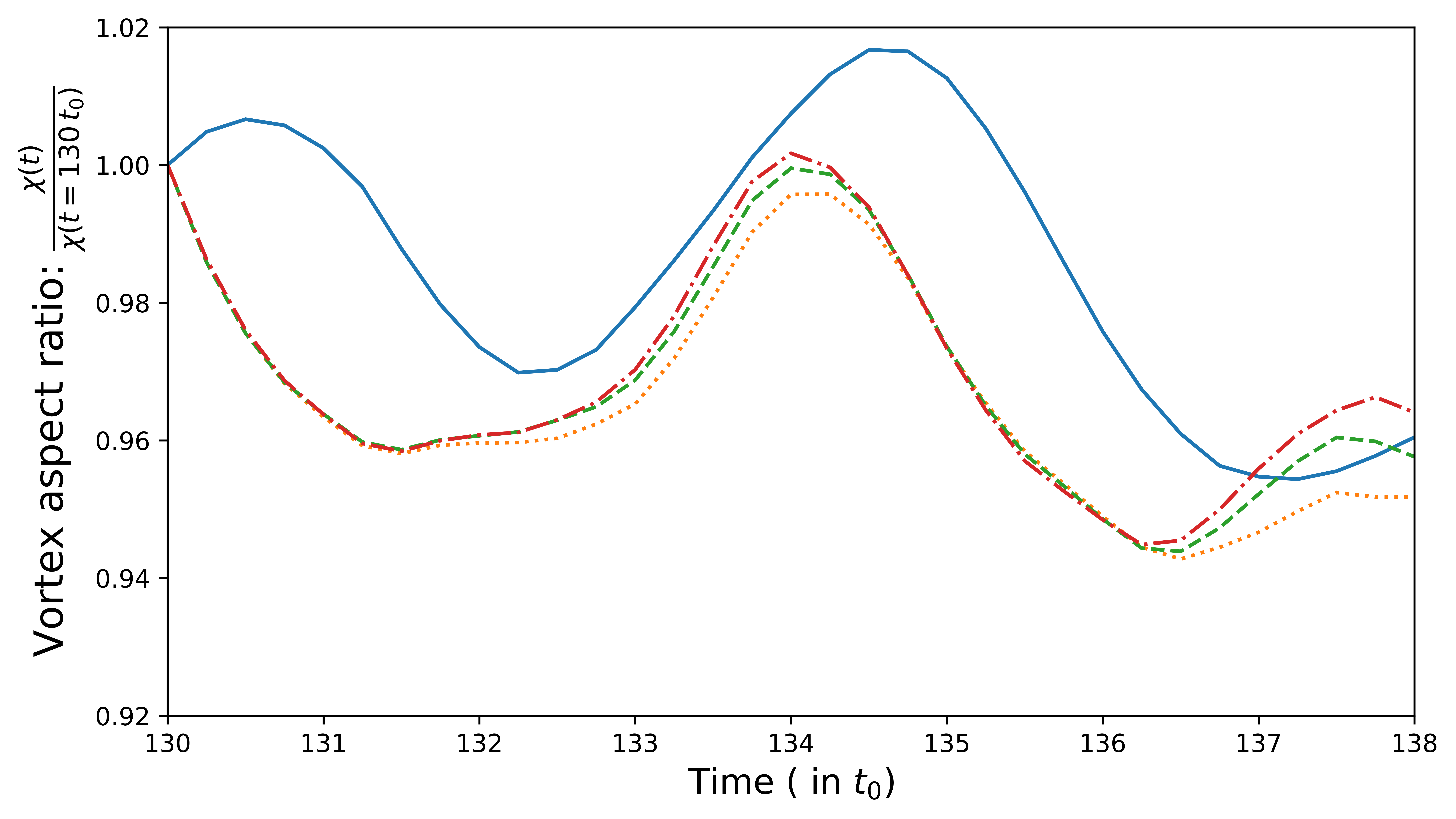}
\includegraphics[width=\hsize]{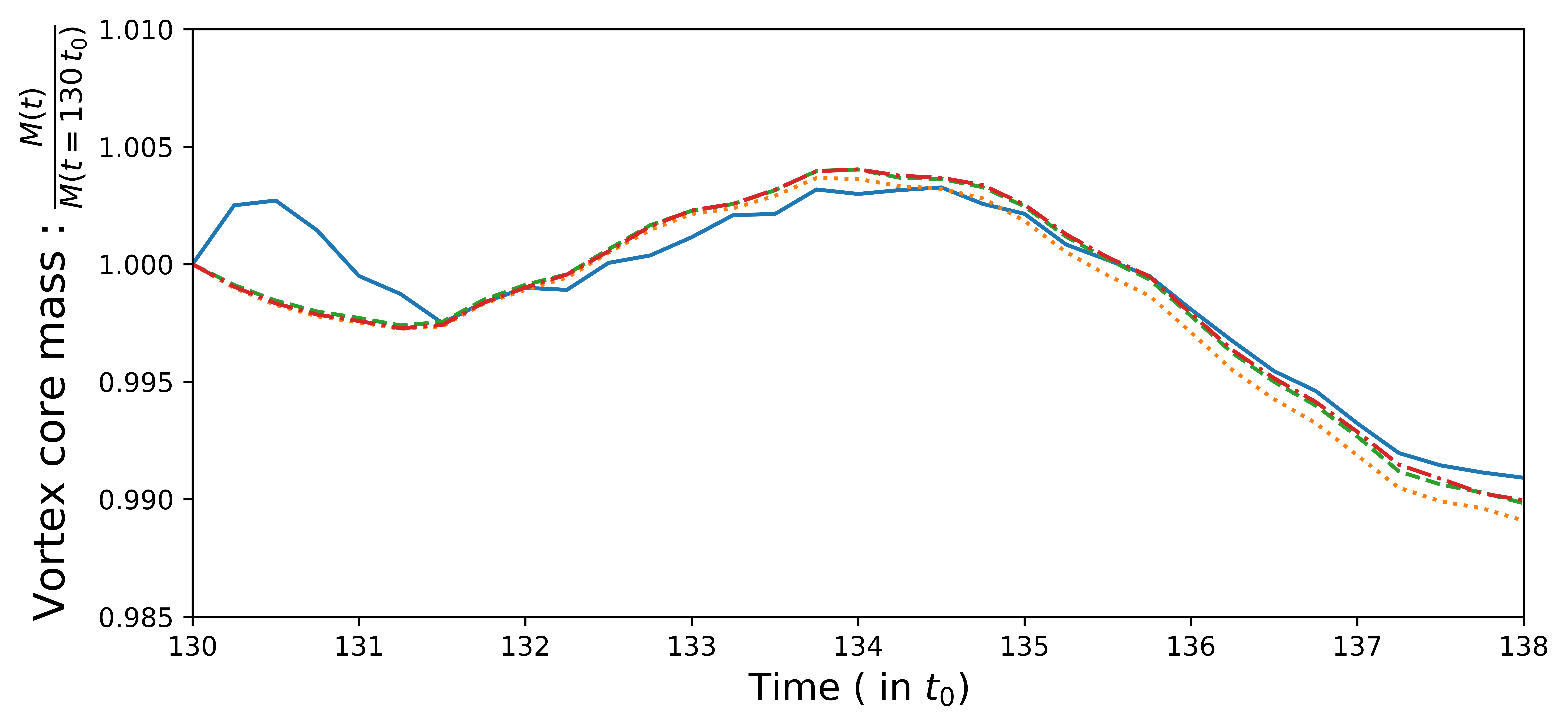}
\includegraphics[width=\hsize]{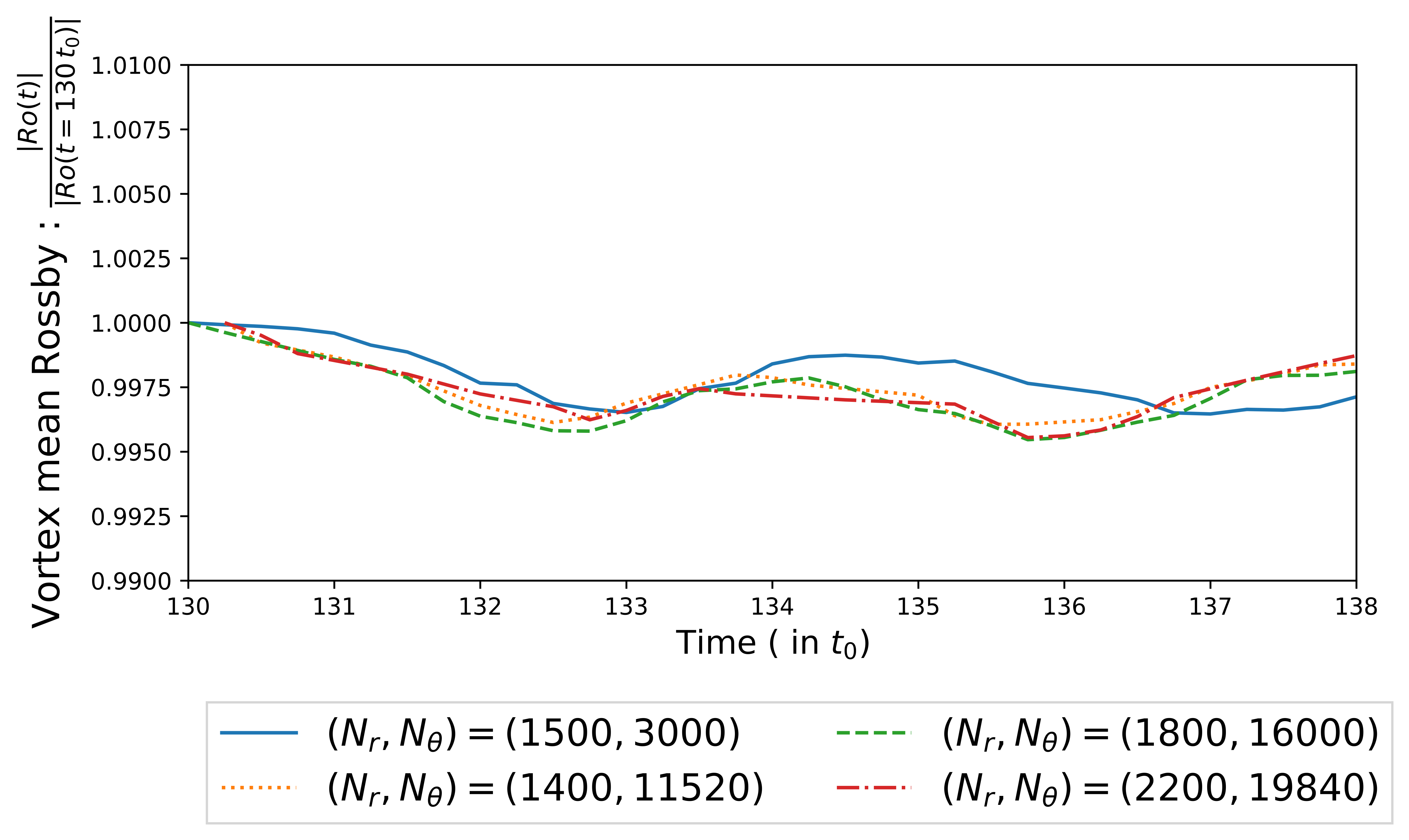}
\caption{\textbf{Numerical convergency for Section \ref{sec:high resolution simulations}}\\
\emph{Top:} Vortex mass evolution.
\emph{Middle:} Rossby number evolution.
\emph{Bottom:} Aspect ratio evolution.\\
All quantities were normalised with their respective value at $t=130\,t_0$.}
\label{fig:numerical convergency 2}
\end{figure}

Our simulations were tested against numerical convergency.
We gathered in Table \ref{tab:numerical convergency1} the resolutions we used for the tests related to Sections \ref{sec:evolution}, \ref{sec:vortex stability} and \ref{sec:high resolution simulations}.

\subsection{Comments for Sections \ref{sec:evolution} and \ref{sec:vortex stability}}\label{subsec:convergency1}

As an example, we show in Figure \ref{fig:numerical convergency 1} the evolution of the mass and the Rossby number for a vortex without SG for four different resolutions (case $q_0=\infty$ of Section \ref{sec:evolution}).
We observe that, for the two lowest resolutions, there is a similar mass loss of 15\% in only 250 orbits and a mean Rossby number increase of $\sim6$\%.
If only looking at \emph{Res1} and \emph{Res2,} it seems that simulations are in the numerical convergent regime. However, for \emph{Res3} there is less mass decay (only ~7\%) and the mean Rossby number evolution is the same as \emph{Res4} up to the 115$^{th}$ orbit. 
After that the Rossby number increases.
The situation is completely different for the highest resolution, \emph{Res4}, where there is almost no mass loss and the Rossby number remains stationary.
Even if not shown here, we highlight that when SG is included, vortices decay is stronger compared to simulations without SG, at LR.
The above mass losses, with and without SG, are due to a density contrast decrease, while the vortex parameters don't change.

We conclude that: \textbf{(1)} working only with the lowest resolutions can be misleading, and we could get the impression that the numerical scheme is convergent when it is not; and \textbf{(2)} our highest resolution for Section \ref{sec:evolution} is in the convergent regime and leads to a quasi-stationary vortex.
Based on these conclusions we suggest that numerical simulations on vortices should resolve at least $\sim70$ and $25$ times the pressure scale height in the radial and azimuthal directions, respectively. 
If this condition is not satisfied, spurious decay of the vortices is likely.

\subsection{Comments for Section \ref{sec:high resolution simulations}}\label{subsec:convergency2}

We show in Figure \ref{fig:numerical convergency 2} the evolution of the aspect ratio, the mass and the Rossby number for the self-gravitating vortex, hosted in a non-isothermal disc and studied in Section \ref{sec:high resolution simulations}.
Compared to the previous subsection, we show an additional plot exhibiting the Rossby-number aspect ratio at the $Ro=-0.01$ level.
This allows us to check how the vortex-core morphology is affected when resolution exceeds the threshold $R_H=H/3Q$. Convergence tests were performed, but were strongly time-consuming;
they were carried out only during eight orbital periods. Starting from the 130$^{th}$ orbit as an initial state, we performed a 2D cubic interpolation to get the desired resolution.
We highlight that for \emph{Res1,} the distance $H/3Q$ is well resolved only in the radial direction, while in the azimuthal direction it is resolved only 1.3 times.
For the three other, resolutions above quantity is resolved at least five times in both directions.
We observed that, for all resolutions, there is no notable difference in the vortex mass and in the evolution of the mean Rossby number.
\emph{Res2}, \emph{Res3,} and \emph{Res4} share the same aspect ratio evolution, with a small divergence from the 136$^{th}$ orbit.
The situation is different for \emph{Res1}; even if the tendency is the same as for HR simulations, we notice from the beginning that there is a 2\% difference.
This suggests that vortex-core morphology is affected by SG when small scales are resolved.
In conclusion: \textbf{(1)} vortex core morphology is affected when $H/3Q$ is well resolved in both directions; and \textbf{(2)} our simulations fit the convergent regime.

\section{Potential and forces due to SG}

\begin{figure*}
\centering
\begin{minipage}{\hsize}
\resizebox{\hsize}{!}
{\includegraphics[trim = 0.8cm 0cm 1.7cm 0.75cm, clip, width=0.98\hsize]{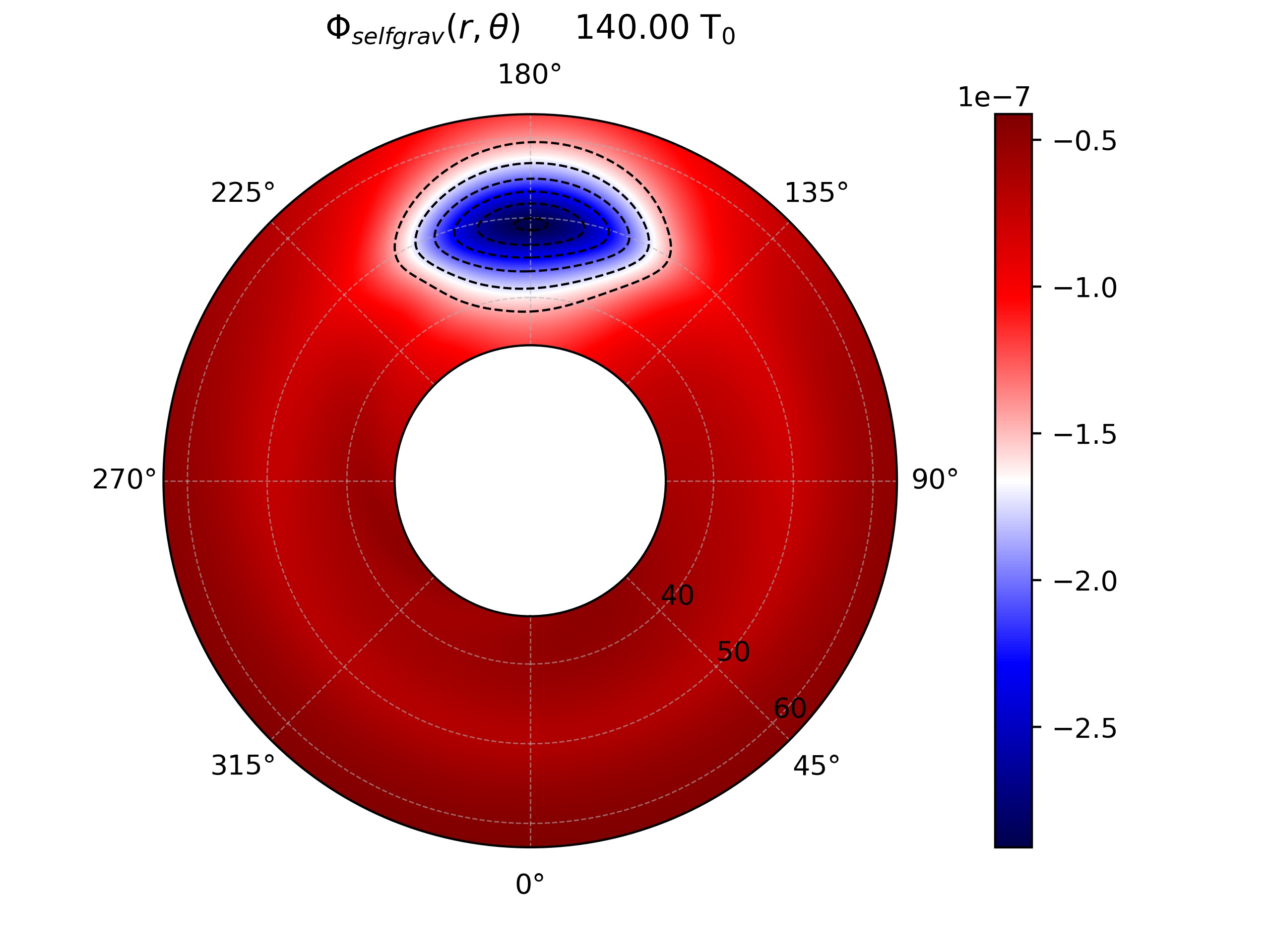}
\includegraphics[trim = 0.8cm 0cm 1.7cm 0.75cm, clip, width=0.98\hsize]{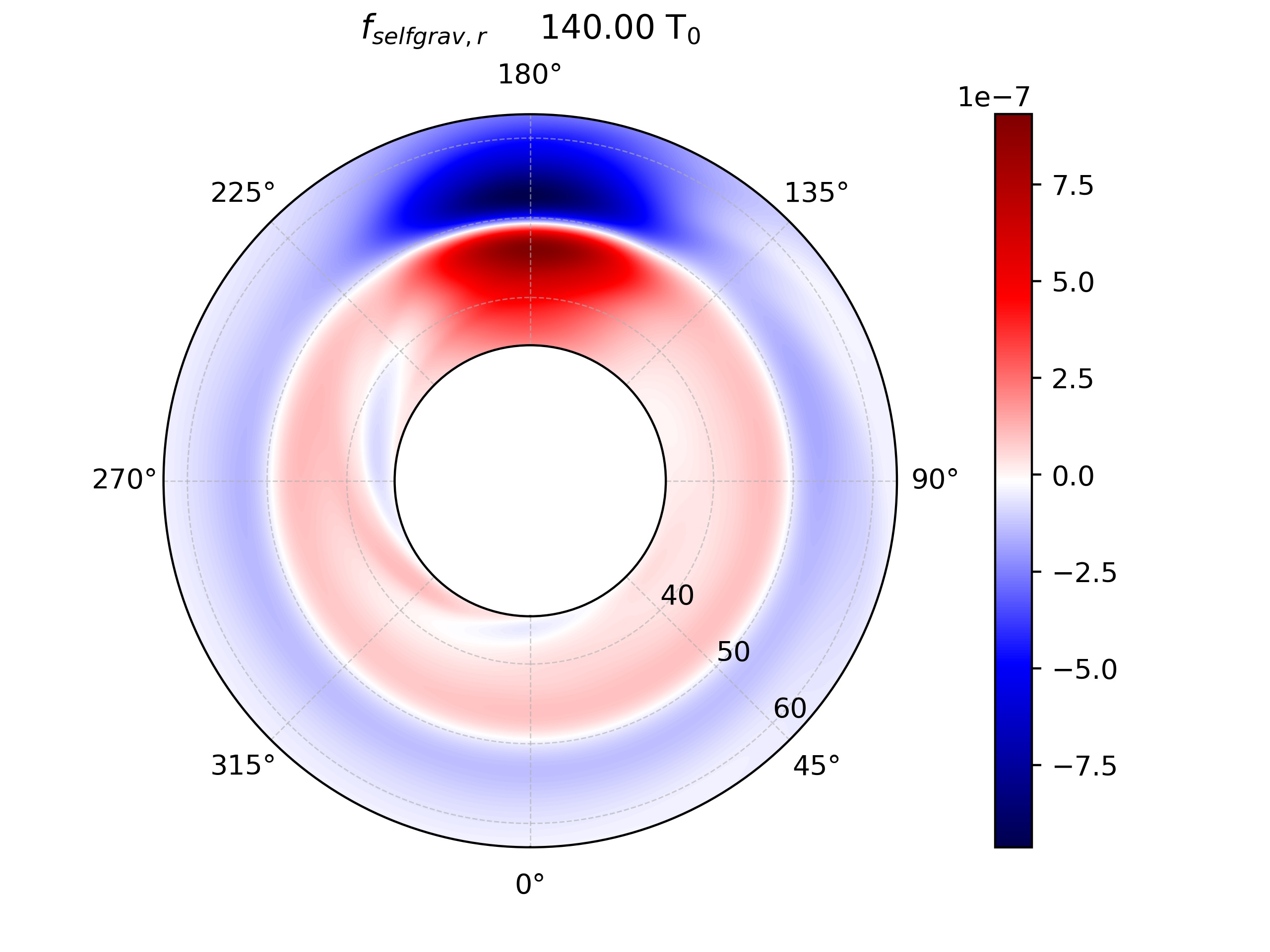}
\includegraphics[trim = 0.8cm 0cm 1.7cm 0.75cm, clip, width=0.98\hsize]{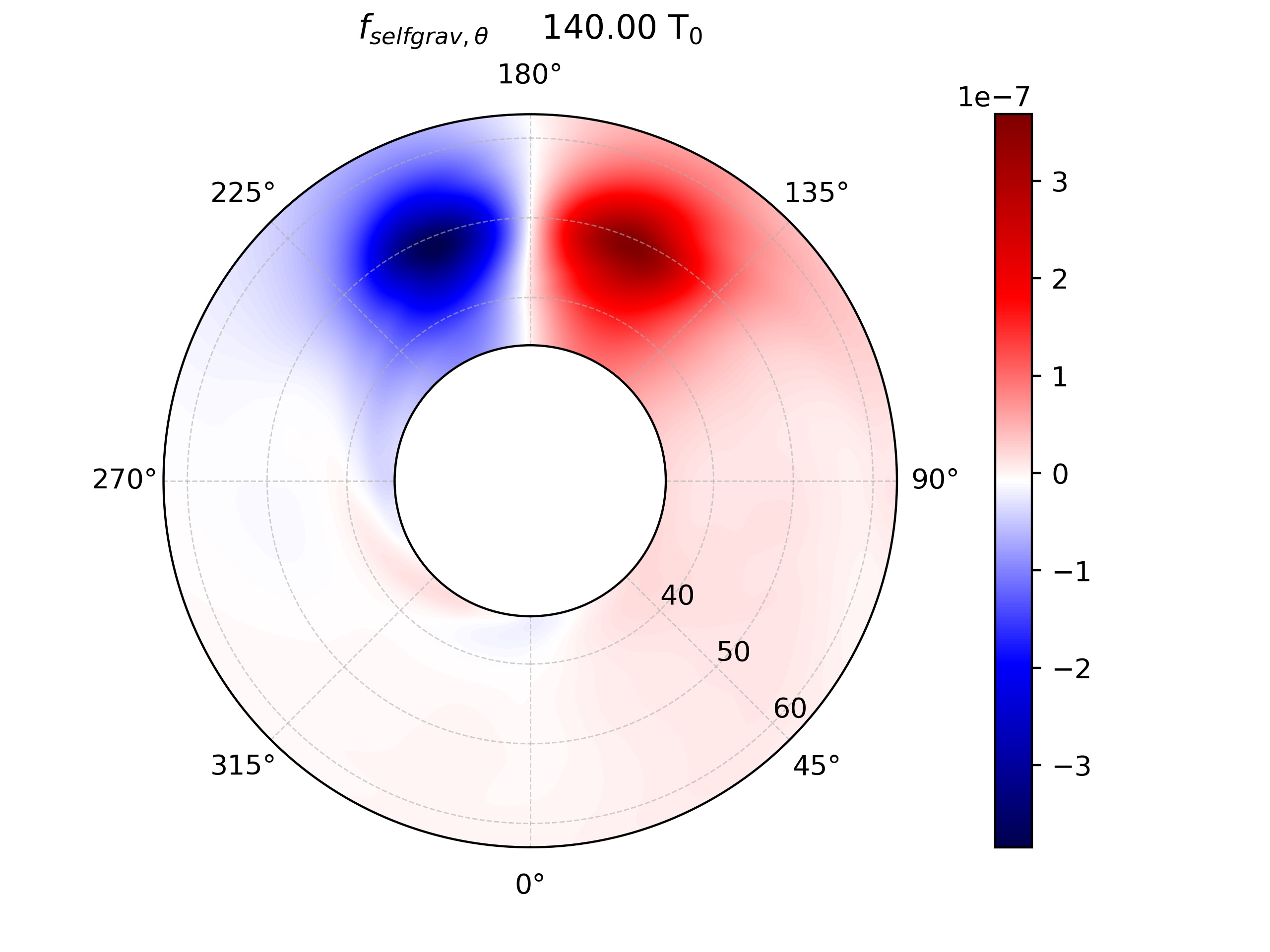}
}
\end{minipage}
\caption{\textbf{SG: potential and forces at $t=140\,t_0$.}
\emph{Left:} SG potential ($\Phi_{SG}$) and iso-contours. \emph{Middle:} SG radial force ($f_{r,SG}$). \emph{Right:} SG azimuthal force ($f_{\theta, SG}$).}
\label{fig:SG potential and forces}
\end{figure*}

Figure \ref{fig:SG potential and forces} shows, at $t=140 t_0$, the spatial distribution of the potential and the force components of SG for the non-isothermal run of Section \ref{sec:high resolution simulations}.

%
%

\end{appendix}

\end{document}